\documentclass[longauth,twocolumns]{aa}  
\usepackage{graphicx}
\usepackage{txfonts}
\usepackage{color}
\usepackage{footnote}
\makesavenoteenv{tabular}
\usepackage{natbib,twoopt}
\usepackage[breaklinks=true]{hyperref} 
\bibpunct{(}{)}{;}{a}{}{,} 
\makeatletter
\newcommandtwoopt{\citeads}[3][][]{\href{http://adsabs.harvard.edu/abs/#3}%
{\def\hyper@linkstart##1##2{}%
\let\hyper@linkend\@empty\citealp[#1][#2]{#3}}}
\newcommandtwoopt{\citepads}[3][][]{\href{http://adsabs.harvard.edu/abs/#3}%
{\def\hyper@linkstart##1##2{}%
\let\hyper@linkend\@empty\citep[#1][#2]{#3}}}
\newcommandtwoopt{\citetads}[3][][]{\href{http://adsabs.harvard.edu/abs/#3}%
{\def\hyper@linkstart##1##2{}%
\let\hyper@linkend\@empty\citet[#1][#2]{#3}}}
\newcommandtwoopt{\citeyearads}[3][][]%
{\href{http://adsabs.harvard.edu/abs/#3}
{\def\hyper@linkstart##1##2{}%
\let\hyper@linkend\@empty\citeyear[#1][#2]{#3}}}
\makeatother 

\begin{document} 

   \title{VLTI-MATISSE $L$- and $N$-band aperture-synthesis imaging of the unclassified B[e] star FS Canis Majoris
   \thanks{Based on observations collected at the European Organisation for Astronomical Research in the Southern Hemisphere within the commissioning of the VLTI-MATISSE instrument (ID 60.A-9257(E)).   }    
  }

\author{K.-H.~Hofmann\inst{1} 
\and A.~Bensberg\inst{2}
\and D.~Schertl\inst{1}
 \and G.~Weigelt\inst{1}
\and S.~Wolf\inst{2} 
\and A.~Meilland\inst{3}
\and F.~Millour\inst{3}
\and L.B.F.M.~Waters\inst{4,5}
\and S.~Kraus\inst{6} 
\and K.~Ohnaka\inst{7}
\and B.~Lopez\inst{3}
\and R.G.~Petrov\inst{3}
\and S.~Lagarde\inst{3}
\and Ph.~Berio\inst{3} 
\and F.~Allouche\inst{3}
\and S.~Robbe-Dubois\inst{3}
\and W.~Jaffe\inst{8}
\and Th.~Henning\inst{9} 
\and C.~Paladini\inst{10}
\and M.~Sch\"oller\inst{11}
\and A.~M\'erand\inst{11}
\and A.~Glindemann\inst{11}
\and U.~Beckmann\inst{1}
\and M.~Heininger\inst{1}
\and F.~Bettonvil\inst{12}
\and G.~Zins\inst{10}
\and J.~Woillez\inst{11}
\and P.~Bristow\inst{11}
\and P.~Stee\inst{3}
\and F.~Vakili\inst{3}
          \and R.~van~Boekel\inst{9} 
         \and M.R.~Hogerheijde\inst{8,13}
         \and C.~Dominik\inst{13}
         \and J.-C.~Augereau\inst{14} 
         \and A.~Matter\inst{3}                                
         \and J.~Hron\inst{15} 
          \and E.~Pantin\inst{16}  
         \and Th.~Rivinius\inst{10}
\and W.-J.~de~Wit\inst{10}
          \and J.~Varga\inst{8,17}
          \and L.~Klarmann\inst{9} 
          \and K.~Meisenheimer\inst{9}
\and V.~G\'amez Rosas\inst{8}
\and L.~Burtscher\inst{8}
\and J.~Leftley\inst{3} 
\and J.W.~Isbell\inst{9}       
\and G.~Yoffe\inst{9} 
\and E.~Kokoulina\inst{3}
\and W.C.~Danchi\inst{3,18}
\and P.~Cruzal\`ebes\inst{3} 
\and A. Domiciano de Souza\inst{3}
\and J.~Drevon\inst{3}
\and V.~Hocd\'e\inst{3}
\and A.~Kreplin\inst{6} 
\and L.~Labadie\inst{19}
\and C.~Connot\inst{1}
\and E.~Nu{\ss}baum\inst{1}
\and M.~Lehmitz\inst{9} 
\and P.~Antonelli\inst{3}
\and U.~Graser\inst{9}
\and C.~Leinert\inst{9}  
  }

       \institute{Max Planck Institute for Radio Astronomy, Auf dem H\"ugel 69, 53121 Bonn, Germany  \\     \email{khh@mpifr-bonn.mpg.de}
\and University of Kiel, Institute of Theoretical Physics and Astrophysics, Leibnizstrasse 15, 24118 Kiel, Germany,
\and Laboratoire Lagrange, Universit\'e C\^ote d'Azur, Observatoire de la C\^ote d'Azur, CNRS, Boulevard de l'Observatoire, CS 34229, 06304 Nice Cedex 4, France
\and Institute for Mathematics, Astrophysics and Particle Physics, Radboud University, P.O. Box 9010, MC 62 NL-6500 GL Nijmegen, the Netherlands
\and SRON Netherlands Institute for Space Research, ﻿Sorbonnelaan 2, NL-3584 CA Utrecht, the Netherlands
\and School of Physics, Astrophysics Group, University of Exeter, Stocker Road, Exeter EX4 4QL, UK
\and Departamento de Ciencias F\'isicas, Facultad de Ciencias Exactas, Universidad Andr\'es Bello, Fern\'andez Concha 700, Las Condes, Santiago, Chile
\and Leiden Observatory, Leiden University, Niels Bohrweg 2, NL-2333 CA Leiden, the Netherlands
\and Max Planck Institute for Astronomy, K\"onigstuhl 17, D-69117 Heidelberg, Germany
\and European Southern Observatory, Casilla 19001, Santiago 19,  Chile
\and European Southern Observatory, Karl Schwarzschild Strasse 2,  85748 Garching, Germany
\and ASTRON, Dwingeloo, Netherlands
\and Anton Pannekoek Institute for Astronomy, University of Amsterdam, Science Park 904, 1090 GE Amsterdam, The Netherlands
\and Univ. Grenoble Alpes, CNRS, IPAG, 38000, Grenoble, France
\and Department of Astrophysics, University of Vienna, T\"urkenschanzstrasse 17, A-1180 Vienna, Austria
\and AIM, CEA, CNRS, Universit\'e Paris-Saclay, Universit\'e Paris Diderot, Sorbonne Paris Cit\'e, F-91191 Gif-sur-Yvette, France
\and   Konkoly Observatory, Research Centre for Astronomy and Earth Sciences, Konkoly Thege Mikl\'os\'ut 15-17, H-1121 Budapest,  Hungary  
\and Astrophysics Science Division, NASA/GSFC, Greenbelt, MD 20771, USA
\and I. Physikalisches Institut, Universit\"at zu K\"oln, Z\"ulpicher Str. 77, 50937, K\"oln, Germany
    }

 \date{Received June 21, 2021; accepted November 14, 2021}

 
 \titlerunning{VLTI-MATISSE imaging of FS~CMa}

\authorrunning{K.-H.~Hofmann et al.\ }
 
  \abstract
   { \object{FS~Canis~Majoris} (FS~CMa, HD~45677) is an unclassified B[e] star surrounded by an inclined dust disk. The evolutionary stage of FS~CMa is still debated. Perpendicular to the circumstellar disk, a bipolar outflow was  detected. Infrared aperture-synthesis imaging  provides us with a unique opportunity to study the disk structure.
   }
   {Our aim is to study the intensity distribution of the disk of FS~CMa in the mid-infrared $L$ and $N$~bands.
   }
   {We performed aperture-synthesis imaging of FS~CMa with the MATISSE instrument (Multi AperTure mid-Infrared SpectroScopic Experiment) in the low spectral resolution mode to obtain images in the $L$ and $N$~bands. We computed radiative transfer models that reproduce the $L$- and $N$-band intensity distributions of the resolved disks.  
   }
   {We present  $L$- and $N$-band aperture-synthesis images of FS~CMa reconstructed in the wavelength bands of 3.4--3.8 and 8.6--9.0 $\mu$m. In the $L$-band image, the inner rim region  of an inclined circumstellar disk and the central object can be seen with a spatial resolution of 2.7~milliarcsec (mas). An inner disk cavity with an angular diameter of  $\sim$6\,$\times$\,12\,mas is resolved.  The $L$-band disk consists of a bright northwestern (NW) disk region and a much fainter southeastern (SE) region.   The images suggest that we are looking at the bright inner wall of the NW disk rim, which is on the far side of the disk. In the $N$~band, only the bright NW disk region is seen.  In addition to deriving the inclination and the inner disk radius, fitting the reconstructed
brightness distributions via radiative transfer modeling allows one to constrain the innermost disk structure, in particular the shape of the inner disk rim.

   }   
   {}

  \maketitle

\section{Introduction}    \label{introduction} 

B[e] stars are emission line stars with  low-excitation permitted emission lines (e.g., \ion{Fe}{ii}), forbidden lines of [\ion{Fe}{ii}] and [\ion{O}{i}], and an infrared excess from circumstellar dust (see 
\citeads{1998ASSL..233....1Z} 
and 
\citeads{1998ASSL..233..277L}). 
 This  B[e] star definition has the disadvantage that  B[e] stars are not a homogeneous group of stars, but it consists of different types of stars of different evolutionary stages. 
FS~CMa (HD 45677, MWC 142) belongs to the subgroup of unclassified B[e] stars. Its distance is $620^{+41}_{-33}$ pc \citepads{2020A&A...638A..21V}. The evolutionary stage of FS~CMa is still debated. It is not known whether FS~CMa is a young star, a star near the main sequence, or a more evolved star  
(\citeads{
1998ASSL..233....1Z, 
 2020ApJ...888....7L} and references). 
 \citetads{2020A&A...638A..21V} conclude that FS~CMa together with several other famous B[e] stars are most likely not pre-main-sequence stars (see list of these stars in Appendix A of their paper).

FS~CMa is famous for its large, inclined  circumstellar disk, which was studied with several  infrared interferometers \citepads{2006ApJ...647..444M,2015A&A...581A.107M,2017A&A...599A..85L,2019A&A...632A..53G}. 
\citetads{2014A&A...564A..80K,2020A&A...636A.116K}  reported the first $H$-band aperture-synthesis images  of the inclined, asymmetric disk of FS~CMa.  
\citetads{2020ApJ...888....7L} resolved a bipolar outflow perpendicular to the disk plane. Aperture-synthesis imaging with the new MATISSE interferometry instrument (Multi AperTure mid-Infrared SpectroScopic Experiment) at the Very Large Telescope Interferometer (VLTI) of the European Southern Observatory provides us with the unique opportunity to image the disk of FS~CMa with unprecedented spatial resolution in the $L$ and $N$ bands.

In this paper we present the first milliarcsecond-resolution mid-infrared $L$- and $N$-band aperture-synthesis imaging of  FS~CMa  and radiative transfer modeling for the interpretation of the images. In Section~\ref{observations}, we describe the observations and data reduction. In Sect.~\ref{imaging}, we present the mid-infrared $L$- and $N$-band aperture-synthesis images.  In Section~\ref{sec:data_an}, we show our modeling of the disk structure of FS~CMa based on the reconstructed high-resolution images. The results are summarized in Sect.~\ref{summary}.

\section{Observations and data reduction}     \label{observations} 

FS~CMa was observed with the four Auxiliary Telescopes (ATs) of the VLTI  \citepads{2007NewAR..51..628S} and
the mid-infrared MATISSE instrument \citepads{2014Msngr.157....5L,2016SPIE.9907E..0AM,2018SPIE10701E..09P,2021arXiv211015556L} during the MATISSE  commissioning run between December 2 and 15, 2018, as summarized in Table~\ref{listA1} in Appendix~\ref{obs}. The observations were carried out in the mid-infrared $L$ and $N$~bands  with low spectral resolution of $R\sim$30.  
The observations were performed with seven different AT quadruplet configurations with projected baseline lengths ranging between 10.5 and 137.0\,m. Figure~\ref{uv}  in the Appendix~\ref{obs} shows the $uv$ plane coverage of all observations of the FS~CMa observing run. The spectrally dispersed $L$- and $N$-band interferograms were recorded simultaneously. The detector integration time (DIT)  was 75\,ms  in the $L$~band and 20\,ms  in the $N$~band. The calibrator stars  are listed in Table~\ref{listA2}.
The $L$-band interferograms were obtained in the science-photometry (SciPhot) mode, that is, interferograms and photometric data were recorded simultaneously.
On the other hand, the $N$-band interferograms were obtained in the high-sensitivity (HighSense) mode, that is, interferograms and photometric data were recorded sequentially.

Each $L$- and $N$-band MATISSE observation  usually consists of at least four interferogram data sets corresponding to four different configurations of the Beam Commuting Device (BCD, see \citealt{2008SPIE.7013E..1GM} and \citealt{2014Msngr.157....5L}).
These four data sets are recorded without chopping.
Additionally, each  observation consists of eight more data sets recorded with chopping.
In the $N$~band, these eight chopping data sets contain the photometric data from the four telescopes (HighSense mode) recorded with two different configurations 
of the BCDs. During the recording of  these eight chopping data sets, additional $L$-band interferograms with photometric data (SciPhot mode) are recorded with the same two BCD configurations.

The data obtained during all nights of this run, except one night, were used for the subsequent data reduction and image reconstruction.  From the four nights with the AT configuration A0-B2-J2-C1, the data of the night 2018-12-05 with the worst seeing conditions were not used because a reliable calibration was not possible (see Table~\ref{listA1} in Appendix~\ref{obs} for more details).

\subsection{Data reduction of the $L$-band data} \label{datareductionL}
The $L$- and $N$-band data were reduced with the MATISSE Data Reduction Software (DRS) of pipeline version 1.5.3. After the data reduction, (a) four sets of  interferometric data (closure phases and visibilities) are obtained from the data sets corresponding to all four BCD configurations without chopping, and (b) two sets of raw  interferometric data  from the eight chopping data sets recorded with two different BCD configurations.
The transfer functions of the raw interferometric data of the target were calibrated using the calibrator stars listed in Table~\ref{listA2}. 
In Appendix~\ref{dataredL}, we show examples of observed  closure phases and visibilities.  The estimation of the visibility and closure phase errors is also described in Appendix~\ref{dataredL}.

 \begin{figure}
 \centering
    \hspace{-0mm} \vspace{0mm}
     \includegraphics[height=175mm,angle=0]{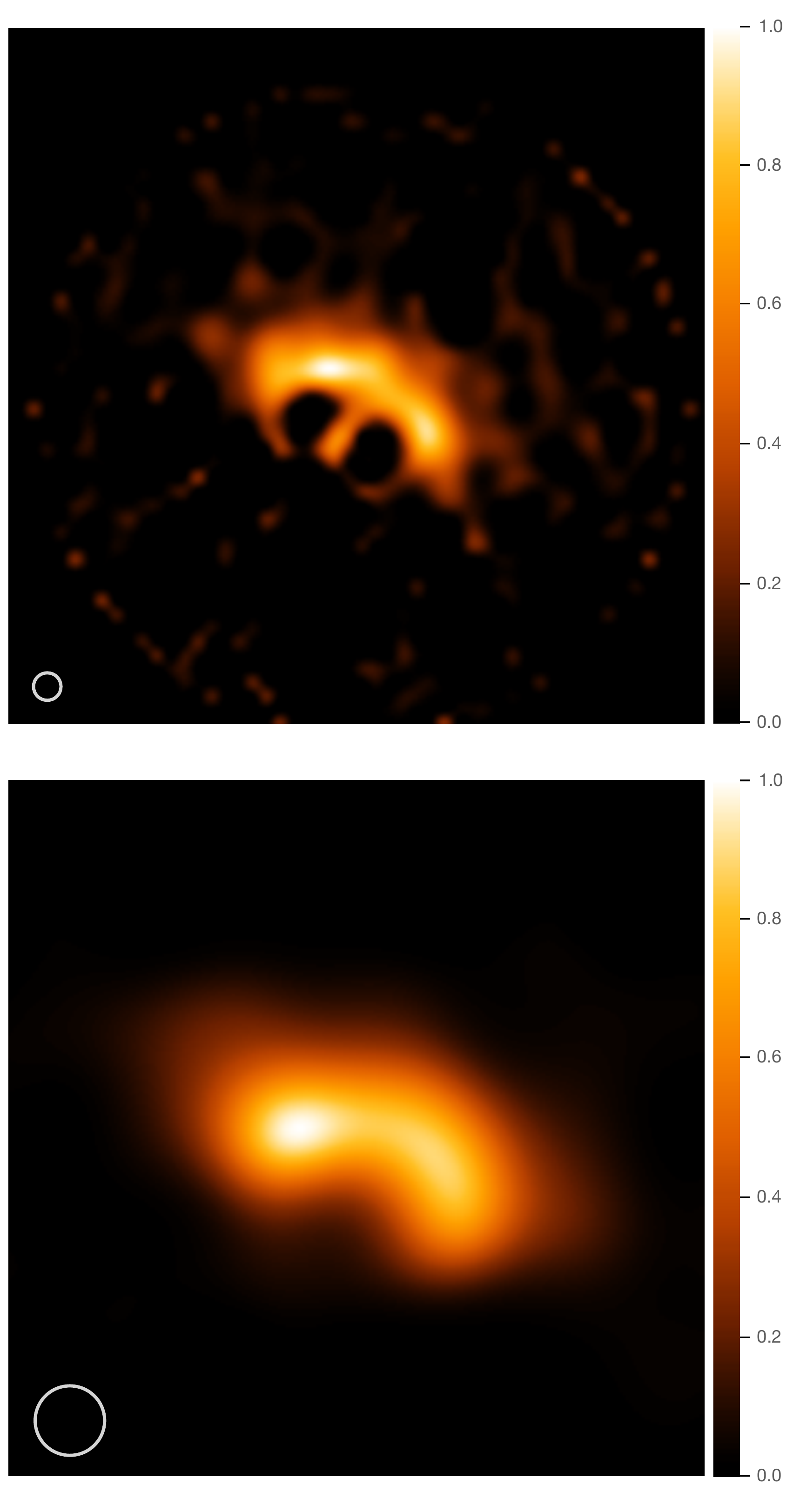}
      \caption[]{\small
              $L$- and $N$-band MATISSE aperture-synthesis images of FS~CMa: (Top) Reconstructed $L$-band image (3.4--3.8~$\mu$m). The inner dark disk cavity in the  $L$-band image has an angular diameter of $\sim$~6\,$\times$\,12~mas (i.e., $\sim$~3.7\,$\times$\,7.4~au for a distance of 620~pc). 
              (Bottom) Reconstructed $N$-band image (8.6--9.0~$\mu$m). The FOV is 64~mas (39.7~au for a distance of 620~pc). North is up and east to the left. The inserted circles represent the 2.7 and 6.6~mas diameters of the $L$- and $N$-band resolution beams. See Sect.~\ref{imaging} for more details.
    }
    \label{reconstructions}
\end{figure}
\begin{table*}                                                     
\tiny
\caption{Image reconstruction parameters and quality of the reconstructed FS~CMa images}              
\label{list1}                                                       
\centering                                              
\begin{tabular}{ccrrccccccccccc}                                
\hline\hline                                             
wavelength r.$\tablefootmark{a}$  &bin$\tablefootmark{b}$ &CP$\tablefootmark{c}$ &V$^2$$\tablefootmark{d}$ &CP err.$\tablefootmark{e}$ &V$^2$ snr $\tablefootmark{f}$ &FOV$\tablefootmark{g}$ &npix$\tablefootmark{h}$ &cost$\tablefootmark{i}$ & reg.$\tablefootmark{j}$ &opt.$\tablefootmark{k}$ & weight$\tablefootmark{l}$ &mask$\tablefootmark{m}$  &CP res$\tablefootmark{n}$  &V$^2$res$\tablefootmark{n}$   \\         
\hline                                                                                     
3.4--3.8\,$\mu$m &1 &11355    &17033   &2.9$^{\circ}$   &18.4  & 80  & 64    & 3  & 6  & 2  & 0.5 & 30--42 &2.9$^{\circ}$  &0.009 \\
8.6--9.0\,$\mu$m &11 &2084      &3079     &10.0$^{\circ}$ &11.5  & 360 & 128 & 2  & 3  & 1  & 0.0 & 160--190 &7.5$^{\circ}$  &0.021 \\
           \hline    \end{tabular}  
\tablefoot{
\tablefootmark{a} {Wavelength range used for image reconstruction.}
\tablefootmark{b} {Number of spectral channels binned (spectral binning; see Sect.~\ref{datareductionN} for details).}
\tablefootmark{c} {Number of closure phases. }
\tablefootmark{d} {Number of visibilities.}    
\tablefootmark{e} {Mean closure phase error of the data.} 
\tablefootmark{f} {Mean signal-to-noise ratio of the visibilities.}
\tablefootmark{g} {FOV of the reconstruction in mas.}
\tablefootmark{h} {Pixel grid npix$\,\times\,$npix used in the image reconstruction.}
\tablefootmark{i} {Number of the cost function as defined in  \cite{2014A&A...565A..48H} and the Appendix~\ref{imarec}.}    
\tablefootmark{j} {Regularization function number as defined in the same paper.} 
\tablefootmark{k} {Optimization routine number as defined in the Appendix~\ref{imarec}.} 
\tablefootmark{l} {Power of the $uv$ density weight (see  \citealt{2014A&A...565A..48H}).}
\tablefootmark{m} {Radii used for the image space mask in mas.}
\tablefootmark{n} {Closure phase (CP) and squared visibility (V$^2$) residuals derived from the measured closure phases and visibilities, and from the closure phases and visibilities  calculated from the reconstructed images.}
}   
\end{table*} 

\subsection{Data reduction of the $N$-band data} \label{datareductionN}
 In order to reduce the  noise in the data, the $N$-band interferograms were spectrally binned, that is,  neighboring spectral channels were averaged before calculating the average closure phases and visibilities. In Appendix~\ref{dataredN}, we show calibrated FS~CMa $N$-band closures phases and visibilities without spectral binning and with three different spectral binnings of 9, 11, and 17 spectral channels. Spectral binning of 7 (e.g., seven neighboring spectral channels in the spectrally dispersed interferograms are averaged) does not reduce the spectral resolution of the data because of the spectral oversampling of the dispersed interferograms recorded in the MATISSE low spectral resolution mode. Data processing without spectral binning (i.e., binning 1) is just shown for illustration in Appendix~\ref{dataredN}, but is usually not used because of the severe noise. The transfer functions of the derived four interferometric data sets were calibrated with the calibrator stars listed in Table~\ref{listA2}.   The estimation of the visibility and closure phase errors is described in Appendix~\ref{dataredN}.

\section{Aperture-synthesis image reconstruction of the dust disk of FS~CMa}   \label{imaging}  
The MATISSE image reconstruction package $IRBis$ \citepads{2014A&A...565A..48H} was applied to the calibrated interferometric data of FS~CMa. This software package is part of the MATISSE data reduction pipeline\footnote{https://www.eso.org/sci/software/pipelines/}. For the reconstruction process, the two built-in minimization engines, the six regularization functions, and the three cost functions were applied (details are described in \citealt{2014A&A...565A..48H} and Appendix~\ref{imarec}). The start image of each image reconstruction run was a circular Gaussian fit to the measured visibilities of the target.
We used  this Gaussian as a prior for the image reconstruction. From the reconstructions that were calculated, the best ones were selected using the image quality measure {\it qrec}. The quality measure $qrec$ is calculated from the reduced $\chi^2$ values of the closure phases and visibilities, and a so-called residual ratio (see \citealt{2014A&A...565A..48H} for more details). Because of the inhomogeneous $uv$ plane coverage, we weighted the  $uv$ point measures with the  power of 0.5 or zero of the inverse $uv$ plane density function (see Table~\ref{list1}; power zero means equal weights for all $uv$ points).

\subsection{$L$-band images of FS~CMa} \label{imagingL}
The field-of-view (FOV) and pixel grid used for image  reconstruction in the $L$~band were 80\,$\times$\,80~mas and 64\,$\times$\,64 pixels, respectively. The image reconstruction was perfomed in the wavelength range  3.4--3.8~$\mu$m.  All six calibrated interferometric data sets (four corresponding to the four BCD configurations without chopping, and two to the two BCD configurations with chopping) of each observation were used in the image reconstruction process, as described in Section~\ref{datareductionL}. Figure~\ref{reconstructions} (top) shows the obtained $L$-band reconstruction. The reconstruction parameters and quality measures  are listed in Table~\ref{list1}. The reconstructed image has a spatial resolution of $\sim$2.7~mas (it was not convolved to reduce the resolution).
In Fig.~\ref{LbanddataRec} in Appendix~\ref{dataComparison}, the measured closure phases and visibilities of an example observation and the ones derived from the reconstructed image (Fig.~\ref{reconstructions} top) are displayed. This comparison is a useful method to investigate whether the closure phases and visibilities of the observations are in acceptable agreement with the closure phases and visibilities of the reconstruction. 

The $L$-band image shows a central object and a ring-like, elliptical circumstellar structure that looks like an inclined disk consisting of a bright NW disk region and a much fainter SE region. The first aperture-synthesis image of FS CMa was reported by \citetads{2014A&A...564A..80K,2020A&A...636A.116K}, who imaged the disk of FS CMa in the $H$ band. In this $H$-band image, the NW disk region and the central object can be seen, whereas the faint SE $L$-band region is not visible. The inner disk cavity in the $L$-band image has a size of $\sim$6\,$\times$\,12~mas. The asymmetric intensity distribution suggests that the NW region is bright because we are looking at the bright, inner wall of the NW disk rim. Therefore, this intensity distribution suggests that the NW rim is on the far side.

Reconstructed interferometric infrared images can consist of both real target structures and reconstruction noise caused by gaps in the $uv$ plane coverage and calibration errors.
To estimate this noise in our reconstructed images, we computed error maps of the reconstructed images presented in Figs.~\ref{Lerrormap} and \ref{Nerrormap}.
For this goal, we first performed computer simulations, in which we used the same $uv$ coverage as the one of the reported MATISSE observations.
As computer simulation targets, we chose the model images shown in Figs.~\ref{Lerrormap}a and \ref{Nerrormap}a  (which are similar to the model images shown in Figs.~\ref{fig:bestfit_mod_51} and \ref{fig:bestfit_mod_nband}, but with the central object intensity increased to the central object intensity of the reconstructed image).
The computed visibilities and closure phases were degraded by simulated noise corresponding to the noise in the MATISSE observations.
Figures~\ref{Lerrormap}b and \ref{Nerrormap}b show the obtained reconstructed images of the computer simulations.
The error maps  shown in Figs.~\ref{Lerrormap}c and \ref{Nerrormap}c are computed as the absolute value of the difference between the computer simulation targets (Figs.~\ref{Lerrormap}a and \ref{Nerrormap}a) and the reconstructions of the computer simulations (Figs.~\ref{Lerrormap}b and \ref{Nerrormap}b).
The $L$-band error map presented in Fig.~\ref{Lerrormap}c suggests that the clumpiness in the disk structure is partially degraded by the aforementioned reconstruction noise.
This is consistent with the fact that there are differences between the positions of the brightest clumps in Kluska's $H$-band image \citepads{2014A&A...564A..80K,2020A&A...636A.116K} and  the positions of the brightest clumps in the MATISSE image.
Furthermore, the computer simulation clearly shows that the elongation of the reconstructed central object is at least partially a  bridge-like reconstruction artifact caused by the neighborhood of the bright disk rim.
This bridge effect is even partially seen in the original computer object in Fig.~\ref{Lerrormap}a.
However, a close suspected companion star \citepads{2006MNRAS.367..737B}  or  a complex circumstellar environment may also partially contribute to the elongated image of the central object.

\subsection{$N$-band images of FS~CMa} \label{imagingN}
For the $N$-band image reconstruction, the  FOV and pixel grid used  were 360\,$\times$\,360~mas and 128\,$\times$\,128 pixel, respectively. The image reconstruction was performed in the wavelength range of 8.6--9.0~$\mu$m, where the $N$-band data have their highest quality.  All four calibrated interferometric data sets (corresponding to the four BCD configurations without chopping) of each observation were used for image reconstruction, as described in Section~\ref{datareductionN}. Figure~\ref{reconstructions} (bottom) shows the obtained $N$-band reconstruction. The reconstruction parameters and quality measures are listed in Table~\ref{list1}. The reconstruction has a spatial resolution of $\sim$6.6~mas. Figure~\ref{NbanddataRec} in Appendix~\ref{dataComparison} shows the comparison between the measured closure phases and visibilities of a representative measurement and the ones derived from the reconstruction (Fig.~\ref{reconstructions} bottom).

\section{Radiative transfer modeling of the FS~CMa images}  \label{sec:data_an}
The reconstructed images presented in Sect.~\ref{imaging}  provide the basis to constrain the  disk structure of FS~CMa.
In Sect.~\ref{subsec:setup}, the applied model and radiative transfer method are briefly outlined. In Sect.~\ref{subsec:fitting}, the fitting process for the $L$-band image as well as the comparison with the $N$-band image are described. For this, we are using a well-studied basic disk model. Subsequently, we extend this model motivated by the remaining differences between the simulated and reconstructed images.

\subsection{Model description} \label{subsec:setup}

\paragraph{Basic disk model 1}
We start with a basic model of a circumstellar disk that has been successfully used for fitting of observations of several protoplanetary disks (e.g.,~\citealp{Wolf2003};~\citealp{Sauter2009};~\citealp{Brunngraeber2016}). The stellar parameters are fixed using values derived by \cite{Vioque2018}. These and the disk inclination determined by \citetads{2020A&A...636A.116K} are summarized in Tab.~\ref{tab:Parameters}.
\begin{table}[!ht]
  \centering
    \caption{Selected stellar and disk parameters derived from earlier observations of FS~CMa.}
    \label{tab:Parameters}
    \begin{tabular}{lcc}
    \hline
    \hline
    \rule{0pt}{2ex}
      Parameter & Symbol & Value\\
      \hline
      \rule{0pt}{3ex}
      Distance (pc)\tablefootmark{1} & d & $620^{+41}_{-33}$\\
      \rule{0pt}{1ex}
      Source temperature (K)\tablefootmark{1} & $T_{\rm eff}$  &  $16500^{+3000}_{-750}$\\
      \rule{0pt}{1ex}
      Source luminosity ($L_{\odot}$)\tablefootmark{1} & log$\;L_{\star}$  &  $2.88^{+0.32}_{-0.17}$\\
      \rule{0pt}{1ex}
      Inclination $(^{\circ})$\tablefootmark{2} & $i$  & $45\pm 8$\\
      \rule{0pt}{1ex}
      Position angle $(^{\circ})$\tablefootmark{2} & PA  & $76\pm 11$\\
      \hline
    \end{tabular}
    \tablefoot{\tablefoottext{1}{\cite{Vioque2018};}
                \tablefoottext{2}{\citetads{2020A&A...636A.116K}}}
\end{table}
\\
In addition to the model referred to above, an exponential decay of the density in the outer regions based on considerations by \cite{Lynden-Bell1974}, \cite{Kenyon1995}, and \cite{Hartmann1998} is applied. The density distribution is described by
\begin{equation}
    \label{eq:dens_dist}
    \rho_{\rm g} (r,z) = \frac{\Sigma_{\rm g}(r)}{\sqrt{2\pi}\;h_{\rm g}(r)} \exp\left[ - \frac{1}{2} \left(\frac{z}{h_{\rm g}(r)}\right)^2\right].
\end{equation}
Here  $(r,z)$ denote cylindrical coordinates, $\Sigma_{\rm g}$ the vertically integrated surface density, and $h_{\rm g}$ the scale height, which can be written as
\begin{equation}
    \label{eq:vert_dens}
    \Sigma_{\rm g}(r) = \sqrt{2\pi}\;\rho_0\;h_{\rm ref}\;\left(\frac{r}{R_0}\right)^{-\gamma} \exp \left[-\left(\frac{r}{R_{\rm trunc}}\right)^{2-\gamma}\right]
\end{equation}
and
\begin{equation}
    \label{eq:scale_height}
    h_{\rm g}(r) = h_{\rm ref} \left(\frac{r}{R_0}\right)^\beta,
\end{equation}
with the density scaling parameter $\rho_0$ determined by the total disk mass, $R_0$ the radius of the reference scale height $h_{\rm ref}$, and $R_{\rm trunc}$ the truncation radius at which the exponential surface-density drop-off sets in. The parameters $\beta$ and $\gamma$ determine the disk flaring and radial density profile.
\paragraph{Dust properties}
Given the uncertain nature of FS~CMa -- a pre-main-sequence star or an evolved star -- we first tentatively adopt the mixture of astrononomical silicate and graphite. If FS~CMa is a pre-main-sequence star, the dust chemical composition of its disk should be similar to that of the interstellar medium. For this reason, we first use a dust mixture consisting of
 62.5\% astronomical silicate and 37.5\% graphite, which follows the $\frac{1}{3}$--$\frac{2}{3}$ relation for parallel and perpendicular orientations by \cite{DraineMalhotra1993}. On the other hand, if FS~CMa is an evolved star, its disk should be oxygen-rich, consisting of silicate alone without graphite. We describe models with this dust chemistry later in Sect.~\ref{subsec:FluxRatio}. The grain size distribution is described by the typical power law $n(s) \propto s^{-q}$ of \cite{MathisRumplNordsieck1977} with $q=3.5$ and grain sizes between $5\;$nm and $250\;$nm. Since this also includes small dust grains that are unusual for stars with high UV emission, we make sure that all discussed models in the following sections have temperatures below the respective sublimation temperature. The impact of larger dust grains is also discussed later in Sect.~\ref{subsec:FluxRatio}. To derive the required optical properties, we apply the wavelength-dependent refractive indices by \cite{Draine1984}, \cite{Laor1993}, and \cite{Weingartner2001}.
\paragraph{Radiative transfer}
To analyze the reconstructed images from the observations, we calculate corresponding model images based on the previously described model. For this purpose, we apply the publicly available 3D Monte-Carlo radiative transfer code \texttt{POLARIS} \citep{Reissl2016}\footnote{https://www1.astrophysik.uni-kiel.de/$\sim$polaris/}. \texttt{POLARIS} calculates the temperature distribution of the disk using the immediate correction method of~\cite{BjorkmanWood2001} together with the instant reemission method of~\cite{Lucy1999}. A raytracing algorithm is then used to calculate the thermal emission of the dust grains based on the temperature distribution. Finally, the scattered stellar light and scattered reemitted radiation are added using an anisotropic Monte-Carlo scattering method.

\begin{figure*}[!htb]
    \centering
    \includegraphics[width=0.3\textwidth]{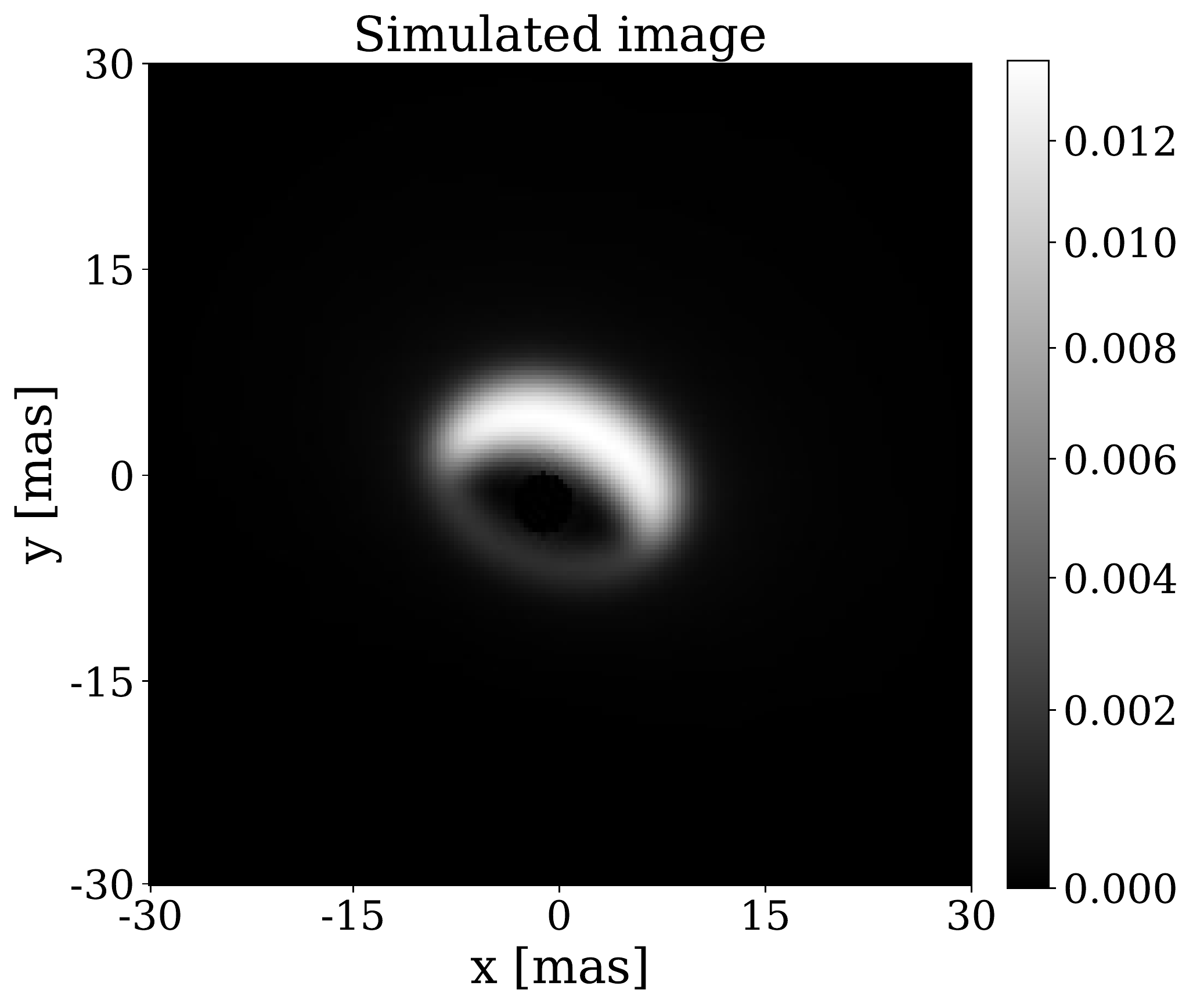}
    \includegraphics[width=0.3\textwidth]{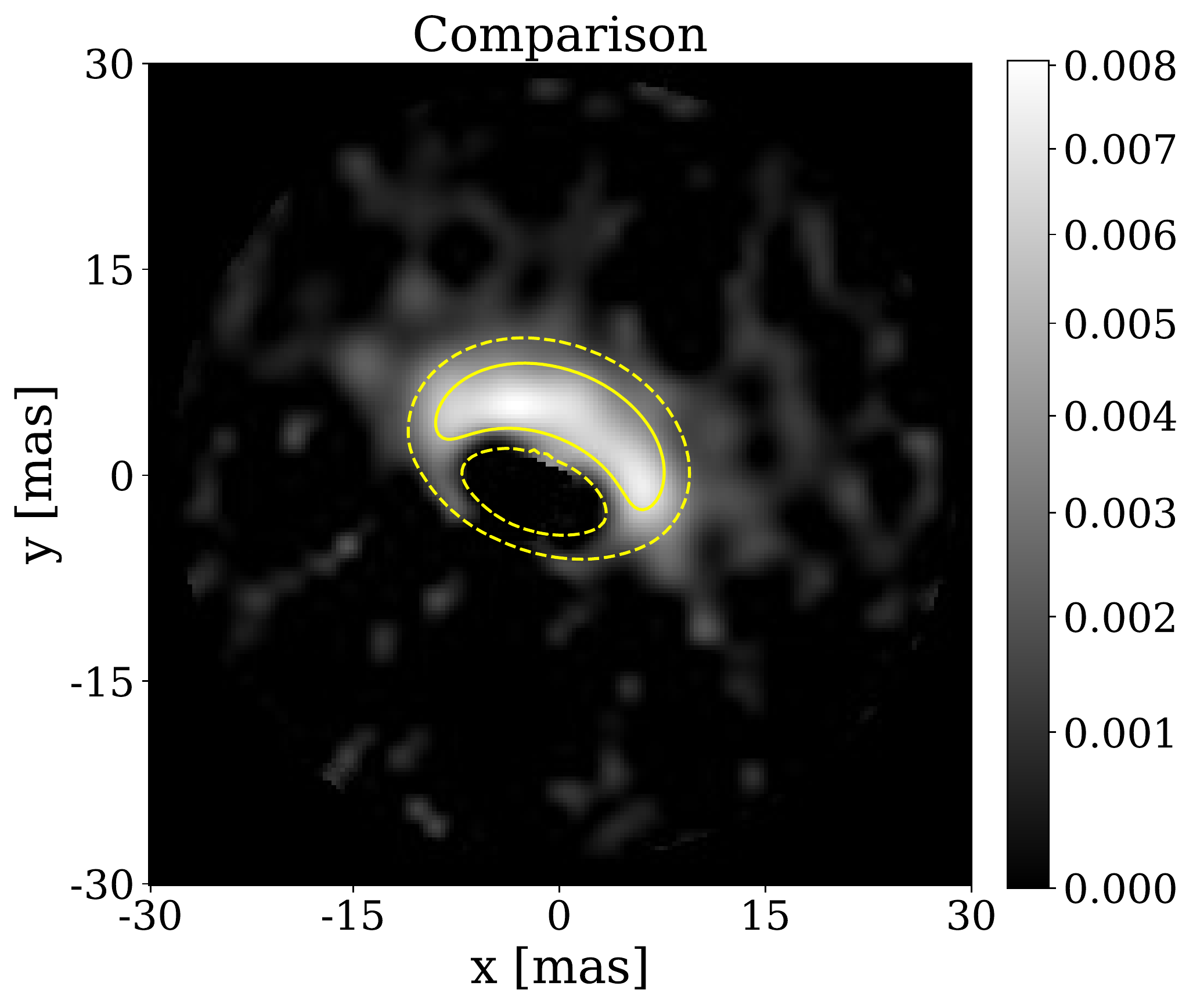}
    \includegraphics[height=0.254\textwidth]{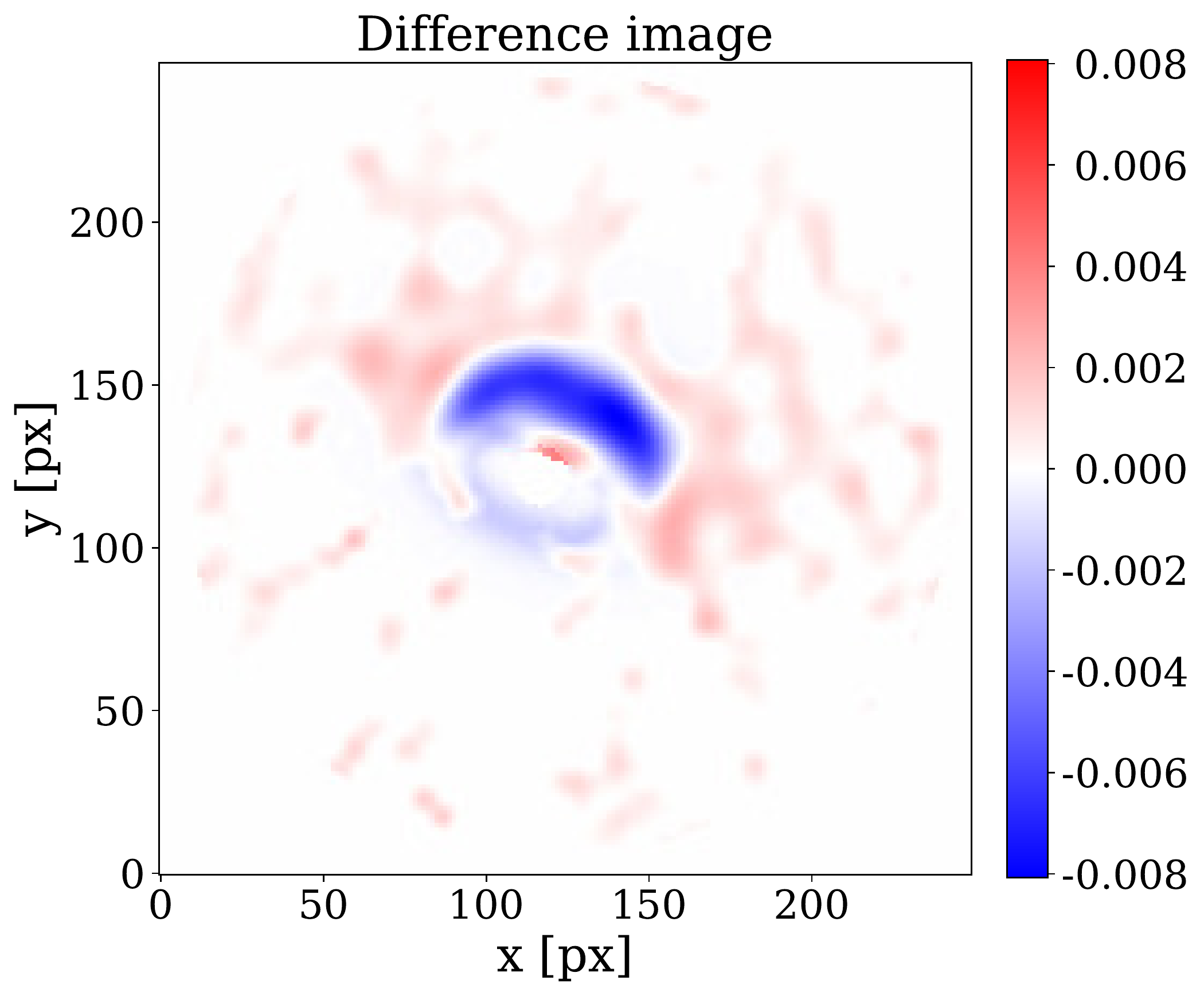}
    \begin{center}
        \begin{tabular}{lcc}
        \hline
        \hline
        \rule{0pt}{2ex}
          Parameters of model 1 & Symbol & Value\\
          \hline
          \rule{0pt}{3ex}
          Inner radius (au) & $R_{\rm in}$ & 5 \\
          \rule{0pt}{1ex}
          Outer radius (au) & $R_{\rm out}$ & 1000 \\
          \rule{0pt}{1ex}
          Truncation radius (au) & $R_{\rm trunc}$ & 200 \\
          \rule{0pt}{1ex}
          Total dust mass ($M_{\odot}$) & $M_{\rm dust}$ & $10^{-3}$\\
          \rule{0pt}{1ex}
          Disk flaring & $\beta$ & $1.2$\\
          \rule{0pt}{1ex}
          Radial density exponent& $\gamma$ & $1.1$\\
          \rule{0pt}{1ex}
          Scale height (au) & $h_{100}$ & 20\\
          \rule{0pt}{1ex}
          Inclination ($^{\circ}$) & $i$ & 42\\
          \rule{0pt}{1ex}
          Position angle ($^{\circ}$) & PA & 20\\
          \hline
        \end{tabular}
     \end{center}
     \caption{Comparison of the observed $L$-band image with the  image of the basic disk model 1: (left) intensity distribution of the $L$-band model  disk image, (middle) overlay of the observed $L$-band image with a contour plot of the model image,  (right)  difference image (observed image minus model image). Contour lines show 50\% (solid) and 10\% (dashed) levels of the maximum flux. The best-fit parameter values are compiled in the table below. The central star is masked in reconstructed and model images to focus on the difference caused by the disk structure. The flux ratio of central object and disk is discussed in Sect.~\ref{subsec:FluxRatio}.}
    \label{fig:first_sim}
\end{figure*}

\subsection{Fitting process} \label{subsec:fitting}
Due to the higher spatial resolution of the reconstructed $L$-band image compared to the $N$-band image, the former has been chosen to guide the fitting. In order to constrain the model parameters, we first analyze their impact on the inner disk structures visible in the reconstructed images. The inner radius $R_{\rm in}$, disk flaring $\beta$, scale height $h_{100}$ at 100\,au, and inclination $i$ turned out to be the most crucial parameters for the observed appearance of the innermost disk region. In contrast to this, the outer radius $R_{\rm out}$, truncation radius $R_{\rm trunc}$, radial density parameter $\gamma$, and total dust mass $M_{\rm dust}$ have hardly any impact and thus cannot be constrained. Therefore, we use parameter values which have been either constrained by former observations of the large-scale disk structure of FS~CMa or can be considered as typical for this class of objects.\\
To quantify the goodness of the fit, we (1) calculate images of the difference between the reconstructed and simulated images. In addition, we (2) consider the absolute mean difference by averaging the absolute value of the difference images, as well as (3) the difference between observed and modeled visibilities and closure phases.
 Relative flux units are used in the reconstructed images and in the simulated images.  In the next step, we mask the extended image of the central star in the reconstructed $L$-band image and in the simulated image to first  focus on reproducing the brightness distribution of the disk. However, unmasked images of the simulation and reconstruction are shown later in Figs.~\ref{fig:bestfit_mod_51} and \ref{fig:bestfit_mod_env}. The masked simulated images are shifted within a range of a few pixels to find the best match with the reconstructed image (i.e., the minimum of the absolute mean difference). In a second step (Sect.~\ref{subsec:cpvResiduals}), we compute the difference between model closure phases and observed closure phases and model visibilities with observed visibilities (called residuals below). The comparison of the total flux of the observed and simulated images is discussed in Sect.~\ref{subsec:totflux}.
\paragraph{Inner radius and inclination}
The appearance of the inner rim of the disk as seen in the $L$-band image is mainly determined by the inner radius and disk inclination. While the inner radius directly affects the size of the inner disk cavity, the inclination determines the eccentricity of the elliptical model image (see Fig.~\ref{fig:first_sim}, left), with potential shadowing by outer disk regions and -- in interplay with the anisotropic scattering function of the dust grains -- the relative contributions of forward and back-scattered radiation. Both parameters are fit simultaneously.\\

We constrain the inner radius $R_{\rm in}$ by varying the inner radius between 3\,au and 7\,au and calculating the absolute mean difference between the reconstructed and model images by averaging the absolute value of the difference images (see Fig.~\ref{fig:first_sim} right) to find the best-fit image.
To constrain the inclination, we perform a series of simulations for inclinations between $35^{\circ}$ and $50^{\circ}$. These are again compared to the reconstructed image by calculating the absolute mean difference between the reconstructed and model images.  Based on these simulations, we deduce first best-fit values for the inclination of $42 \pm 3^{\circ}$ and the inner radius $R_{\rm in} = 5 \pm 0.5\,\text{au}$, respectively.
We note, that the temperature of the dust of the rim does not exceed 1200\,K. The inner radius is therefore not perfectly corresponding to the sublimation temperature of the dust composition, which is 1300--1500\,K for silicate (see, e.g., \citeads{Pollack1994}) and 2500\,K for graphite (see, e.g., \citeads{Kobayashi2011}).
\paragraph{Scale height and flaring parameter}
While the inner radius and inclination are directly related to the inner ellipse seen in the $L$-band model, the scale height $h_{100}$ and the flaring parameter $\beta$ determine the radial extent of the bright inner rim.  We conduct simulations with scale heights between 15\,au and 30\,au in steps of 5\,au and compare the resulting radial extent of the inner wall with that of the reconstructed $L$-band image. Constraining the flaring parameter $\beta$ is less straightforward because it affects various features in the image. Here, we have to compare the overall shape of the simulated images with the reconstructed images.  To cover a sufficiently wide range of flarings, we run simulations with a flaring parameter $\beta$ between 0.5 and 1.5 in steps of 0.1. By increasing the flaring we increase the flux on the far side of the disk. We find that at a certain degree of flaring the disk starts to cover parts of the inner region. Considering this and excluding unrealistically high and low scale heights, we constrain the flaring parameter to 1.2 with a best-fit scale height at 100\,au of around 20\,au.\\
\\
Figure~\ref{fig:first_sim} presents a comparison of the best-fit basic model 1 with the observed image:  (from left to right) the intensity distribution  of the simulated model 1 image, an overlay of the observed  $L$-band image with a contour plot of the model image, as well as the corresponding difference image (the central star is masked in the figure).

It can easily be seen that the inner elliptical gap and the extent of the bright region are well fit. However, the difference image also shows that the brightness distribution does not fit the characteristic asymmetry of the brightness distribution of the disk rim.

\subsection{Modified curved inner rim (models 2 and 3)} \label{subsec:modrin}

\begin{figure*}[!ht]
    \centering
    \includegraphics[width=0.3\textwidth]{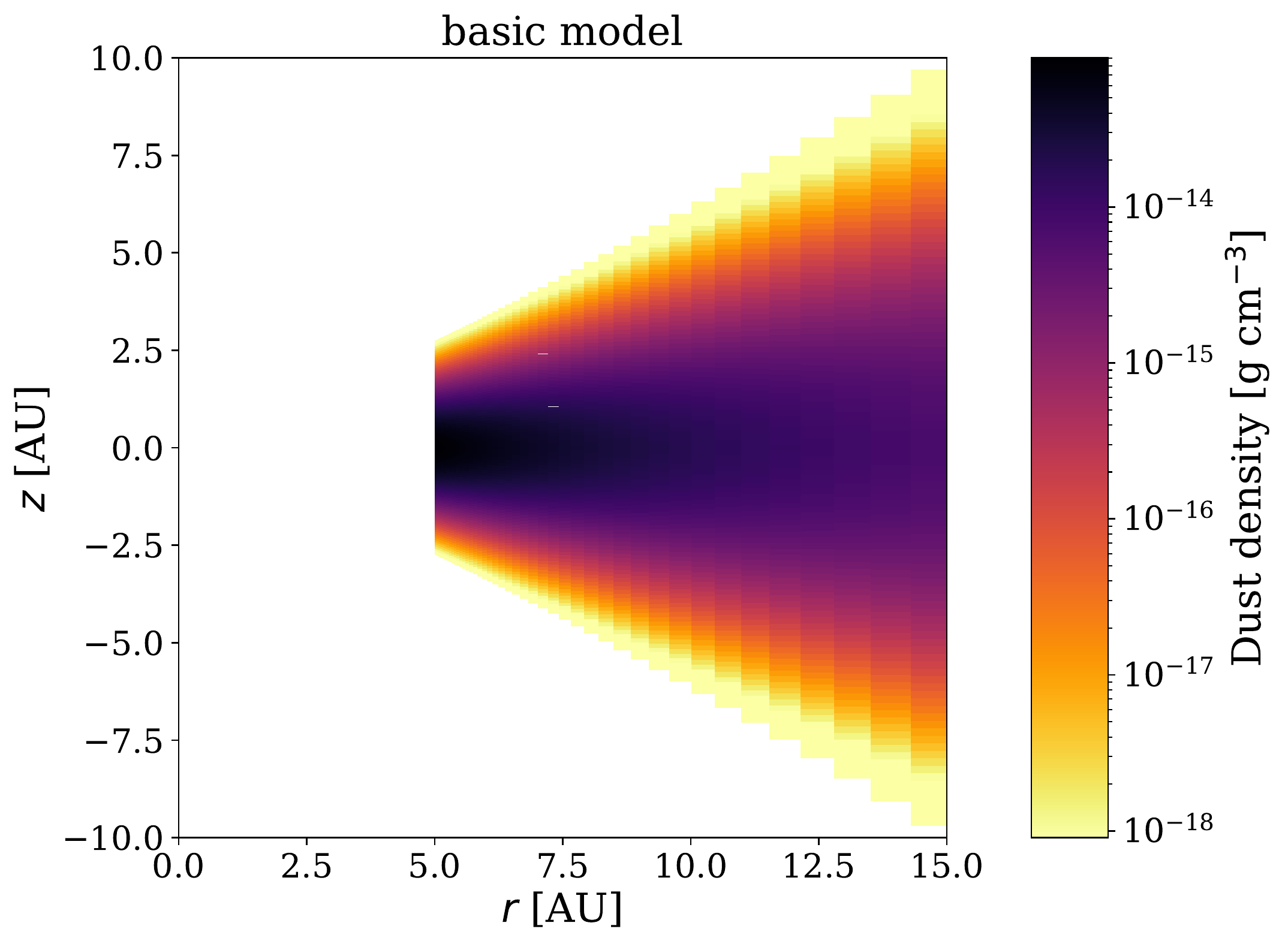}
    \includegraphics[width=0.3\textwidth]{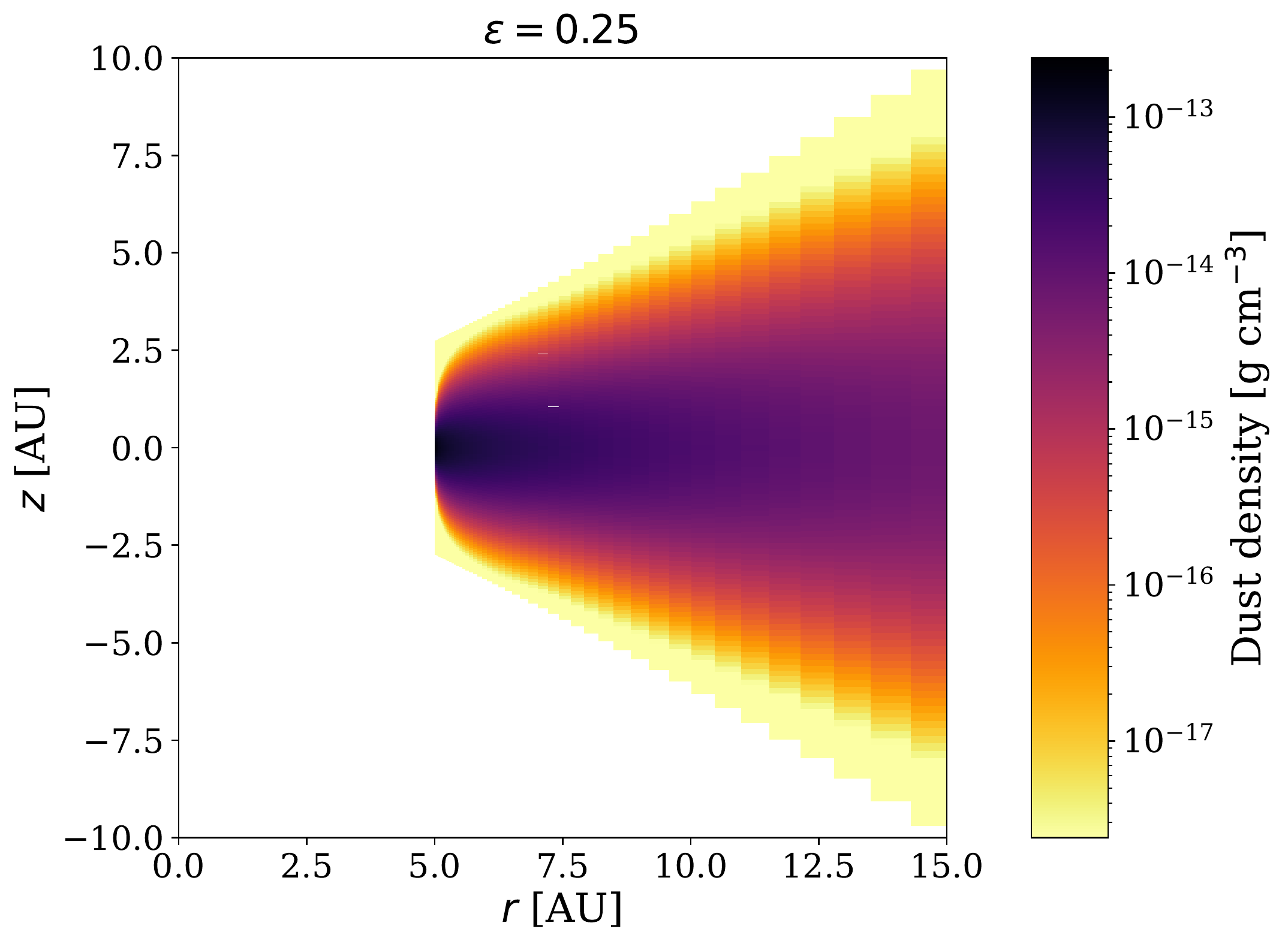}
    \includegraphics[width=0.3\textwidth]{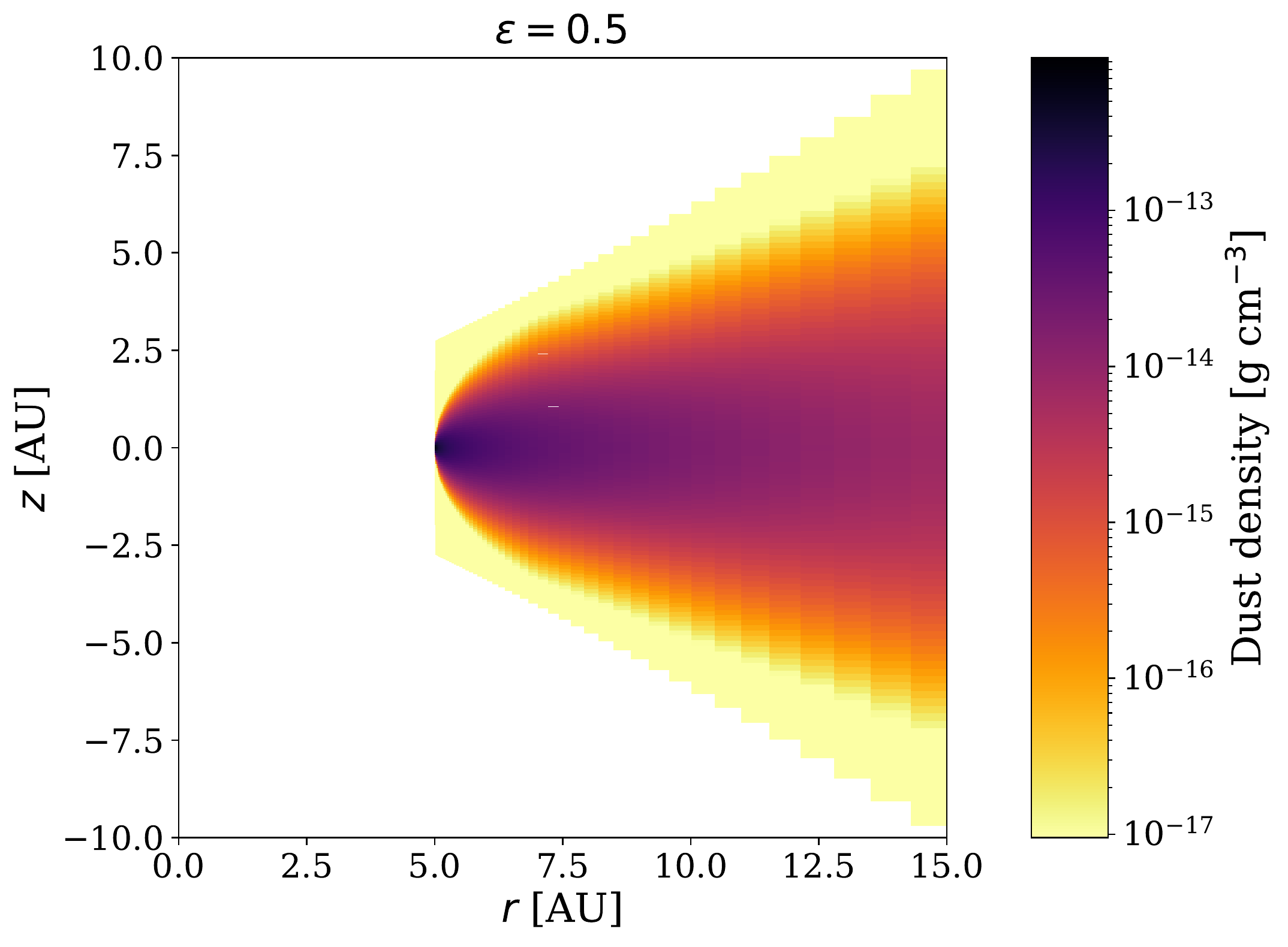}

    \includegraphics[width=0.3\textwidth]{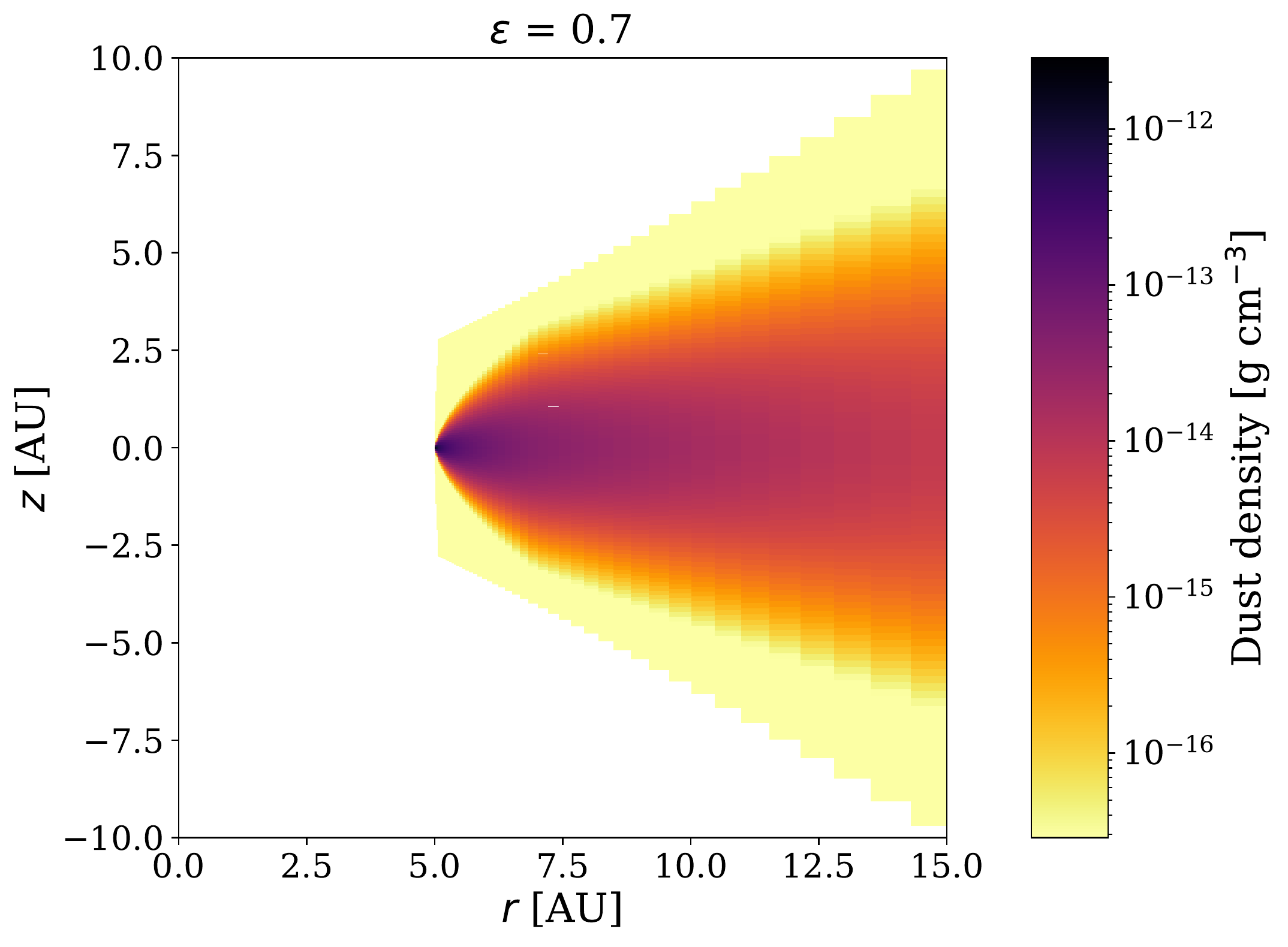}
    \includegraphics[width=0.3\textwidth]{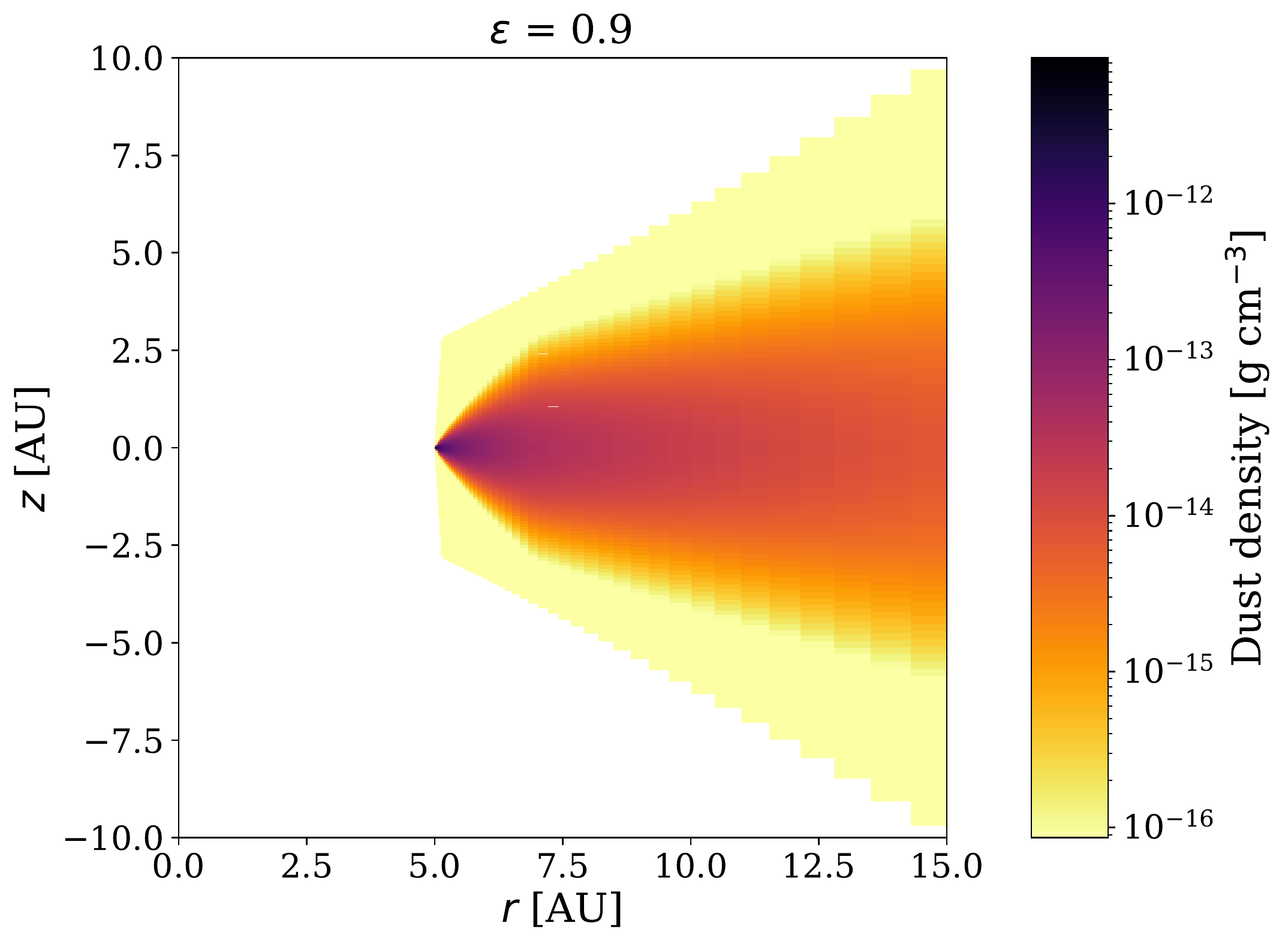}
    \includegraphics[width=0.3\textwidth]{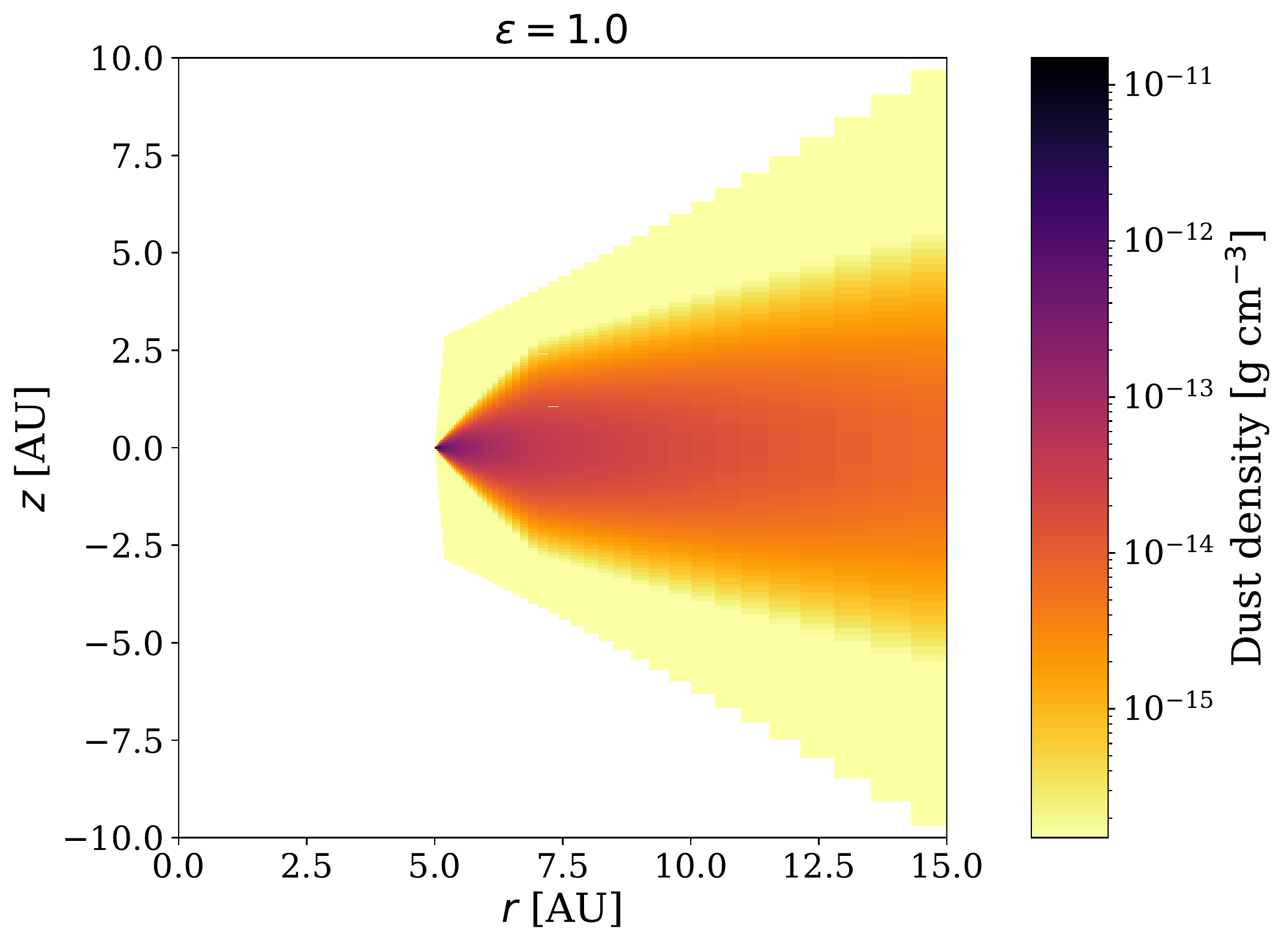}
    \caption{Vertical cuts through the density distributions of the different shapes of the modified inner rim and unmodified (yellow) inner rim for comparison ($R_{\varepsilon}$ fixed at $1.4\,R_{\rm in}$ = 7\,au). The shape parameter $\varepsilon$ is increasing from left to right and top to bottom (0.25, 0.5, 0.7, 0.9, 1.0). The upper left image shows the unmodified inner rim described in Eq.~\ref{eq:dens_dist} for reference.}
    \label{fig:epsilonplot}
\end{figure*}

\begin{figure}[!ht]
    \centering
    \includegraphics[width=.4\textwidth]{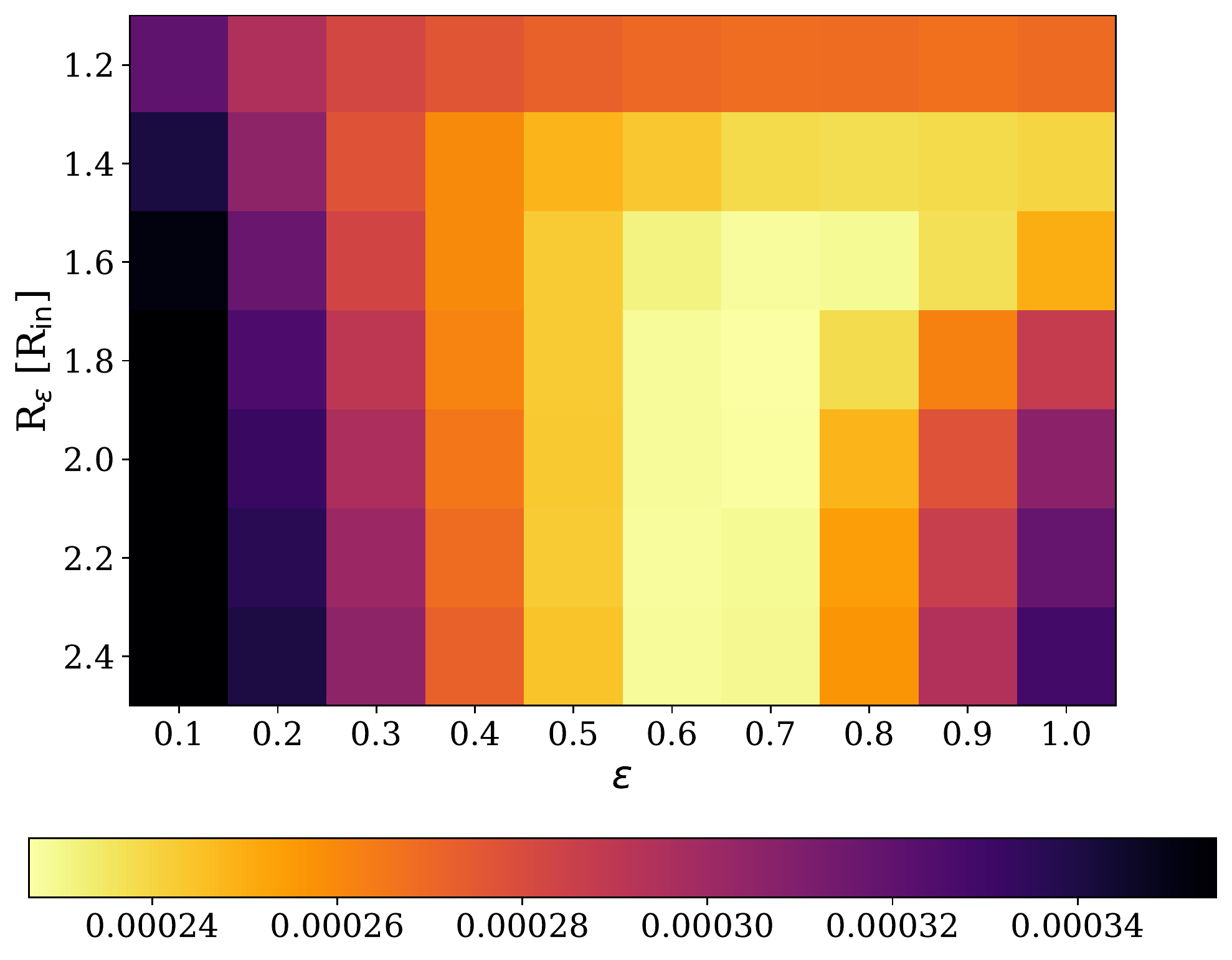}
    \caption{Absolute mean difference between simulated (model 3) and reconstructed $L$-band images for different inner rim parameters $\varepsilon$ and $R_{\varepsilon}$.}
    \label{fig:diffmeangrid}
\end{figure}

\begin{figure*}[!ht]
    \centering
    \includegraphics[width=0.3\textwidth]{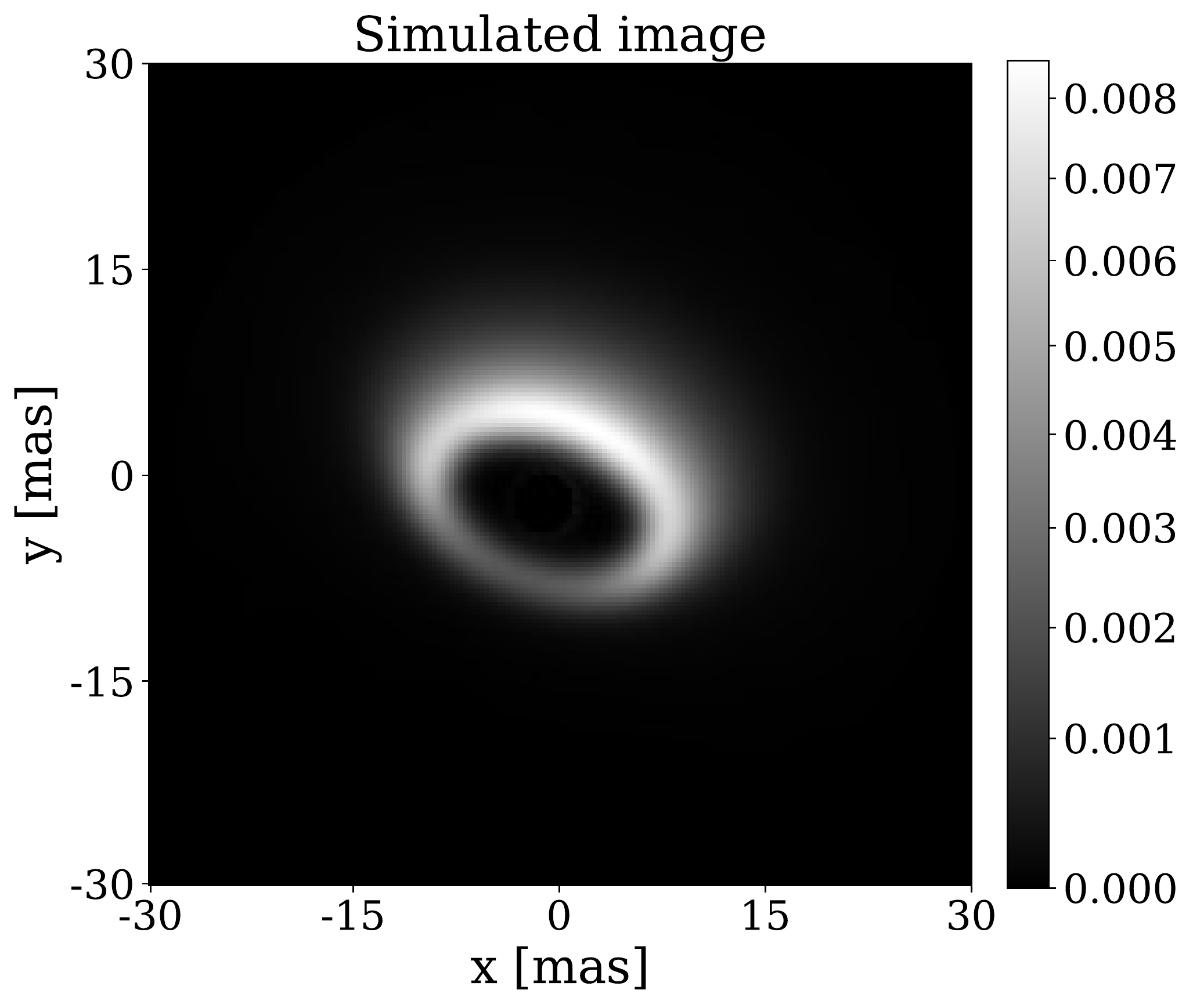}
    \includegraphics[width=0.3\textwidth]{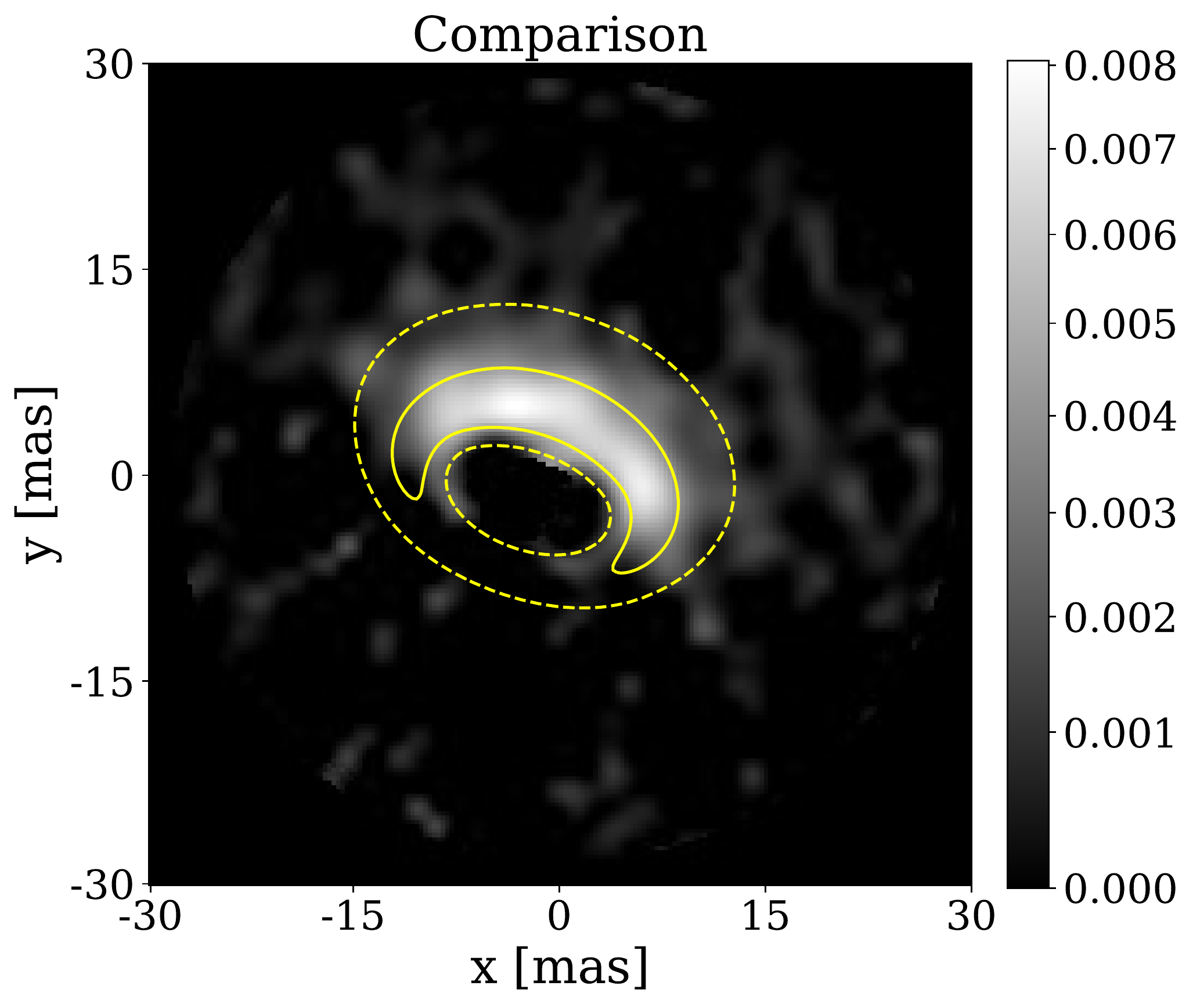}
    \includegraphics[height=0.254\textwidth]{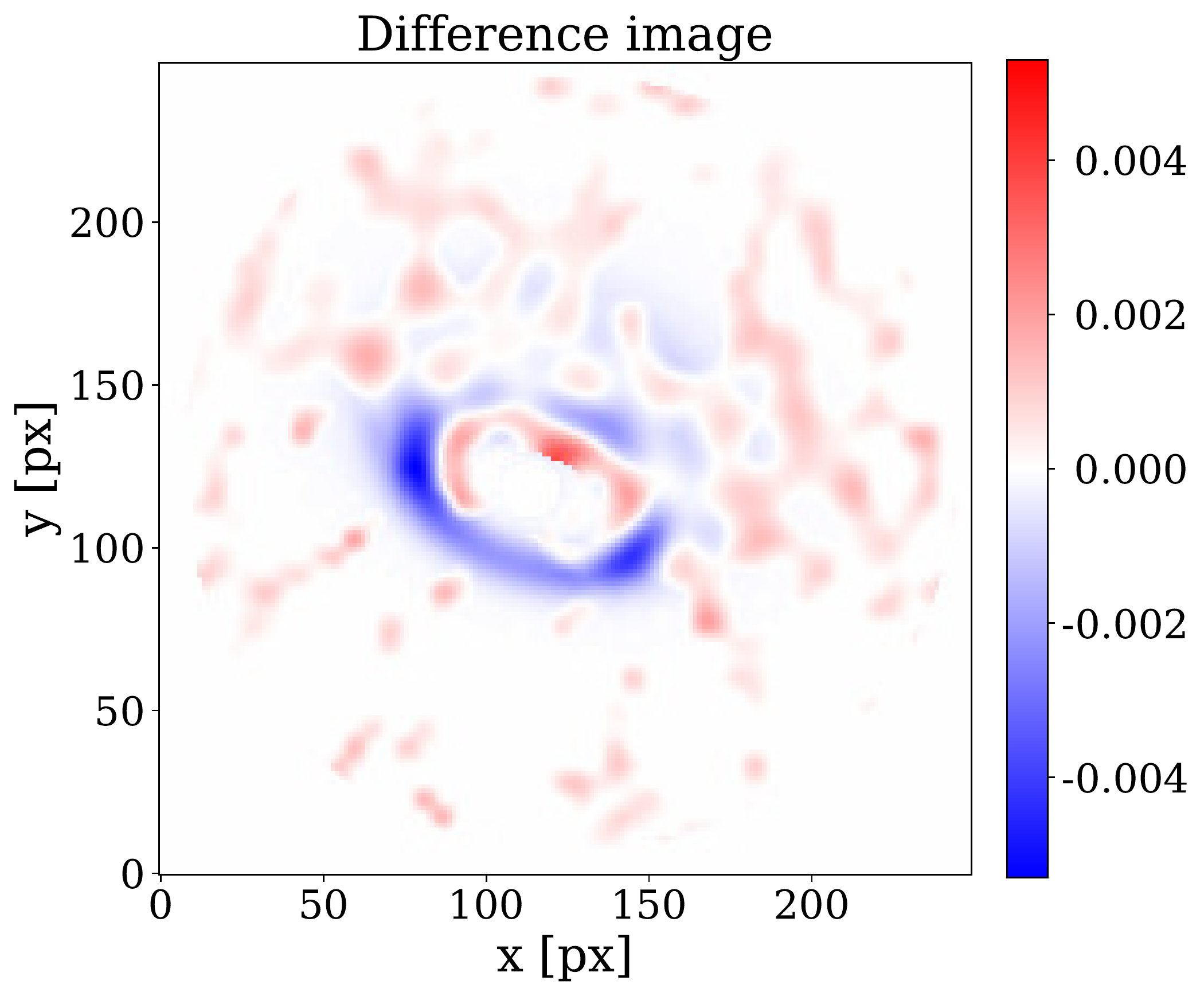}
    \caption{Comparison of the observed $L$-band image with the  image of model 2: (left) best-fit $L$-band model disk image with modified inner rim, (middle) overlay of contour plot of model image and reconstructed $L$-band image, and (right) difference image. Contour lines show 50\% (solid) and 10\% (dashed) levels of the maximum flux. As outlined in Sect.~\ref{subsec:fitting}, the given flux unit is relative. Rim parameter: $R_{\varepsilon} = 1.8\,R_{\rm in}$ = 9\,au, $\varepsilon = 0.7$. The central object is masked in reconstructed and model image to focus on the difference caused by the disk.}
    \label{fig:bestfit_mod}
\end{figure*}

\begin{figure*}[!ht]
    \centering
    \includegraphics[width=0.3\textwidth]{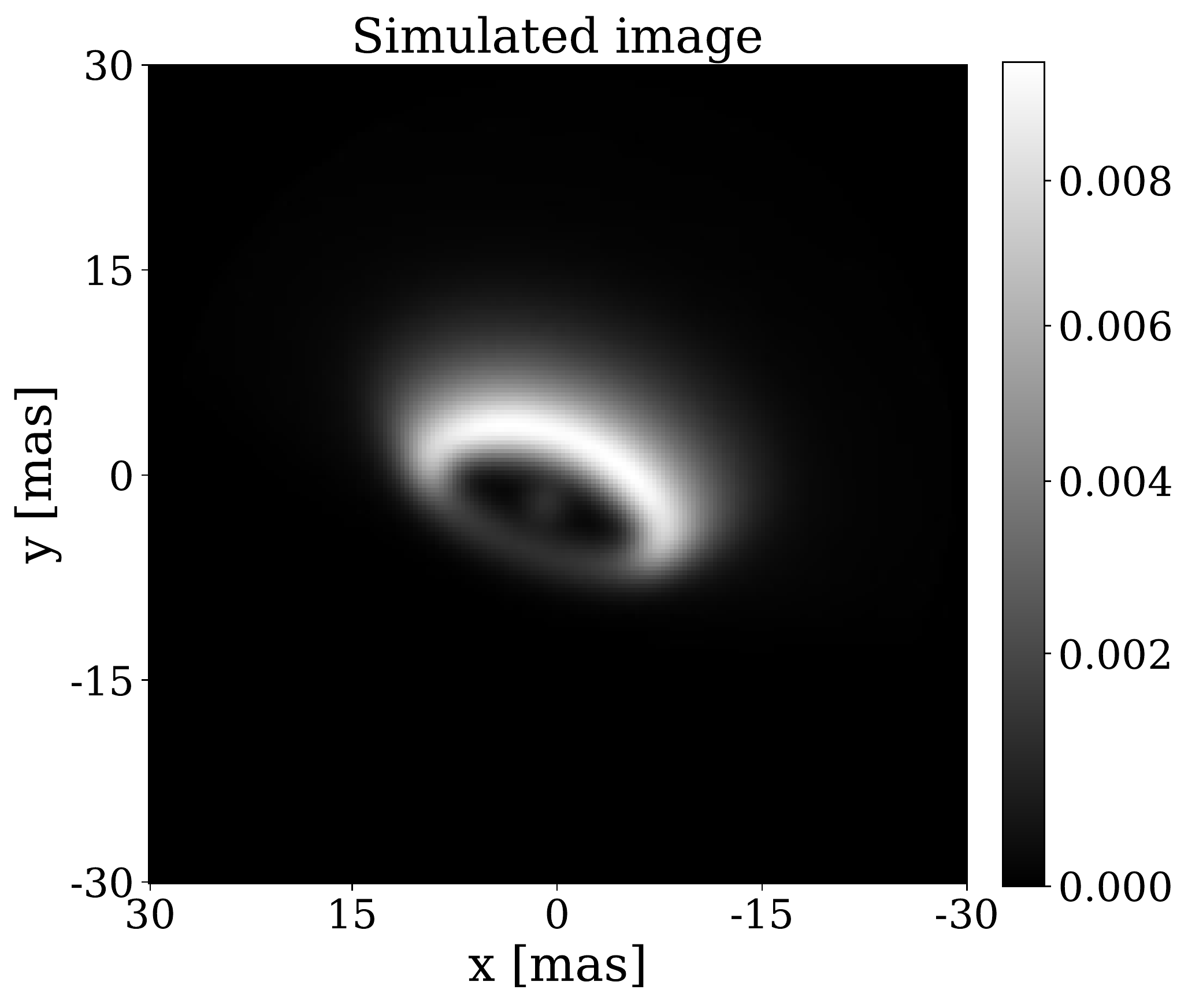}
    \includegraphics[width=0.3\textwidth]{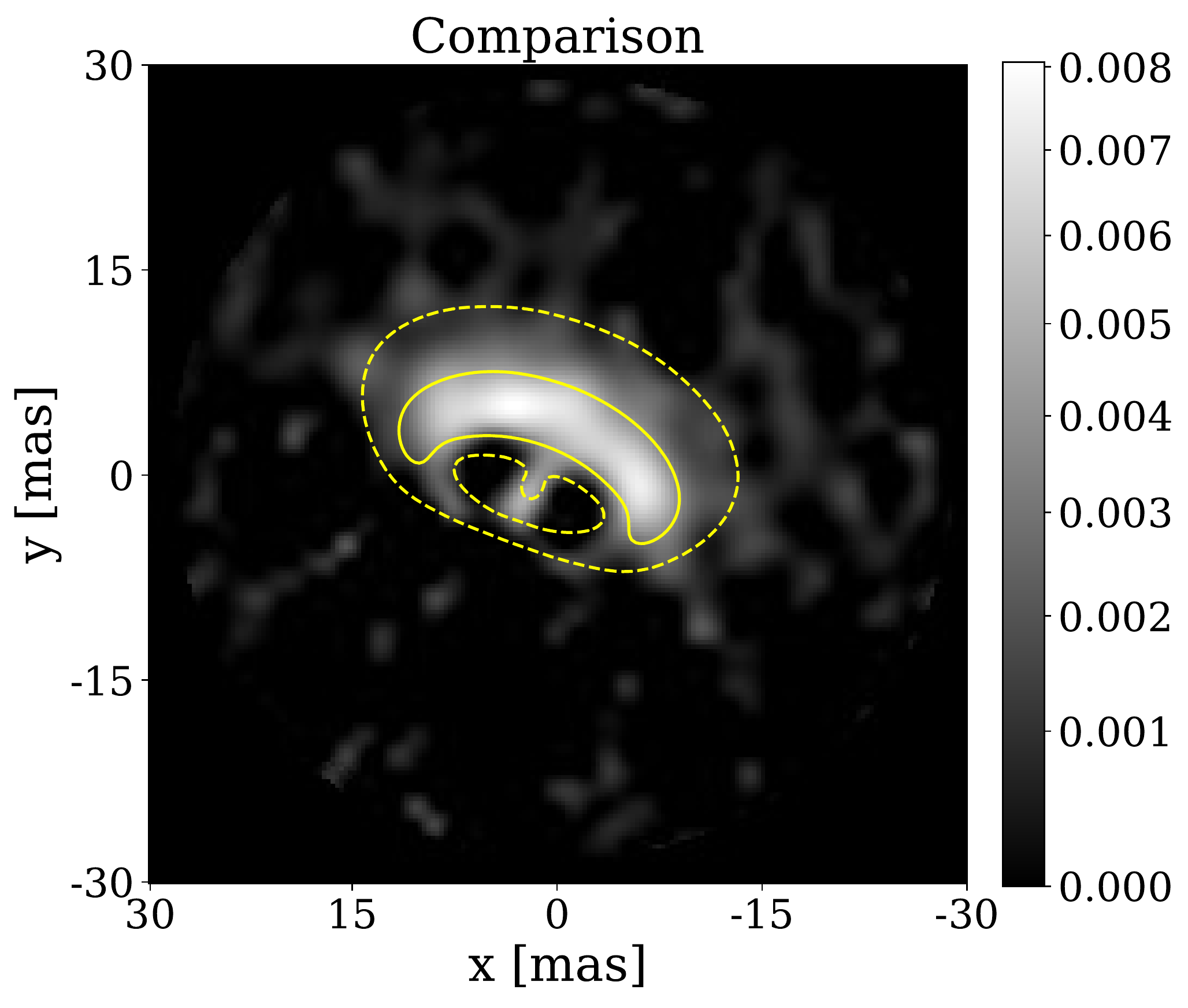}
    \includegraphics[height=0.254\textwidth]{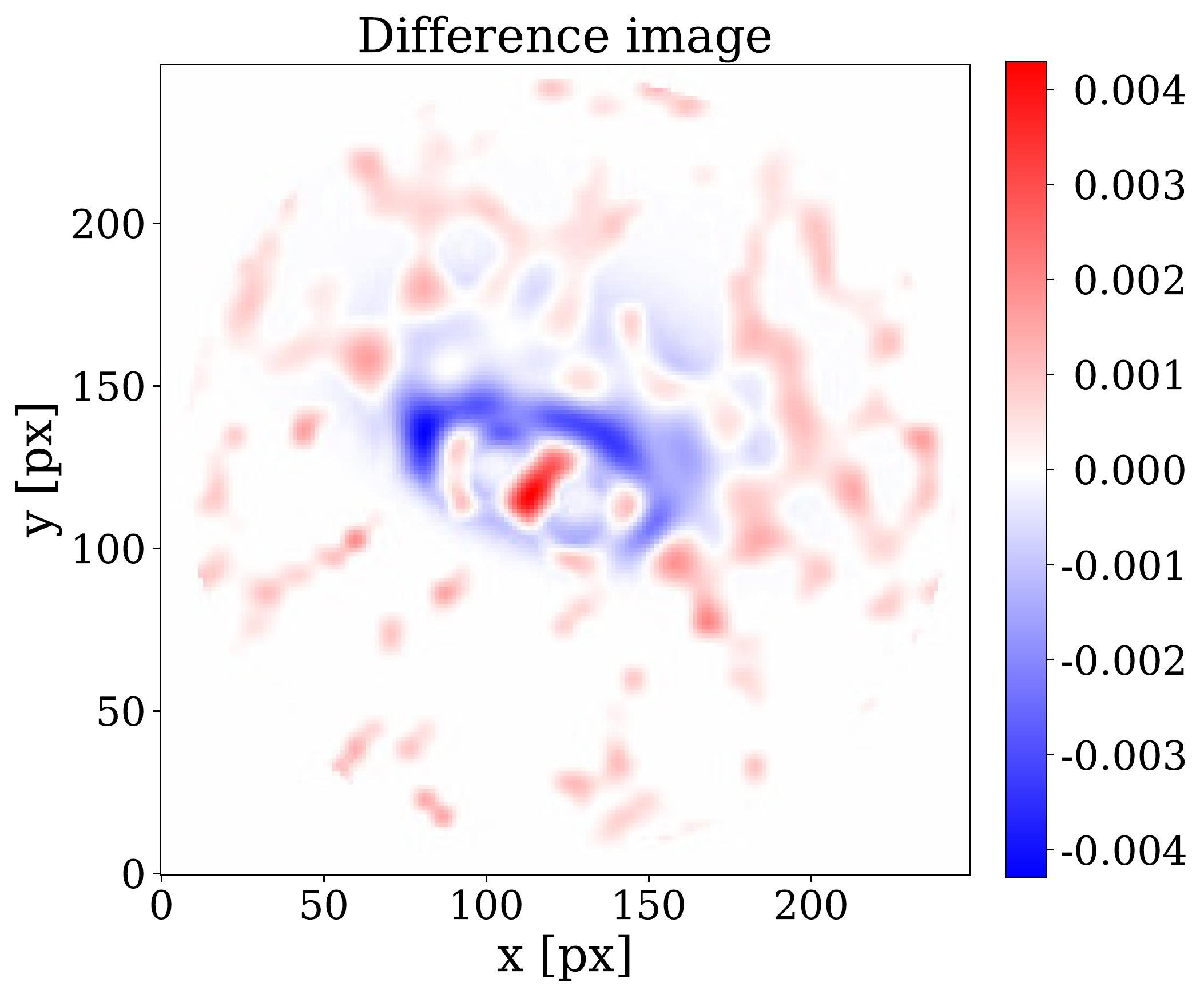}
    \caption{Comparison of the observed $L$-band image with the  image of model 3: (left) best-fit $L$-band model disk image with adjusted inclination and modified inner rim, (middle) model image contour plot over reconstructed $L$-band image, and (right) difference image (observed image minus model). Contour lines show 50\% (solid) and 10\% (dashed) levels of the maximum flux. As outlined in Sect.~\ref{subsec:fitting}, the given flux unit is relative. Rim parameter: $R_{\varepsilon} = 1.8\,R_{\rm in}$ = 9\,au, $\varepsilon = 0.7$. Inclination: $i=51^\circ$. The central object is shown in the model and observed image.}
    \label{fig:bestfit_mod_51}
\end{figure*}

\begin{figure*}[!ht]
    \centering
    \includegraphics[width=0.3\textwidth]{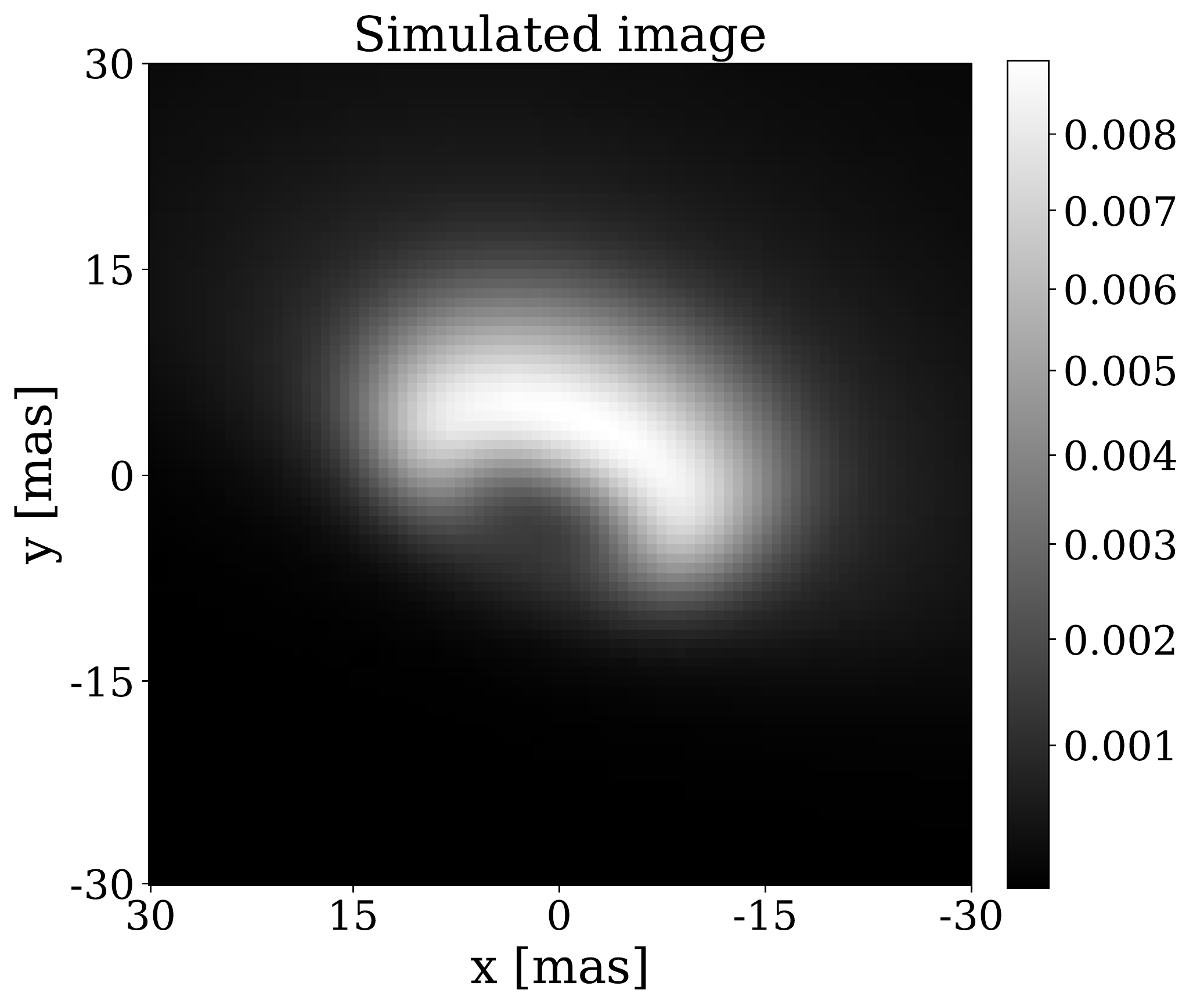}
    \includegraphics[width=0.3\textwidth]{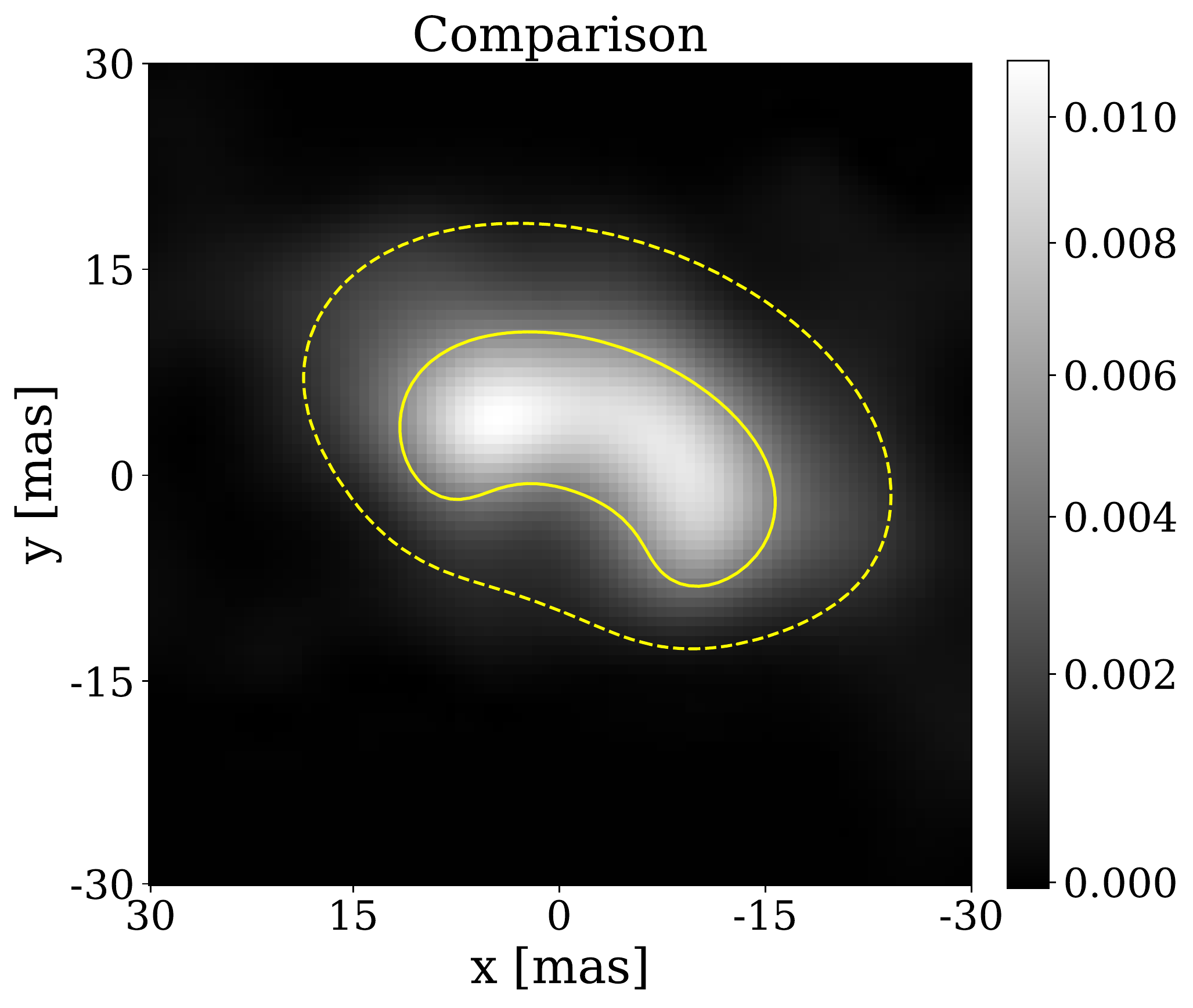}
    \includegraphics[height=0.254\textwidth]{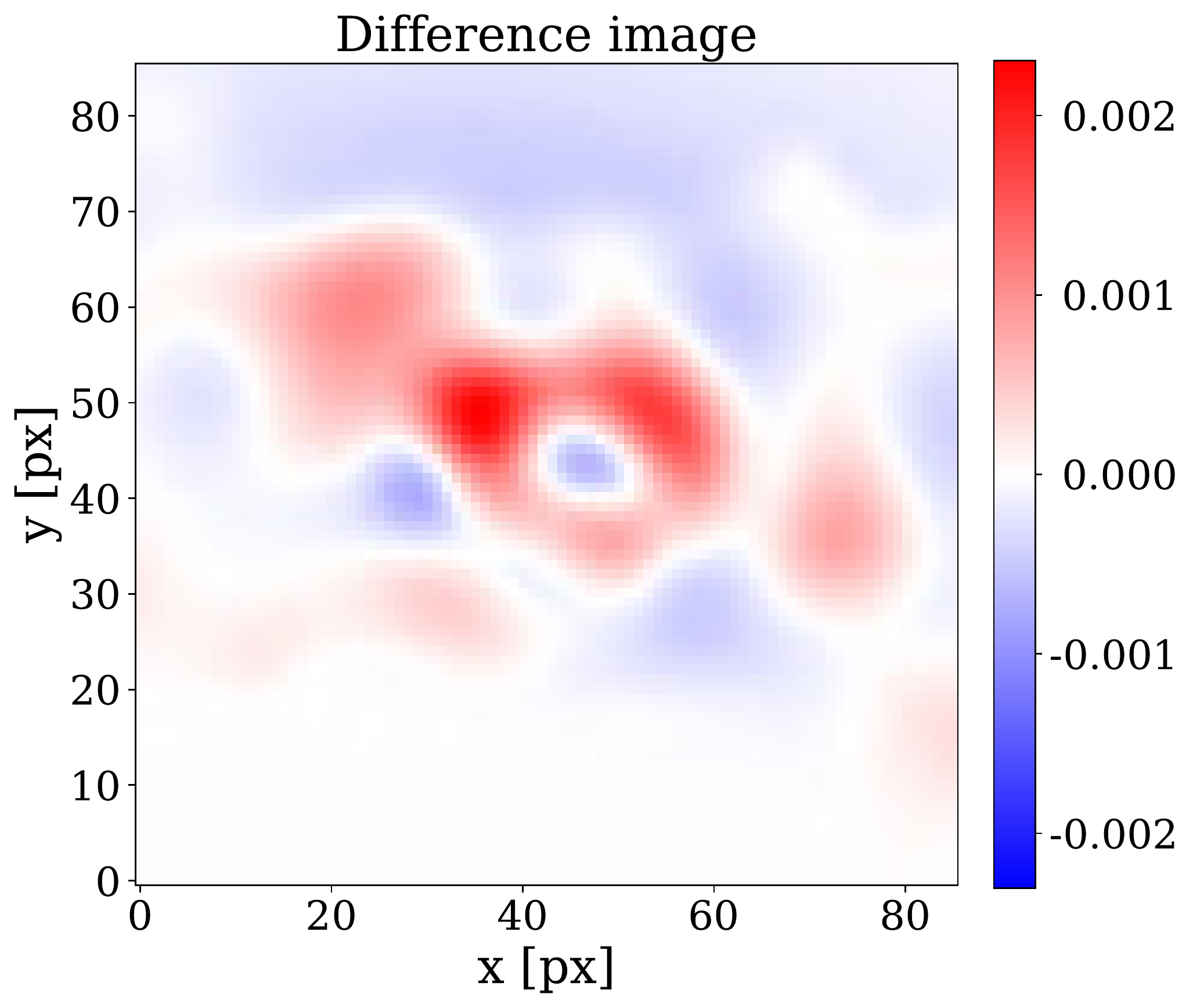}
    \caption{Comparison of the observed $N$-band image with the  $N$-band image of model 3: (left) best-fit $N$-band model image with modified inner rim, (middle)  model image contour plot over reconstructed $N$-band image,  and (right) difference image. Contour lines show 50\% (solid) and 10\% (dashed) levels of the maximum flux. As outlined in Sect.~\ref{subsec:fitting}, the given flux unit is relative. Rim parameters: $R_{\varepsilon} = 1.8\,R_{\rm in}$ = 9\,au, $\varepsilon = 0.9$. Inclination: $i=51^\circ$. }
    \label{fig:bestfit_mod_nband}
\end{figure*}

While the basic model 1 described above allows us to roughly reproduce the shape of the image of the disk as well as the radial extent of the bright region, it is not capable of reproducing the asymmetry of the ring-like brightness distribution, as the difference image in Fig.~\ref{fig:first_sim} shows. Thus, the shape of the inner rim has to be modified toward a lower vertical rim extent in our model. Besides the motivation derived from our observational finding, this modification is also in agreement with previous studies based on hydrodynamical simulations and analytical considerations for the inner rim of circumstellar disks (e.g., \citealp{IsellaNatta2005}, \citealp{Mcclure2013}, \citealp{Flock2016}). To match the brightness distribution along the inner disk rim and to take the theoretical studies into account, we expand our model by a simple parameterization of differently curved shapes of the inner rim described by the modified scale height:
\begin{equation}
    \label{eq:modrin}
    h_{\rm g}(r) = h_0(R_{\varepsilon}) \cdot \left(\frac{r - R_{\rm in}}{R_{\varepsilon}-R_{\rm in}}\right)^\varepsilon ,
\end{equation}
where $R_{\varepsilon}$ is the radius up to which the rim is modified, $h_0(R_{\varepsilon})$ is the unmodified scale height of $R_{\varepsilon}$, and $\varepsilon$ is a parameter to describe the different shapes of the inner rim. For illustration, selected examples for the different inner rim shapes are shown in Fig.~\ref{fig:epsilonplot}. We now fix the best-fit parameters described in the section above and determine the best-fit of the brightness distribution of the inner rim by sampling a parameter space of (1.2--2.4)\,$R_{\rm in}$ for $R_{\varepsilon}$ in steps of 0.2\,$R_{\rm in}$ and 0.1--1.0 in steps of 0.1 for $\varepsilon$, respectively. The resulting map of the absolute mean difference for each model can be found in Fig.~\ref{fig:diffmeangrid}. In the following, we shortly discuss the impact of the two rim parameters $\varepsilon$ and $R_{\varepsilon}$.
\paragraph{Inner rim parameter $\varepsilon$}
The parameter $\varepsilon$ allows us to modify the shape and therefore illumination of the inner rim of the model. As the case $\varepsilon = 0.1$ describes a vertical rim that extends up to $R_{\varepsilon}$ with a constant height, increasing $R_{\varepsilon}$ results in a worse fit (see Fig.~\ref{fig:diffmeangrid}). Increasing $\varepsilon$ to 0.5 changes the structure of the inner rim toward a parabolic shape. This configuration does not only result in the required radially more efficient illumination of the inner disk region, but also allows the observer to see more of the inner region on the near side of the disk. Besides the qualitatively better match of the observed inner rim structure, the absolute mean difference between reconstructed and simulated images is decreasing. Values of $\varepsilon$ up to 1 are describing the transition from a parabolic shape to a peaked, wedge-like inner rim, leading to an extension of the bright region to the outer parts. The corresponding simulated images show a broad, extended ring that is inconsistent with the reconstructed images.
\paragraph{Radius of modified inner rim $R_{\varepsilon}$}
The radius of the modified inner rim describes the point up to which the basic model is modified. It is therefore directly related to the shape of the simulated image for different values of the parameter $\varepsilon$. In the case of low values of $\varepsilon$, for example, a vertical rim with constant height, an increase of $R_{\varepsilon}$ results in an increasing mismatch between the simulated and reconstructed images. A similar trend can be found for high values of $\varepsilon$, which result in a bright extended ring, as described above, that increases even more when increasing $R_{\varepsilon}$. For shapes of the inner rim between parabolic and tapered, we see a decrease of the absolute difference up to a value of $R_{\varepsilon} =  1.8\,R_{\rm in}$. This corresponds to an extension of the bright region as well as a blurring of the bright rim. Values of $R_{\varepsilon}>1.8\,R_{\rm in}$ will not improve the model because the modified rim is extending seamlessly into the flared disk in this case, which means changes of $R_{\varepsilon}$ in this regime will not influence the shape of the inner rim. Values of $R_{\varepsilon}$ below 1.2\,$R_{\rm in}$ can also be neglected, because the modification of the rim would be too small to be noticed.\\
\\
According to the absolute mean differences measured for the different parameter sets, we find best-fit values of $R_{\varepsilon} = 1.8\,R_{\rm in}$ and $\varepsilon = 0.7$ (model 2; Fig.~\ref{fig:bestfit_mod}). We note that the model with $\varepsilon = 0.6$ is deviating less than 1\% from the best-fit case and is even best-fitting for larger values of $R_{\varepsilon}$. The simulated image, the contour plot as well as the plot of the differences for the best-fit case are shown in Fig.~\ref{fig:bestfit_mod}.
It can be seen that the difference image now has smaller values. We note that the difference plot of observed image minus model image still shows two regions with higher differences than on average. In these regions, the flux in the simulated image is higher than the measured flux on the near side.\\
To make sure that the modified rim is not affecting the best-fit parameters derived in Sect.~\ref{subsec:fitting}, we repeat the calculations for the entire parameter grid. Comparing the differences confirms the best-fit parameters depicted in the table of Fig.~\ref{fig:first_sim} except for the inclination for which we now find a best-fit value of $51^\circ$. The corresponding simulated image of this model 3, the contour plot as well as the plot of the differences are shown in Fig.~\ref{fig:bestfit_mod_51}.\\
As a further check, we simulate the corresponding $N$-band image of our model and the respective difference images. The best-fit parameter values in this case are $R_{\varepsilon} = 1.8\,R_{\rm in}$ and $\varepsilon = 0.9$. In contrast to the best-fit of the reconstructed $L$-band image, the simulated $N$-band images with $\varepsilon =$ 0.7 or 0.8 are of comparable goodness, as their respective absolute mean differences are also within the measured uncertainties and are deviating by only 3\%. The simulated image, contour plot and the plot of the differences between simulated and observed images of the $N$-band best-fit parameter values are shown in Fig.~\ref{fig:bestfit_mod_nband}. The best-fit parameter values derived on the basis of the reconstructed $L$-band image fits the $N$ band equally well. As in the case of the $L$ band, the differences are on the order of the uncertainties of the reconstructed image. The trends found for the parameters describing the shape of the inner rim as depicted in Fig.~\ref{fig:diffmeangrid} for the $L$-band image can also be confirmed with the $N$-band simulations. Thus, we consider the best-fit model 3 is consistent with the reconstructed $L$- and $N$-band images.

 \begin{figure}
 \centering
    \hspace{-0mm} \vspace{0mm}
     \includegraphics[height=55mm,angle=0]{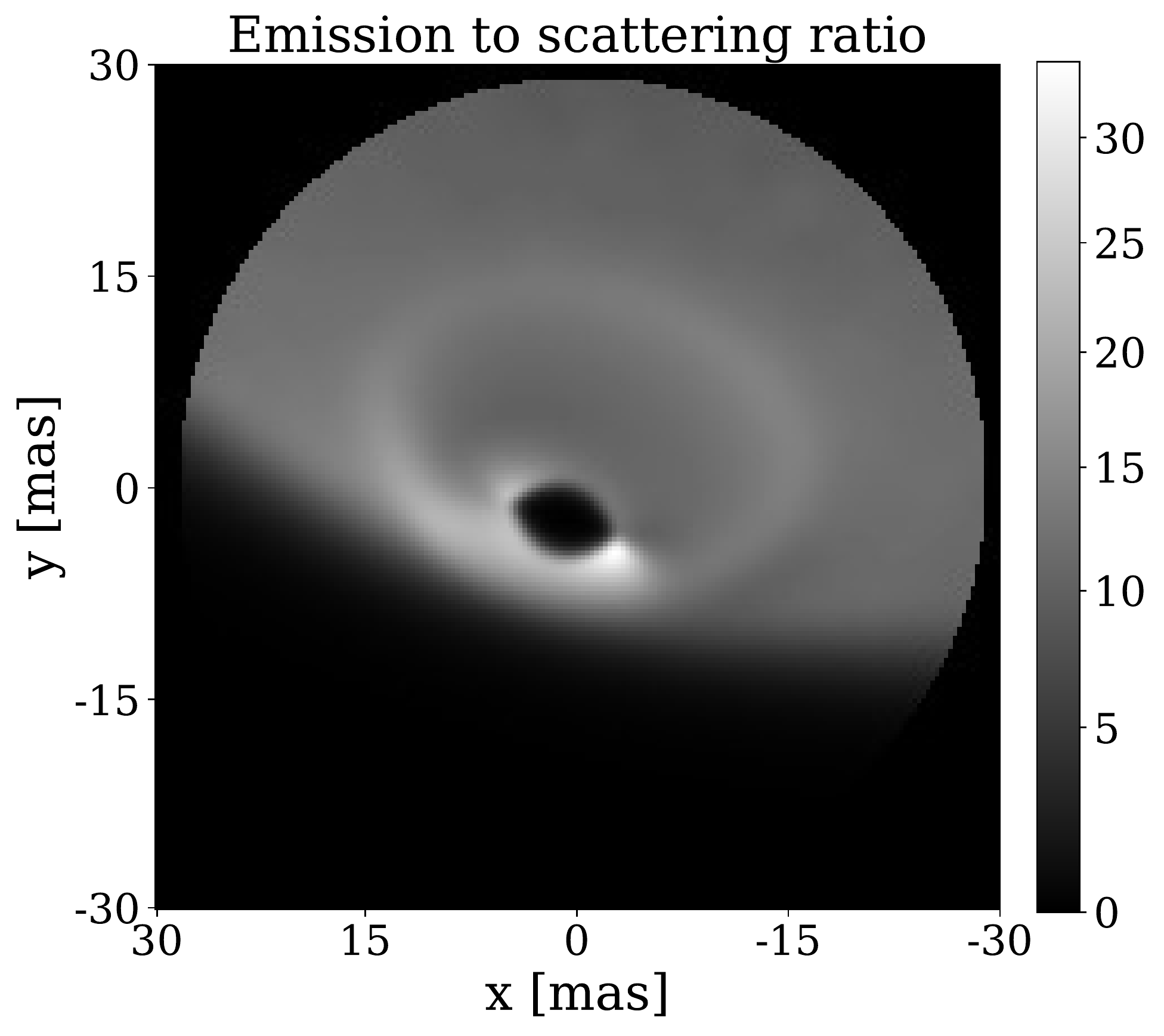}
     \includegraphics[height=55mm,angle=0]{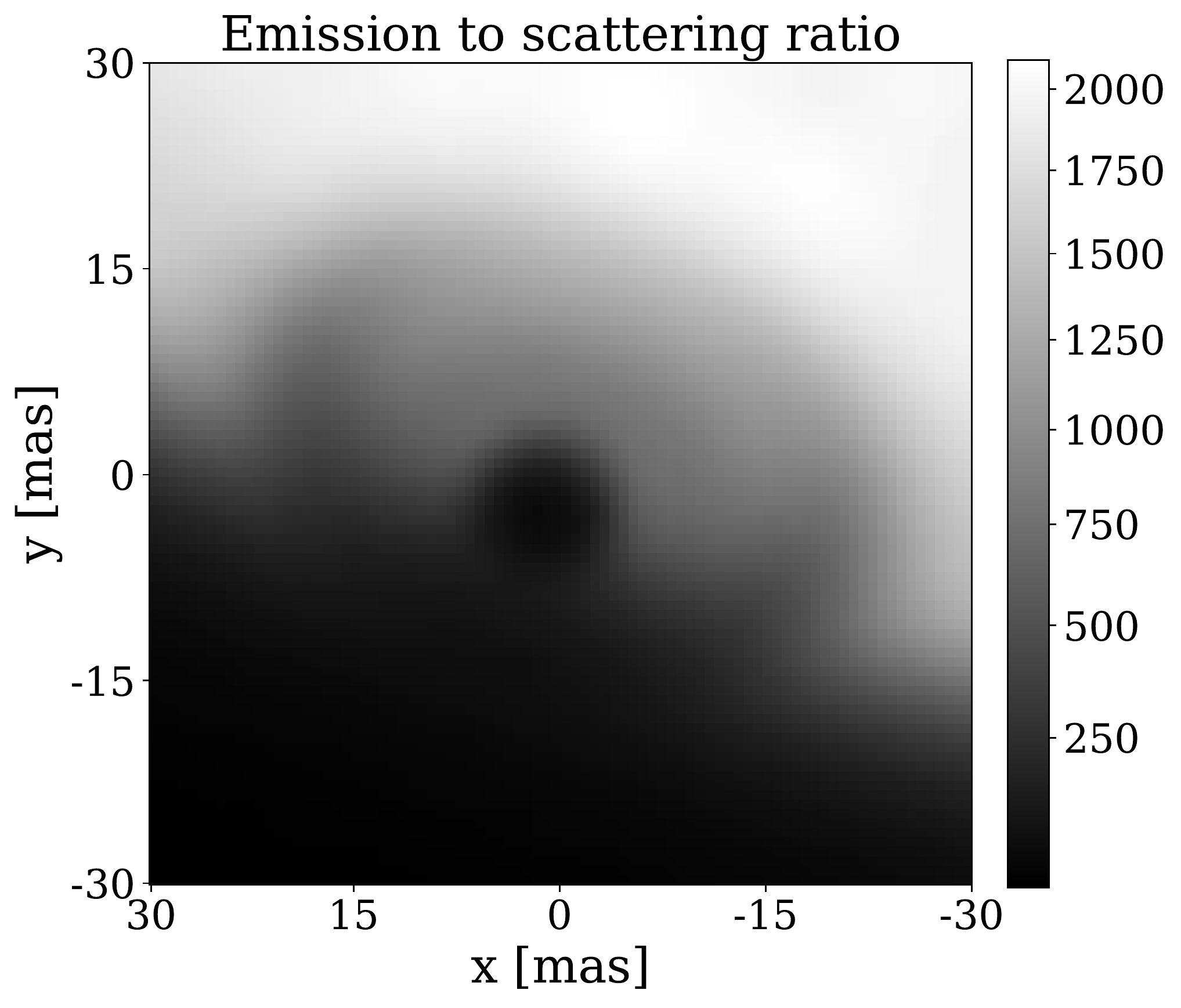}
      \caption[]{\small
              Ratio of thermal to scattered emission of the $L$-band (top) and $N$-band (bottom) model. Direct emission of the star is included in the scattered contribution. The ratio of the total thermal emission to total
              scattered  and direct stellar emission is 10 in case of the $L$-band image and 380 in the case of the $N$-band image.     
    }
    \label{fig:ratio_emission}
\end{figure}

\subsection{Total flux} \label{subsec:totflux}

We measure a total $L$-band model flux of 35 Jy in the model FOV of 80 mas, while we derive an observed flux of 47 Jy (estimated uncertainty about $\pm50\%$) from the FS CMa interferograms in their larger 0.6 arcsec FOV of the $L$-band interferograms. The observed flux was derived by comparing  the integrated flux of  the interferograms of the target and the  calibrator stars. In the $N$-band, we obtain a total model flux of 110 Jy in the model FOV of 360~mas and derived an observed flux of 213 Jy ($\pm50\%$) from the FS CMa $N$-band interferograms in their FOV of 2.3 arcsec. In the model, we are thus covering about 3/4 of the observed  flux in the $L$ band, but only 1/2 of the flux in the $N$ band. The differences may be caused by the flux from regions outside the model FOV.

We note that the thermal model emission always outweighs the scattered contribution by a factor of 10 in the $L$ band and a factor of 300 in the $N$ band, respectively. A detailed image of the ratio between thermal and scattered emission for both bands can be found in Fig.~\ref{fig:ratio_emission}.



\subsection{Model image with pure astrosilicate and a dust envelope with constant density (model 4)} \label{subsec:FluxRatio}

\begin{figure*}[!ht]
    \centering
    \includegraphics[width=0.3\textwidth]{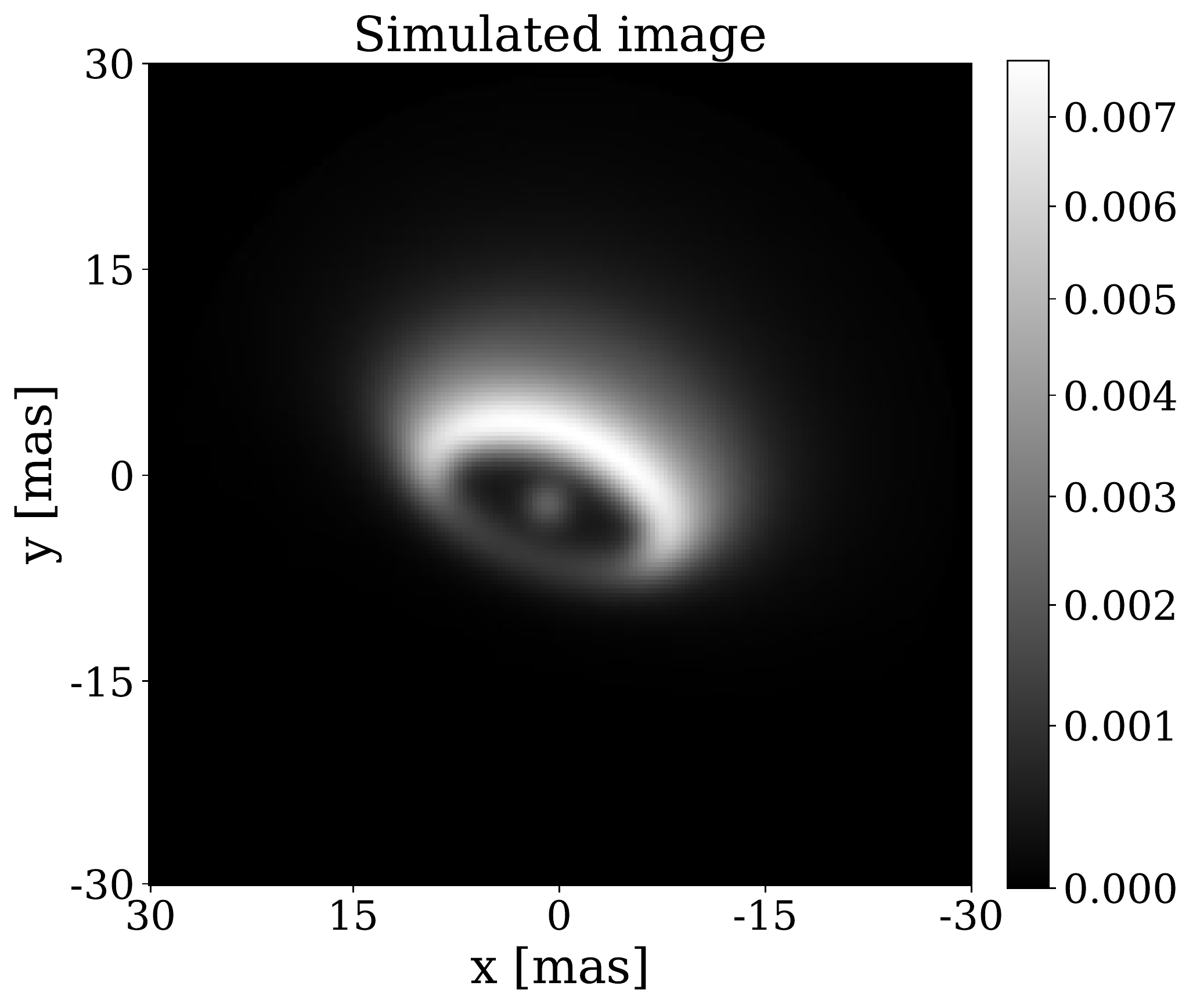}
    \includegraphics[width=0.3\textwidth]{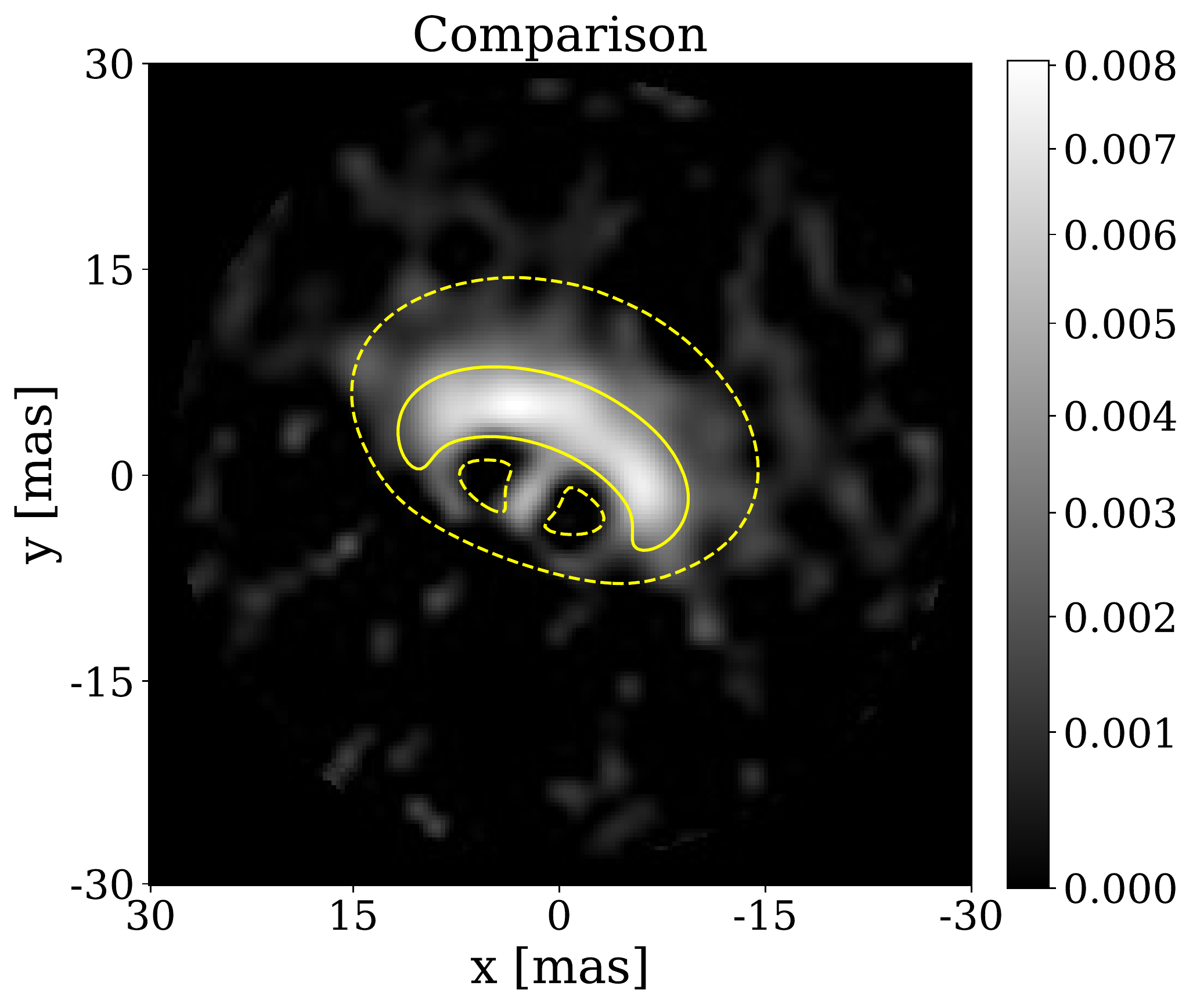}
    \includegraphics[height=0.254\textwidth]{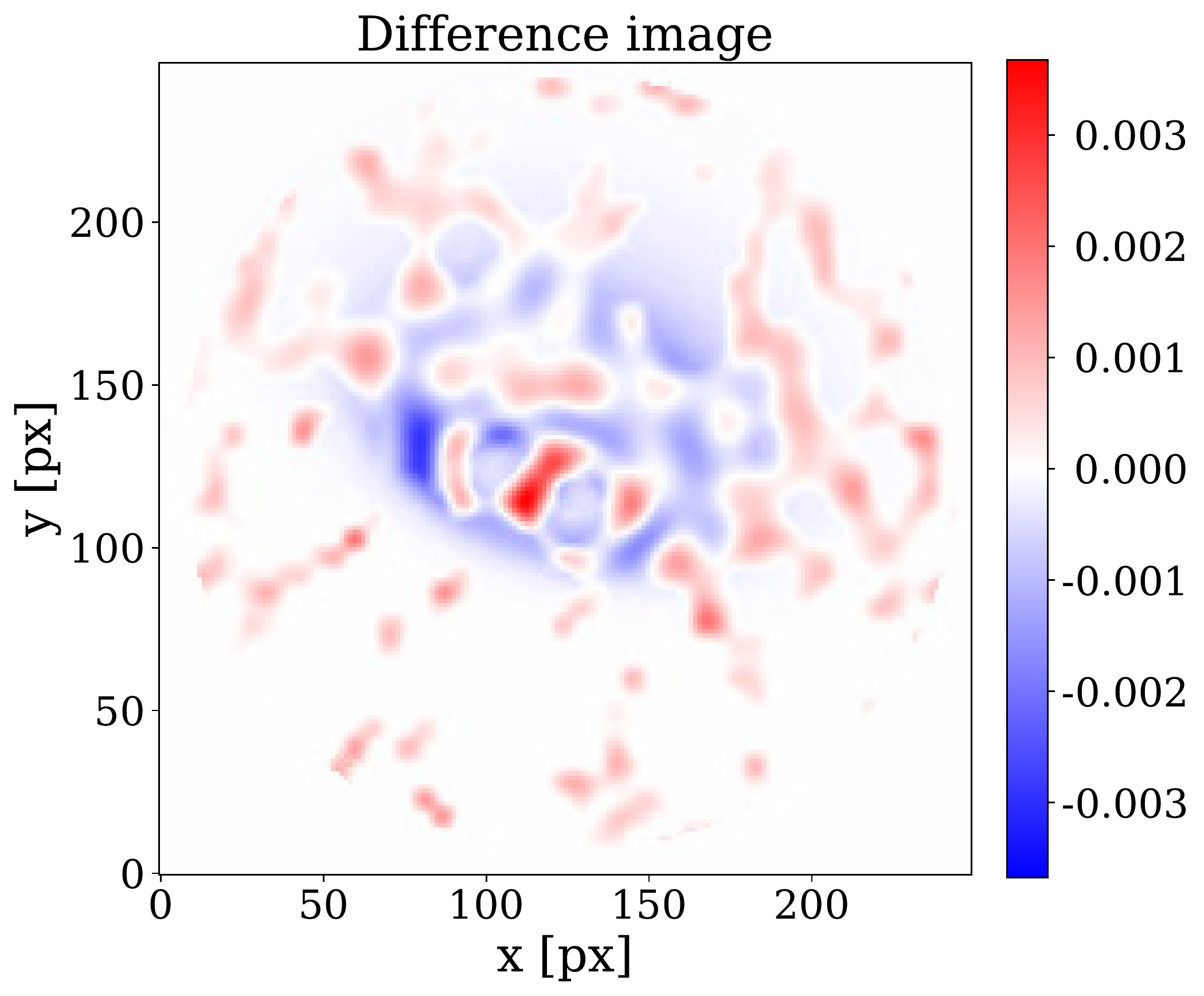}
    \caption{Comparison of the observed $L$-band image with the  image of model 4: best-fit $L$-band model image with pure astrosilicate and a dust envelope with constant density and $\tau=0.1$ (left), contour plot over reconstructed $L$-band image (middle) and difference image (right). Contour lines show 50\% (solid) and 10\% (dashed) levels of the maximum flux. As outlined in Sect.~\ref{subsec:fitting}, the given flux unit is relative. Rim parameter: $R_{\varepsilon} = 1.8\,R_{\rm in}$ = 9\,au, $\varepsilon = 0.7$. Inclination: $i=51^\circ$.}
    \label{fig:bestfit_mod_env}
\end{figure*}

Model 3 presented in Fig.~\ref{fig:bestfit_mod_51} in the Sect.~\ref{subsec:modrin}  is able to reproduce the asymmetric structure of the inclined FS~CMa disk. However, the brightness of the central object in the  observed $L$-band image and the model image disagree. In the observed $L$-band image,  the integrated $L$-band flux of the central object is about $(4.5\pm0.9)\%$ of the total image flux, whereas in the model, the $L$-band flux from the central object is only 0.61\%.

Even though the central star shows photometric variability of more than a magnitude within years as discussed for example in~\cite{deWinter1997}, the stellar parameters have been well constrained by multiple previous studies (e.g.,~\citealp{Vioque2018},~\citeads{2020ApJ...888....7L}).
Increasing the luminosity of the star within the measured uncertainties will increase the flux scattered and reemitted by the disk and therefore hardly affect the star-to-disk brightness ratio. This also holds true for the effective temperature of the star which when increased, shifts the stellar emission toward shorter wavelengths, resulting in a decrease of the star-to-disk brightness ratio in the $L$ and $N$ bands.\\

We therefore turn our attention toward the disk properties to analyze the outlined discrepancies. The luminosity of the disk can be decreased by either reducing the illumination of the inner region or by choosing a dust mixture with different optical properties. For the first, we find that varying the flaring parameter $\beta$ and the scale height $h_{100}$ around the corresponding best-fit values in order to modify the disk illumination is not significantly increasing the star-to-inner rim brightness ratio. Concerning the dust properties, we conduct simulations with grain sizes up to $10\,\mu$m and find that the brightness ratio is decreasing because of an increase of the scattered radiation. Moreover, changing the dust composition to pure astronomical silicate could reduce the flux caused by dust reemission while not affecting the amount of scattered radiation and therefore reduce the brightness ratio. Corresponding simulations of our best-fit model with pure astronomical silicate show that the star-to-disk brightness ratio can be increased by a factor of 2--3 while preserving the shape of the inner rim. This finding might indicate that the dust composition is different from the assumed silicate-graphite mixture. However,  it should also be noted that the measured total flux in this case is even lower and covers only a third of the observed flux.\\

The offset between the visibilities at low spatial frequencies of the simulated image and the observational data 
suggests a constant background flux that is missing in the simulated image. A very faint constant background flux would not be visible in the reconstructed image and was therefore not considered in the fitting process in Sect.~\ref{subsec:fitting} and~\ref{subsec:modrin}. To account for this, we expand our disk model by adding an optically thin dust sphere with an inner radius of 5~au and an outer radius wider than the FOV of the reconstructed image. We test two simple dust density distributions for two different optical depths: a sphere with a) a constant density and b) an 1/r density profile for $\tau=0.1$ and $\tau=1$. 
In case of the $\tau=1$ models, the simulated images show additional flux in regions that show no flux in the reconstructed images. We therefore focus on the optically thin case.
Since there are no significant differences between the simulated images resulting from the two density distributions in the optically thin case ($\tau=0.1$), we choose the simpler, constant density distribution for the subsequent discussion. This faint background, caused by the optically thin dust sphere, together with the assumed pure astrosilicate dust  improved the fit of the model visibilities and contributed to the following new star-disk flux ratio described below. The simulated image, contour plot, and the plot of the differences between simulated and observed images of the $L$-band best-fit model 4 with pure astrosilicate are shown in Fig.~\ref{fig:bestfit_mod_env}. In this model, the integrated flux from the central object increased from 0.61\% (in the model in Fig.~\ref{fig:bestfit_mod_51}) to 1.1\%, but is still smaller than the observed $(4.5\pm0.9)\%$ flux fraction. This means that model 4 with pure astrosilicate dust and an optically thin dust envelope changed the flux ratio of central object and dust disk by almost a factor of two, but the remaining flux ratio discrepancy is still a factor of four. Therefore, we discuss a more promising alternative in Sect.~\ref{subsec:gasdisk}, the possible impact of a gaseous Be star disk.

In the previous sections, we have developed disk models that approximately agree with the observed image of the disk, but the modeled flux fraction of central object (0.61--1.1\% of the total flux of central object plus dust disk) is lower than in the reconstructed image (where it is 4.5\%).  This disagreement may be caused by a compact gaseous disk surrounding the central star  (Sect.~\ref{subsec:gasdisk}).
Because of this gaseous disk, the central object of FS~CMa seems to be much more complicated than just a normal star.  This is probably the reason, why our model, which does not include a gaseous disk, is not able to reproduce the correct flux ratio of central object and dust  disk.  Because of this uncertainty, our main goal is only to model the shape of the dust disk. Modeling of the entire complex system of dust disk and central object is out of the scope of this paper.\\

In this complex situation, it became clear that the just applied modeling criterion of using the difference images of model and reconstructed images has the advantage that we can find the best-fit disk image by ignoring the reconstructed image of the central object.  For this goal, we first masked the central object in previous sections before we calculated the average absolute value of the difference images to obtain an integral measure of the observation-model difference. In the next Sect.~\ref{subsec:cpvResiduals}, we use another important criterion to find the best-fit image, namely the differences between modeled and observed closure phases and visibilities.

\subsection{Comparison of the closure phases and visibilities derived from the model intensity distribution to the measured closure phases and visibilities}  \label{subsec:cpvResiduals}

 
Instead of calculating the differences between reconstructed and the modeled images to constrain model parameters (as in the previous sections), we now compute the average absolute value of the difference  between the observed and  model visibilities and closure phases called ``model-observation residuals".


The model-observation residuals of the closure phases and the visibilities for the $L$-band model 3 image (Fig.~\ref{fig:bestfit_mod_51}) 
 are $43^\circ$ and 0.040, respectively. These big residuals are not surprizing because we know already from the previous sections that the model flux ratio of the central object and the disk disagrees with the observations.

These $L$-band model-observation residuals are very big in comparison to another important type of residuals, the ``reconstruction-observation'' residuals, which are the residuals between the visibilities and closure phases of the reconstructed image  (Fig.~\ref{reconstructions}) and the observed visibilities and closure phases. The reconstruction-observation residuals are a measure of the quality of the reconstructed image. 
The reconstruction-observation residuals for the $L$-band  image in Fig.~\ref{reconstructions} are only $2.9^\circ$  and 0.009 for closure phase and visibilities, respectively (see Table~\ref{list1}).

\begin{figure}
    \hspace{0mm} \vspace{0mm}
     \includegraphics[height=90mm,angle=270]{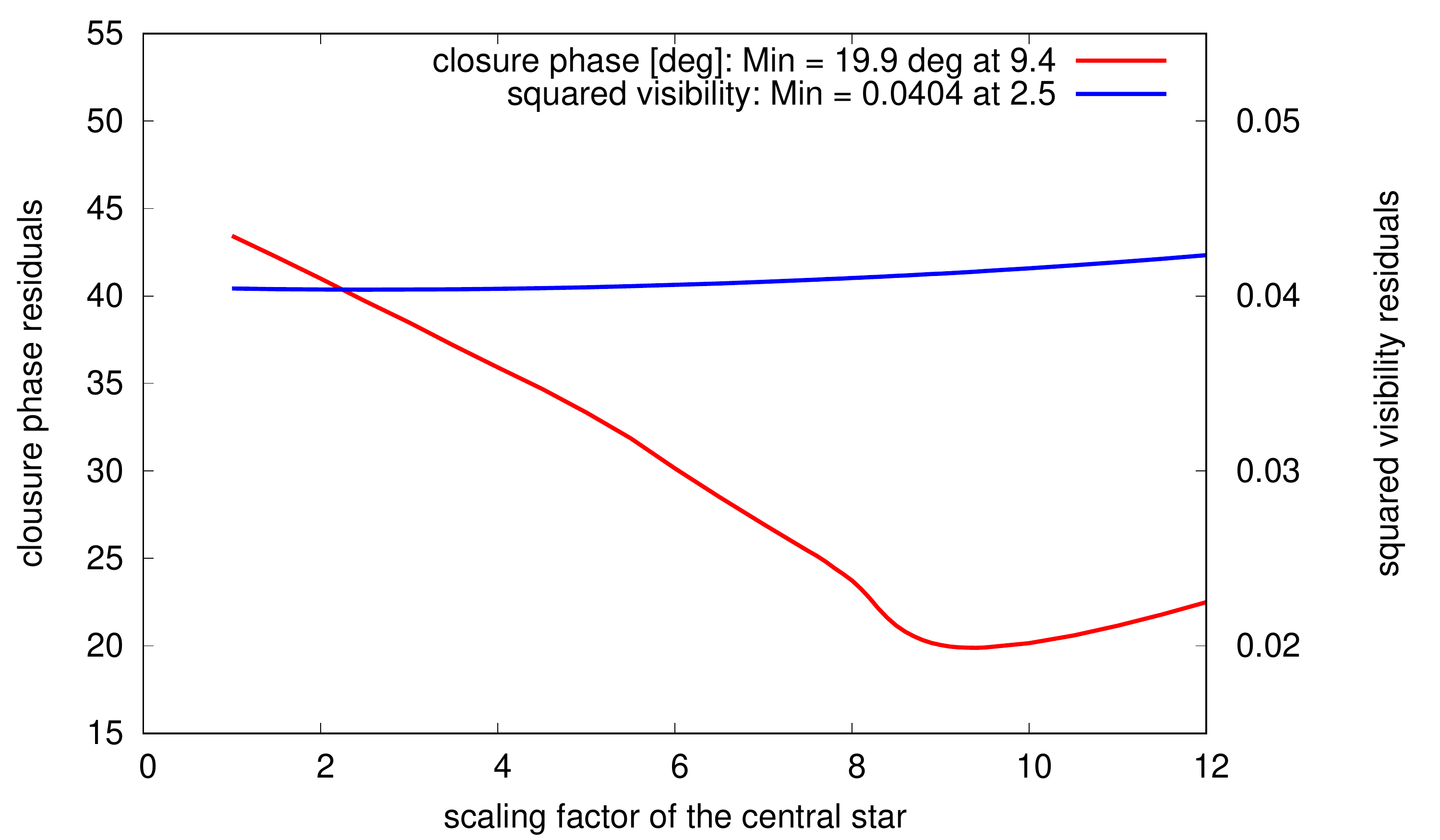}
      \caption[]{\small
               Closure phase and visibility residuals of the best-fit $L$-band model 3 shown in Fig.~\ref{fig:bestfit_mod_51} as function of the central star scaling factor.    }
    \label{fig:ResidualsLbandMod51deg.scale}
\end{figure}

\begin{figure}
    \hspace{0mm} \vspace{0mm}
     \includegraphics[height=90mm,angle=270]{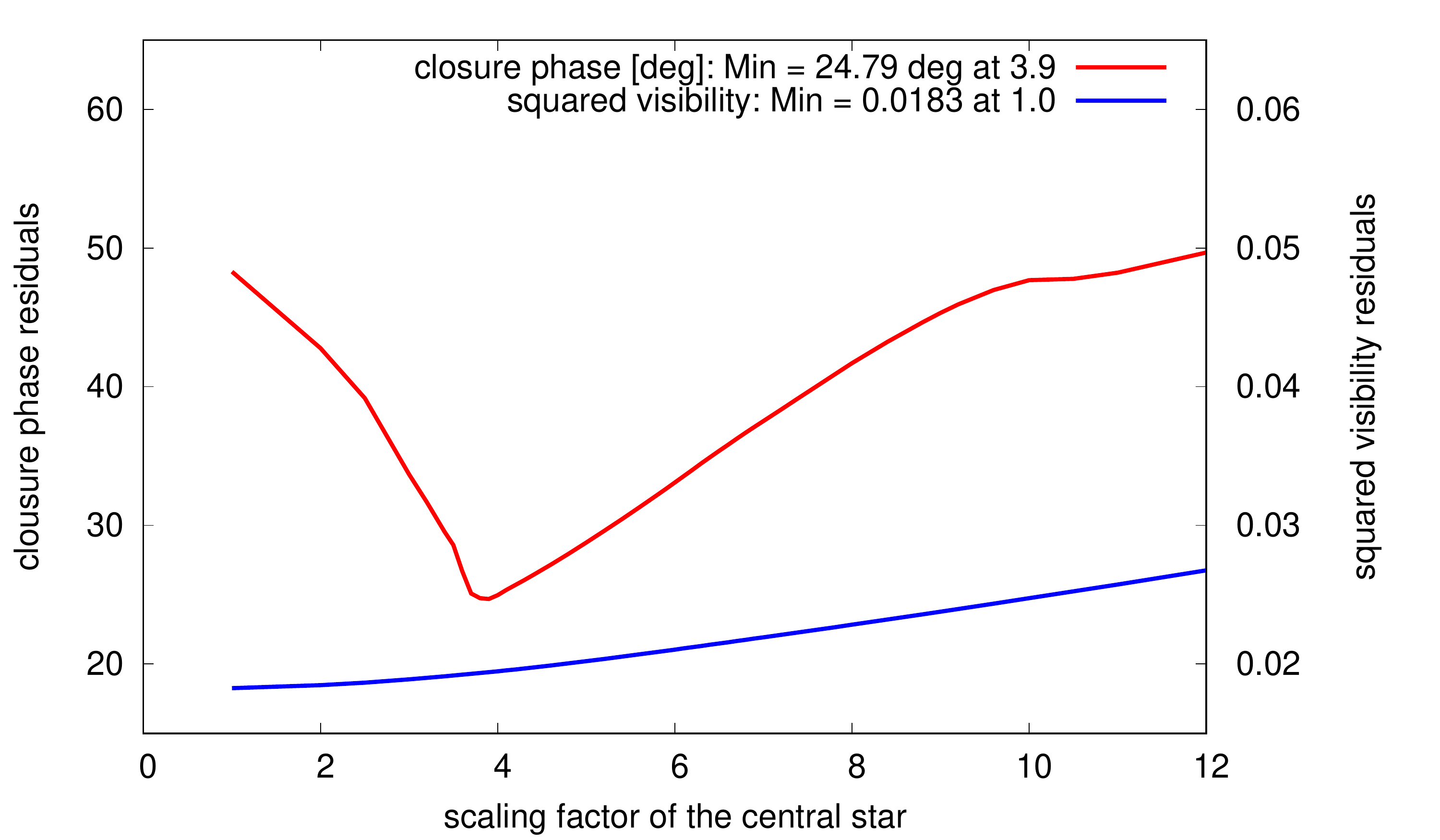}
      \caption[]{\small
              Closure phase and visibility residuals of the modified best-fit $L$-band model 4 shown in Fig.~\ref{fig:bestfit_mod_env} as function of the of the central star scaling factor.
    }
    \label{fig:ResidualsLbandMod51degEnv.scale}
\end{figure}

One of the reasons for the big aforementioned model-observation residuals (closure phase: $43^\circ$, visibility: 0.040)  is the small flux fraction  of the central object (only 0.61--1.1\% of the total flux) of the models 3 and 4 (Figs.~\ref{fig:bestfit_mod_51} and \ref{fig:bestfit_mod_env}). Other reasons are most likely  additional disk asymmetries or clumpiness, which are not included in the modeling.
This strong dependence of the  model-observation residuals on the too small flux fraction of the central object is problematic if we want to use the model-observation closure phase and visibility residuals to constrain the parameters of the model disk alone because these residuals depend on both the required dust disk parameters and the wrong flux ratio of central object and dust disk.  We can constrain the disk parameters only if we avoid the influence of the wrong flux ratio of central model star and disk by compensating the wrong flux ratio, which can approximately be performed by applying a suitable flux scaling factor to the central model star. This scaling factor is only needed as a tool to evaluate the model-observation residuals of the disk and is not used in radiative transfer modeling.
For these reasons, we first compute the dependence of the model-observation residuals on the scaling factor of the model central star.
The results are shown in  Figs.~\ref{fig:ResidualsLbandMod51deg.scale} and \ref{fig:ResidualsLbandMod51degEnv.scale} for our two models 3 and 4 presented in Figs.~\ref{fig:bestfit_mod_51} and \ref{fig:bestfit_mod_env}.

Figure~\ref{fig:ResidualsLbandMod51deg.scale} shows the closure phase and visibility residuals of  the model 3 in Fig.~\ref{fig:bestfit_mod_51} (i.e., model-observation residuals) as a function of different scaling factors for the brightness of the model central star. The minimum closure phase residuals of $19.9^\circ$ are achieved with a scaling factor of 9.4. Figure~\ref{fig:ResidualsLbandMod51degEnv.scale} shows the residuals of model 4 (modified model image displayed in Fig.~\ref{fig:bestfit_mod_env})  as a function of the scaling factors. In this case, the minimum closure phase residuals of $24.7^\circ$ are obtained with a scaling factor of 3.9, whereas the closure phase residuals in the case of no scaling are $48.2^\circ$. The remaining closure phase residuals of $\sim$25$^\circ$ are most likely caused by additional asymmetries 
in the dust disk, which are not taken into account in the models. This discrepancy is also 
shown in Fig.~\ref{LbanddataRecModel}, where the measured interferometric data are plotted together with the interferometric data derived from the two models 3 and 4,
but modified with the above model star scaling factors 9.4 and 3.9, respectively.

In addition to the above model-observation residual method, we used the $\chi^2$ method to determine the average deviation of the model visibilities and closure  phases  from the observed values. In Appendix~\ref{chi2method}, we show that the $\chi^2$ method provides similar scaling factors of 9.0  and 3.7 (instead of the above 9.4 and 3.9) for the two models 3 and 4, respectively.

The model star scaling factors derived from the closure phase residuals are approximately consistent with the modeling results obtained when we used the absolute mean differences between the observed and modeled images as criterion. In this case, the model images have a flux fraction in the central object of only 0.61\% (model 3) or 1.1\% (model 4) of the total flux, whereas the flux fraction of the central object is (4.5 +/-0.9)\% in the observed $L$-band image, corresponding to ratio of 4.5/0.61=7.4  (model 3) and 4.5/1.1=4.1 (model 4). These model-reconstruction deviations are similar to the model central star scaling factors of 9.4 (model 3) and 3.9 (model 4) derived from the closure phase residuals.

\begin{figure}
    \hspace{0mm} \vspace{0mm}
     \includegraphics[height=90mm,angle=270]{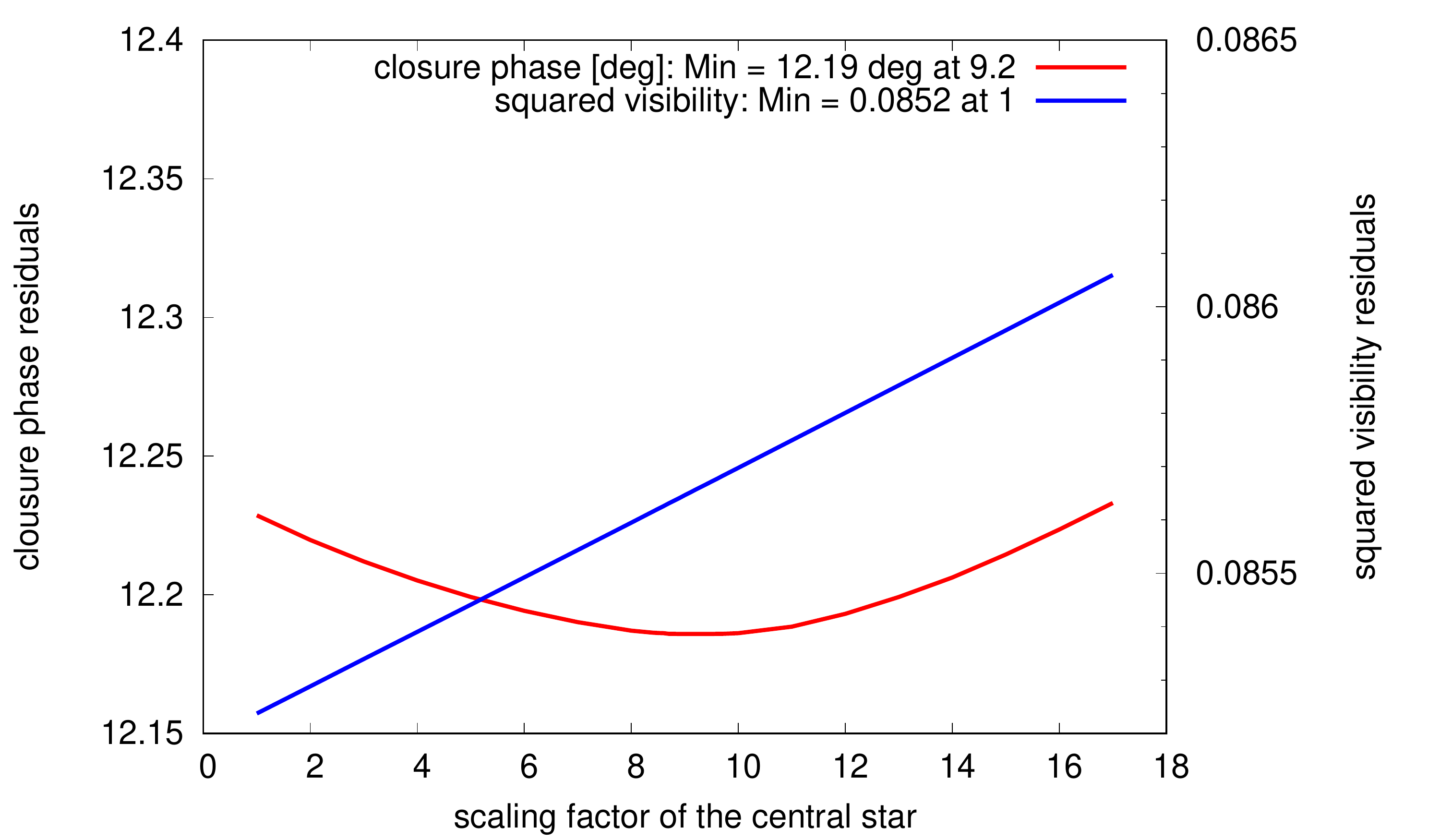}
      \caption[]{\small
             Closure phase and visibility residuals of the modified best-fit $N$-band model shown in Fig.~\ref{fig:bestfit_mod_nband} as a function of the central star scaling factor.     }
    \label{fig:ResidualsNbandMod51deg.scale}
\end{figure}

Figure~\ref{fig:ResidualsNbandMod51deg.scale} shows the model-observation residuals of the  $N$-band model 3 image displayed in Fig.~\ref{fig:bestfit_mod_nband}  as a function of the different scaling factors of the central star. The minimum closure phase residuals of $12.19^\circ$ are obtained with a central star scaling factor of 9.2, whereas the closure phase residuals in the case of no scaling are $12.23^\circ$. The difference between scaling and no scaling of the central star is very small because the integrated stellar intensity is only 0.03\% of the total flux of the best-fit $N$-band model. In Fig.~\ref{NbanddataRecModel},  the measured interferometric data are plotted together with the interferometric data derived from the $N$-band reconstruction shown in Fig.~\ref{reconstructions}  and derived from the model displayed in Fig.~\ref{fig:bestfit_mod_nband} (model 3), but modified with the above scaling factor 9.2. 

We have shown by three  different methods that the model flux ratio of the central star and the total flux (of disk and central object) is about 3--9 times too small. In Sect.~\ref{subsec:gasdisk} we discuss a likely cause for this discrepancy. As we discussed in previous sections, our goal is only to constrain model parameters of the dust disk. The modeling of the complex star-disk system is out of scope of this paper because of the complex central object discussed in Sect.~\ref{subsec:gasdisk}.

Therefore, we focus now on the application of the model-observation closure phase residuals and the aforementioned $\chi^2$ results to constrain the same disk shape parameters $\rm R_{\epsilon}$ and $\epsilon$, which were constrained in Sect.~\ref{subsec:modrin} using the presented difference between the reconstructed image and the model image. A third required free modeling parameter is the aforementioned scaling factor of the brightness of the central star because of the derived flux ratio discrepancy.

\begin{figure}
    \hspace{0mm} \vspace{0mm}
     \includegraphics[height=90mm,angle=270]{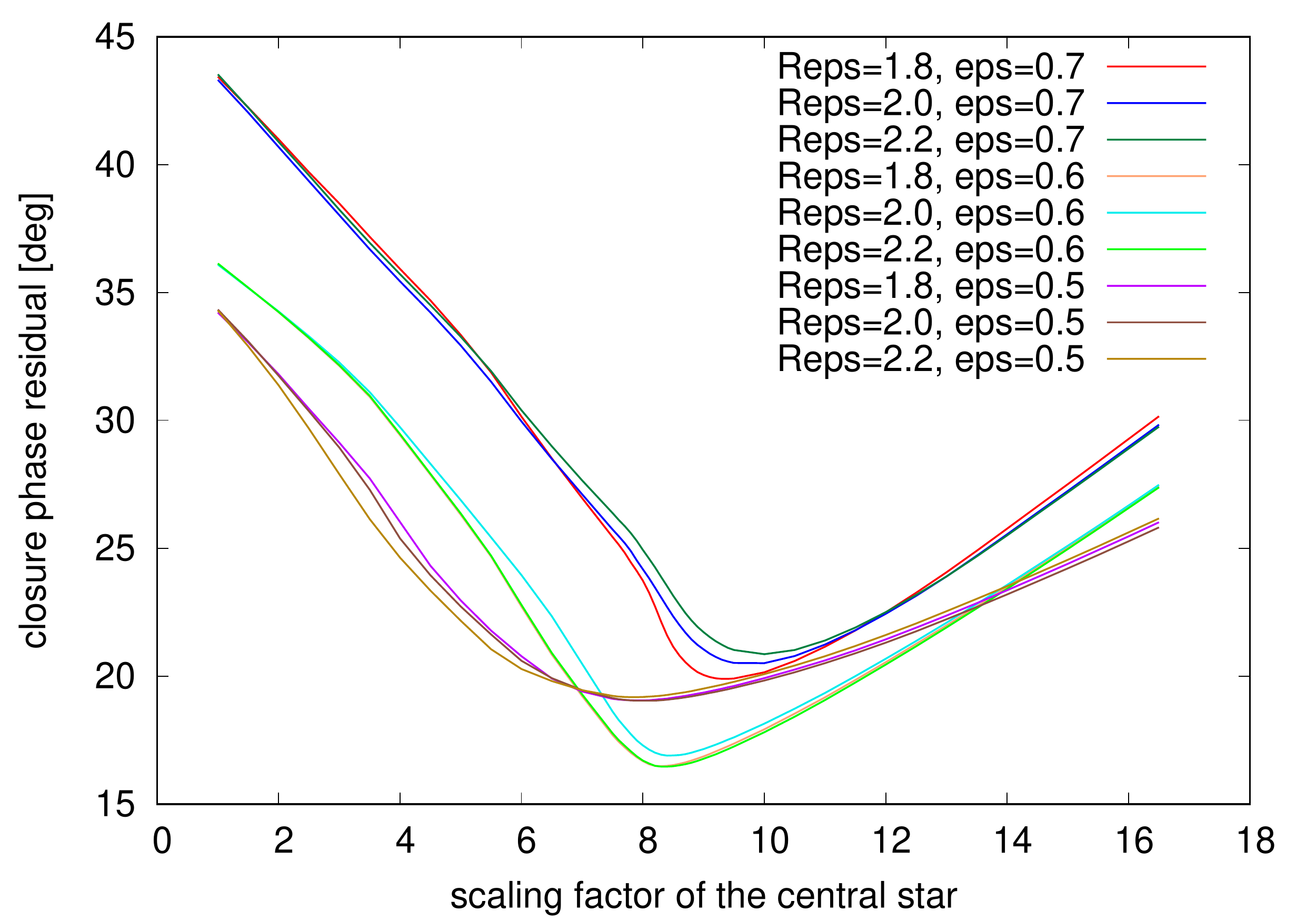}
      \caption[]{\small
               Closure phase residuals of  $L$-band models (different versions of model 3) with different inner rim parameters $\varepsilon$ and $R_{\varepsilon}$  as a function of the central star scaling factor. The best-fit models are those two with $\epsilon$ = 0.6 and $R_{\epsilon}$ = 1.8 and 2.2 $R_{\rm in}$ at the central star scaling factor 8.4. }
    \label{fig:Lbandmodel.scale}
\end{figure}

\begin{figure}[!ht]
    \centering
    \includegraphics[width=.4\textwidth]{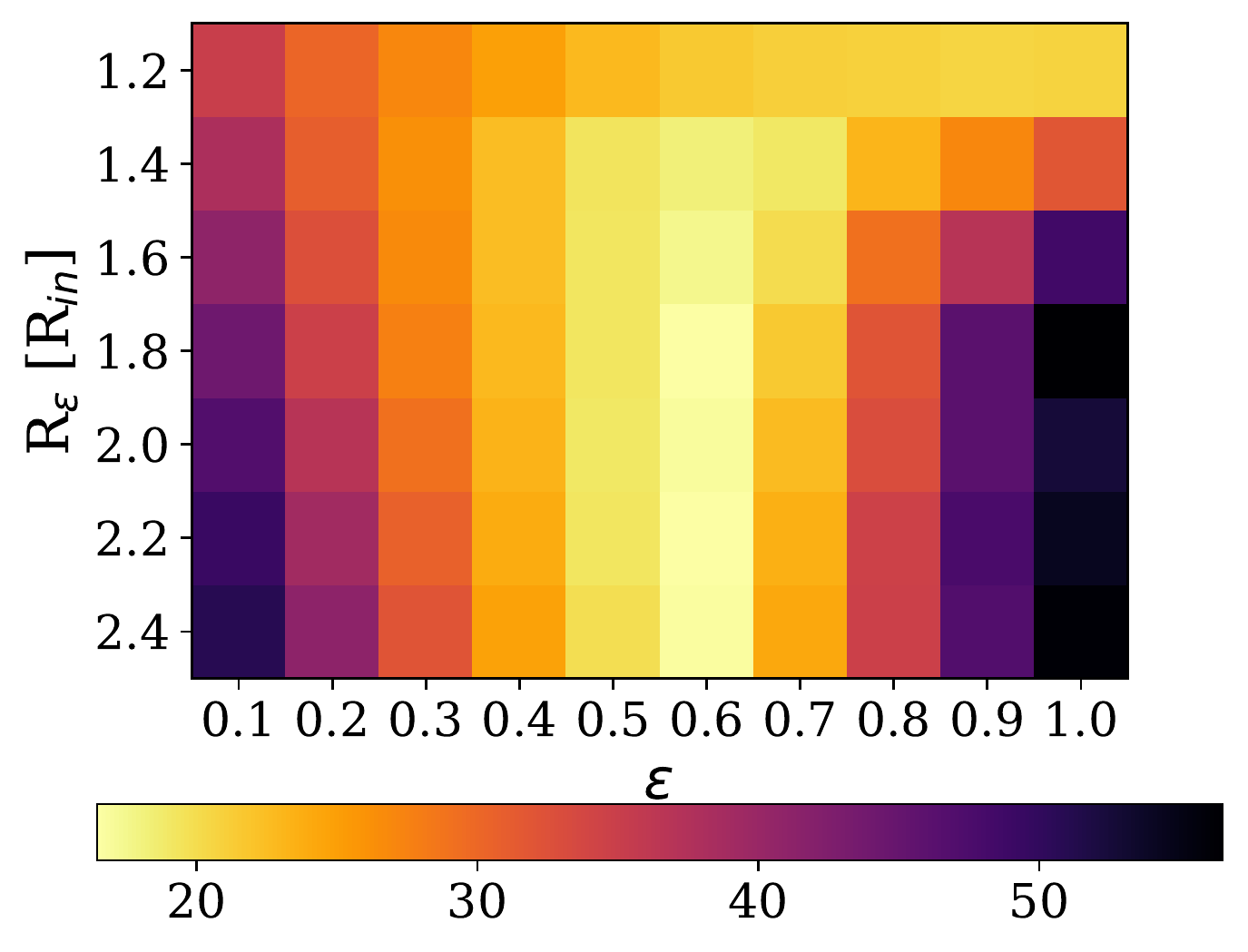}
    \caption{Model-observation closure phase residuals of $L$-band models (different versions of model 3) for different inner rim parameters $\varepsilon$ and $R_{\varepsilon}$ at the best-fit scaling factor 8.4  of the central star.}
    \label{fig:CPresidualgrid}
\end{figure}

Figure~\ref{fig:Lbandmodel.scale} shows the model-observation closure phase residuals of several versions of model 3 (i.e., $\rm R_{\epsilon}$--$\epsilon$ pairs with $\rm R_{\epsilon}$ = 1.8, 2.0, 2.2 and  $\epsilon$ = 0.5, 0.6, 0.7) as a function of the central star scaling factor. The best-fit models, with the smallest closure phase residuals of $16.5^\circ$, have a central star scaling factor of 8.4,  $\epsilon$ = 0.6, and $\rm R_{\epsilon}$ = 1.8 and 2.2.
Furthermore, the $\chi^2$ method presented in Appendix~\ref{chi2method} provides 
the same disk parameters $\epsilon$ and $\rm R_{\epsilon}$ as derived with the residual
method applied in this section.
Figures~\ref{fig:CPresidualgrid} and \ref{fig:CPchi2grid}  show the same  type of $\rm R_{\epsilon}$--$\epsilon$ plot as in Fig.~\ref{fig:diffmeangrid}, but now using the model-observation closure phase residuals or $\chi^2$ results, instead of the absolute mean differences between simulated and reconstructed images. The results of the three fits (in Figs.~\ref{fig:diffmeangrid}, \ref{fig:CPresidualgrid}, and \ref{fig:CPchi2grid}) are similar. They differ in the inner rim parameter value $\epsilon$ only, 0.6 in Figs.~\ref{fig:CPresidualgrid} and \ref{fig:CPchi2grid} instead of 0.7 in Fig.~\ref{fig:diffmeangrid}. In all models presented in this study, the dust disk temperature is below the sublimation temperature. Figures \ref{fig:ResidualsLbandMod51deg.scale} and \ref{fig:ResidualsLbandMod51degEnv.scale} suggest that model 3 can reproduce the disk structure slightly better than model 4 because the closure phase residuals  of model 3 at minimum scaling factor are smaller ($\sim20^\circ$; Fig.~\ref{fig:ResidualsLbandMod51deg.scale}) than the closure phase residuals of model 4 (Fig.~\ref{fig:ResidualsLbandMod51degEnv.scale}; $\sim25^\circ$).

\subsection{Influence of an additional gaseous B[e] star disk on the model brightness ratio between  the central object and  the dust disk of FS CMa}\label{subsec:gasdisk}

In the models discussed above, the model image of the central object is always too faint with respect to the dust disk.  If we assume that the central star is surrounded by an additional  compact (i.e., unresolved) gaseous disk, the model brightness ratio between central object (= star plus gaseous disk) and dust disk would change, as discussed below.
The theory of gaseous  envelopes and disks of Be stars is discussed in  \citetads{1984A&A...136...37L}, \citetads{1986A&A...162..121W},
 \citetads{1991MNRAS.250..432L},  \citetads{2011IAUS..272..325C}, \citetads{2013A&ARv..21...69R}, \citetads{2015MNRAS.454.2107V,2016ASPC..506..135V}.
 The gaseous disks of several Be and B[e] stars were studied with high spatial resolution using optical and infrared long-baseline interferometers  
 \citepads[e.g.,]{
 1986A&A...165L..13T,
 1989Natur.342..520M,
 1993ApJ...416L..25Q,
 1995A&A...300..219S,
 1997ApJ...479..477Q,
 1999A&A...345..203B,
 2006AJ....131.2710T,
 2007A&A...464...81D,
 2009A&A...505..687M,
 2011A&A...526A.107M,
 2011ApJ...729...17T,
 2012ApJ...744...19K
 }.
 
 Below we discuss that the  infrared continuum flux from a gaseous disk can be several times brighter  than the stellar flux, and the brightness of the gaseous disk depends on the wavelength and the viewing angle (e.g., it is brightest for face-on viewing direction; \citeads{2015MNRAS.454.2107V,2016ASPC..506..135V}). Because of this viewing angle dependence, a gaseous disk can change the brightness ratio between the central object (= star plus gaseous disk) and the dust disk of FS CMa.  A gaseous disk can much increase the flux from the central object seen by the observer (inclination angle $\sim$51$^\circ$). However, the illumination of the dust disk by the gaseous disk may be smaller if the gaseous and dust disk are aligned. 

The dependence of the intensity ratio of total flux to stellar flux   on the inclination angle, density, and wavelength  can be  calculated using the DISCO code (https://amhra.jmmc.fr/) based on the \citetads{2015MNRAS.454.2107V,2016ASPC..506..135V} model for classical Be star gaseous disks. The stellar parameters used are given in Tab.~\ref{tab:Parameters}. The value of the inclination angle  is assumed to be 51$^{\circ}$, as derived during the modeling process. The parameters of the gaseous disk model are given in Tab.~\ref{tab:GasDisk}. 

\begin{figure}
    \hspace{-8mm} \vspace{-5mm}
     \includegraphics[height=70mm,angle=0]{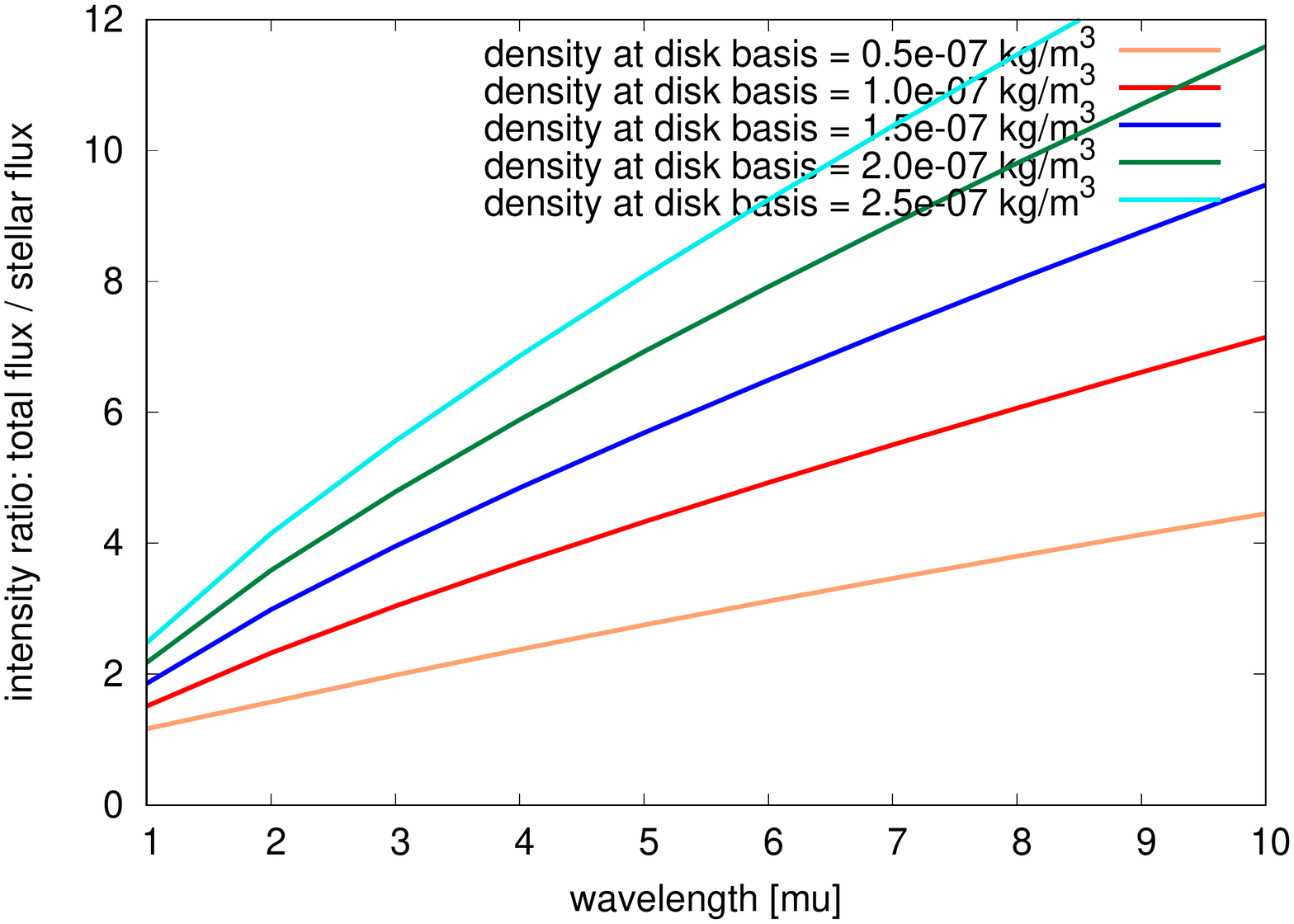}
      \caption[]{\small
              Wavelength dependence of the intensity ratio of the total flux (stellar plus gaseous disk) to the stellar flux for the inclination angle of 51$^{\circ}$ for different gas densities at the base of the gaseous disk. 
    }
    \label{fig:IntensityRatioDensities}
\end{figure}

\begin{figure}
    \hspace{-8mm} \vspace{-5mm}
     \includegraphics[height=70mm,angle=0]{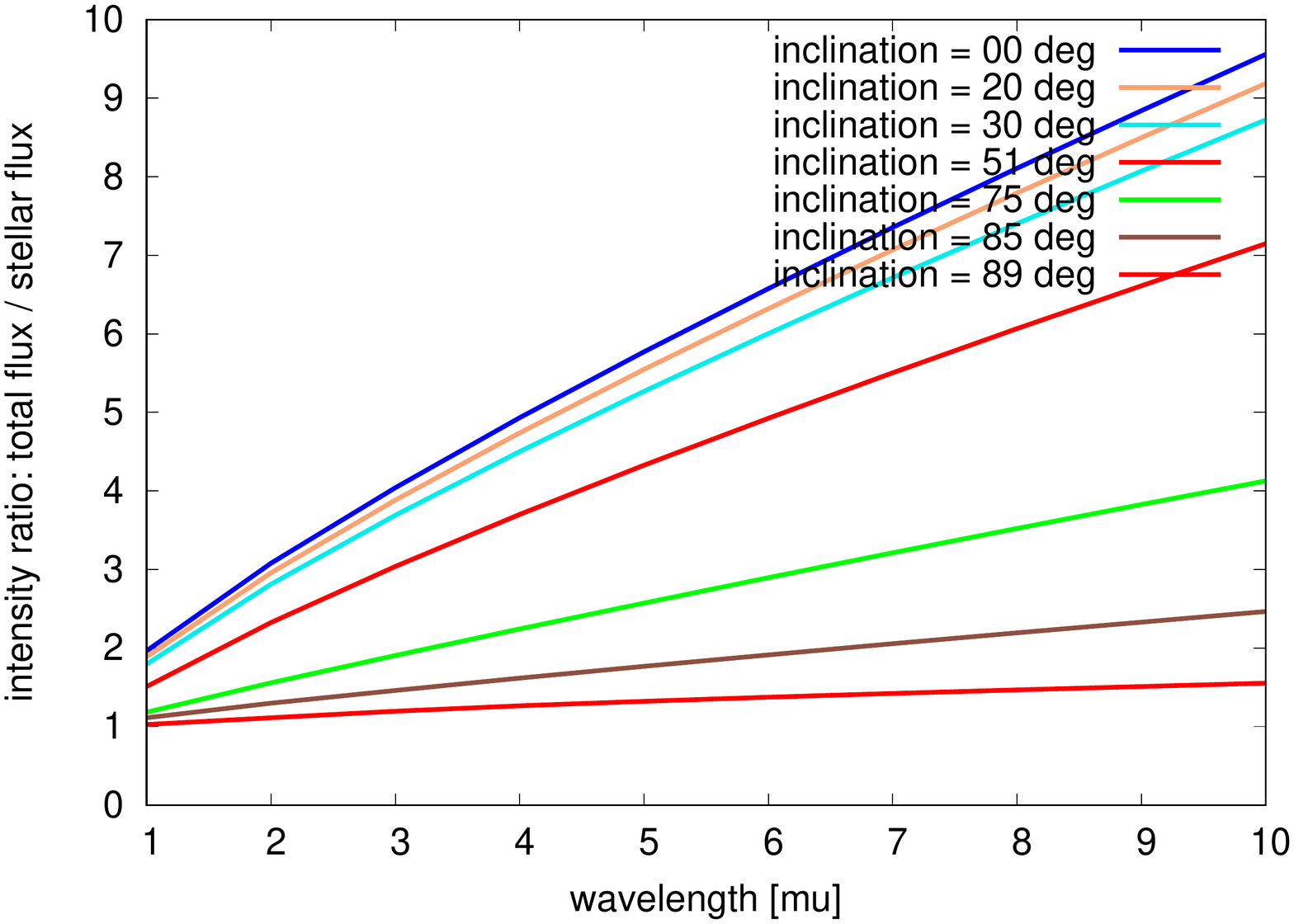}
      \caption[]{\small
             Wavelength dependence of the intensity ratio of total flux (stellar plus gaseous disk) to the  stellar flux for different inclination angles with a gas density at the base of the gaseous disk of 10$^{-7}$\,kg/m$^3$. 
  }
    \label{fig:IntensityRatioInclinations}
\end{figure}

Figure~\ref{fig:IntensityRatioDensities} shows the dependence of the model intensity ratio of total flux (i.e., stellar plus gaseous disk flux) to the stellar flux on wavelength for  different gas densities at the base of the gaseous disk and for the inclination angle of 51$^{\circ}$ of FS~CMa.
Figure~\ref{fig:IntensityRatioInclinations} shows the intensity ratio of the total flux (i.e., stellar plus gaseous disk flux) to the stellar flux as a function of the wavelength for different inclination angles for a gas density at the base of the gaseous disk of 10$^{-7}$\,kg/m$^3$. For example, at $L$-band wavelengths, the total flux is about 3--4 times larger than the stellar flux for the inclination range of 0--51$^{\circ}$ (0$^{\circ}$ corresponds to face-on) of the gaseous disk in contrast to larger inclination angles. This inclination angle dependence of the brightness of such a gaseous disk would have a strong influence on the FS~CMa model image if FS~CMa is surrounded by a bright gaseous disk and the gaseous disk is aligned with the dust disk. In this case, the gaseous disk is seen edge-on (i.e., low brightness) from viewing locations within the dust disk, whereas we see a bright 51$^{\circ}$ inclined disk. Therefore, such a gaseous disk may explain the difference between the modeled and observed flux ratio of the central object and the dust disk because a gaseous disk is not taken into account in our radiative transfer modeling.

\begin{table}[!ht]
  \centering
    \caption{Parameters of the modeled gaseous disk}
    \label{tab:GasDisk}
    \begin{tabular}{lc}
    \hline
    \hline
    \rule{0pt}{2ex}
      \textbf{Parameter}  & \textbf{Value}\\
      \hline
      \rule{0pt}{3ex}
      Disk outer radius ($R_{\odot}$)  & $50$\\
      \rule{0pt}{1ex}
      Disk temperature (K)   &  $9900$\\
      \rule{0pt}{1ex}
      Density at the disk base (kg/m$^3$)  &  $0.5 - 2.5\times\,10^{-7}$\\
      \rule{0pt}{1ex}
      Density power law exponent & $-3.5$\\
      \rule{0pt}{1ex}
      Power law coefficient of disk flaring  & $1.5$\\
      \hline
    \end{tabular}
\end{table}

\section{Summary and conclusions}    \label{summary}

FS~CMa belongs to the subgroup of unclassified B[e] stars. It is not known whether FS CMa is a young Herbig B[e] or a star in a later evolutionary stage.  \citetads{2020A&A...638A..21V} concluded that FS CMa is not a young star. FS CMa is famous for its large, inclined  circumstellar disk and spectacular bipolar outflow perpendicular to the disk plane.  

 We performed aperture-synthesis imaging of FS CMa with the new VLTI-MATISSE interferometry instrument and carried out radiative transfer modeling to interpret the images. The following results were obtained:

\begin{itemize}  
   
 \item We reconstructed the first aperture-synthesis $L$- and $N$-band images of the stellar disk and the central object. Image reconstruction was performed with the IRBis  method. The $L$- and $N$-band images have  resolutions of 2.7~mas and   6.6~mas, respectively.
 
\item  In the $L$-band image, the inner rim region  of the circumstellar dust disk and the central star can be seen. The size of the resolved inner dark disk cavity is $\sim$6$\times$12~mas. The resolved $L$-band disk consists of a bright  NW disk region and a much fainter SE one. The images suggest that we are looking at the bright  inner wall of the NW disk rim, while the SE rim is partially self-shadowed and appears therefore dimmer in the image. Therefore, the  large brightness of NW rim suggests that the NW rim is on the far side. The $N$-band image shows only the bright NW disk region.

\item  The reconstructed images provide the basis for a detailed analysis of the disk stucture, and thus the spatial distribution of the dust in the innermost disk region.
Performing radiative transfer simulations, we derived an inner disk radius of 5~au
and a disk inclination of $51^\circ$.
Most importantly, the good quality of the reconstructed images, in particular the high-resolution $L$-band image, allowed us to constrain the shape of the vertical structure of the inner few au of the inner disk. Using a parameterized disk model, the shape of the rim is described by two parameters characterizing a) the curvature of the disk rim 
and b) the radial range beyond which the density profile is described by the global disk model. In particular, we find a rather flat shape of the disk between the inner disk radius and 9~au.
Regardless of the specific evolutionary state of FS~CMa, this finding provides  direct constraints for further studies of the physical mechanisms resulting in this particular shape, that is, the interaction of the stellar radiation with the gas and dust disk surrounding the central star. However, a detailed analysis taking the specifics of FS~CMa into account is beyond the scope of this study.

\item While the relative brightness distribution can be well fitted, 
the flux ratio between the central object and the net flux of the disk can not be matched
on the basis of a star-to-dust disk model alone.
Depending on the specific model setup, characterized by the chemical composition and size distribution of the dust, even taking a hypothetical envelope with a low optical depth as a source of possible contamination into account,
the latter results in a star-to-disk flux ratio which is too low.  
All model images have a flux fraction in the central object of only $\sim$0.61–1.1\% of the total flux, whereas the flux fraction of the central object is $\sim$4.5\% in the observed $L$-band image. 
To explain this discrepancy, we suggest the possible additional contribution of a compact gaseous B[e] star disk in the vicinity of the star.
Taking the dependence of the brightness on the viewing angle into account, we find that this gaseous disk may be sufficiently bright to explain the observed flux ratio.
More detailed studies of FS~CMa are needed in the future to improve our understanding of its evolutionary stage, suspected companion, brightness variations, and physical properties of its suggested gaseous disk.

\end{itemize}

\begin{acknowledgements} 

MATISSE was designed, funded and built in close collaboration with ESO, by a consortium composed of institutes in France (J.-L. Lagrange Laboratory -- INSU-CNRS -- C\^ote d'Azur Observatory -- University of C\^ote d'Azur), Germany (MPIA, MPIfR and University of Kiel), the Netherlands (NOVA and University of Leiden), and Austria (University of
Vienna).  
We thank all ESO colleagues for the excellent collaboration.  
This work has been supported by the French government through the UCAJEDI Investments in the Future project managed by the National research Agency (ANR) with the reference number ANR-15-IDEX-01. 
The Konkoly Observatory and Cologne University have also provided some support in the manufacture of the instrument.  
K.O. acknowledges the support of the Agencia Nacional de Investigación y Desarrollo (ANID) through the FONDECYT Regular grant 1180066 and 1210652.
S.K. und A.K. acknowledge support from an STFC Consolidated Grant (ST/V000721/1) and ERC Starting Grant (Grant Agreement No. 639889). 
The research of J.V. and M.H. is supported by NOVA, the Netherlands Research School for
Astronomy. T.H. acknowledges support from the European Research Council under the Horizon 2020 Framework Program via the ERC Advanced Grant Origins 83 24 28. 
A.G. acknowledges support from the European Research Council (ERC) under the European Union’s Horizon 2020 research and innovation programme under grant agreement No 695099 (project CepBin). 
P.A. acknowledges support from the Hungarian NKFIH OTKA grant K132406, and from the European Research Council (ERC) under the European Union’s Horizon 2020 research and innovation programme under grant agreement No 716155 (SACCRED). 
J.S.B. acknowledges the support received from the UNAM PAPIIT project IA 101220 and from the CONACyT project 263975. 
AD thanks support from FAPESP \#2019/02029-2. 
FN acknowledges FAPESP for support through proc. 2017/18191-8.

This research has made use of the services of the ESO Science Archive Facility. 
This publication makes use of the SIMBAD database operated at CDS, Strasbourg, France.
This research has made use of the Jean-Marie Mariotti Center (JMMC) — MOIO AMHRA service at  https://amhra.jmmc.fr/.
 We thank the referee for helpful suggestions.

 \end{acknowledgements}

 \bibliographystyle{aa} 
 \bibliography{ref} 

 
\begin{appendix}

\section{Observations}    \label{obs}
\begin{figure}
 \centering
    \hspace*{-20mm} \vspace*{-0mm}
     \includegraphics[height=80mm,angle=0]{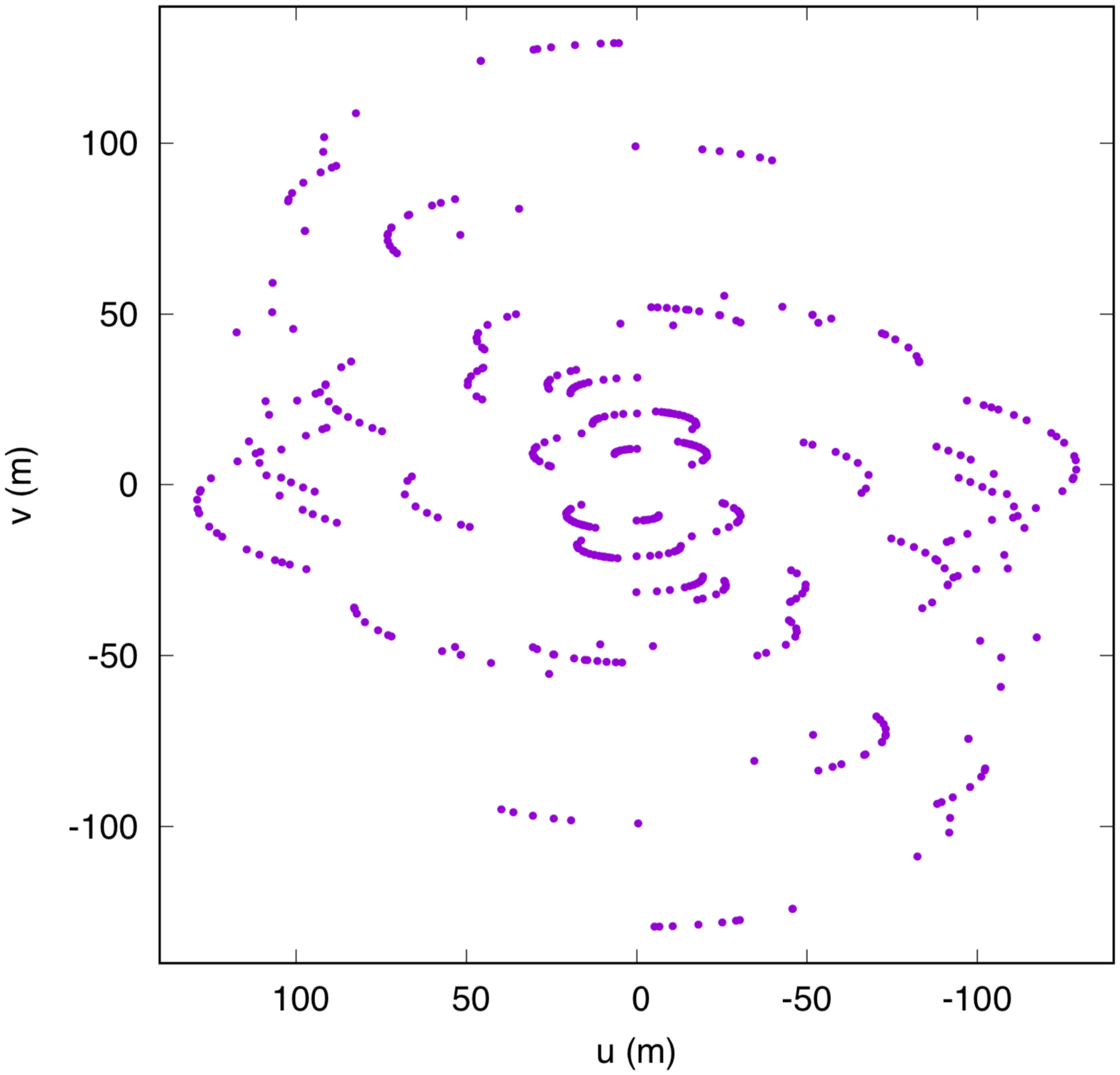}
      \caption[]{\small
              $uv$ coverage of the FS~CMa observations. 
    }
    \label{uv}
\end{figure}

The VLTI-MATISSE observations of FS~CMa are summerized in Table~\ref{listA1}. The $uv$ coverage is shown in Fig.~\ref{uv}. The calibrator stars used in this project are listed in Table~\ref{listA2}.

\begin{table*}                                                     
\normalsize
\caption{Summary of the VLTI-MATISSE observations of FS~CMa.}              
\label{listA1}                                                       
\centering                                              
\begin{tabular}{lrrrlrrr}                                
\hline\hline                                             
date            &n$\tablefootmark{a}$   &UT start$\tablefootmark{b}$ &UT  end$\tablefootmark{c}$    &array  &JD   &seeing$\tablefootmark{d}$    &$\tau_{0}$$\tablefootmark{e}$ \\         
\hline 
2018-12-02  &3                      &06:25                       &08:18                         &A0 B2 J2 C1  &2458454.277 &1.10                        &3.47\\
2018-12-03  &2                      &04:00                       &05:13                         &A0 B2 J2 C1  &2458455.178 &0.49                        &4.73\\
2018-12-05  &2                      &07:14                       &07:36                         &A0 B2 J2 C1  &2458457.308 &1.17                        &2.44\\
2018-12-06  &5                      &04:00                       &08:33                         &A0 B2 J2 C1  &2458458.333 &0.53                        &4.07\\
2018-12-07  &5                      &02:44                       &07:28                         &A0 B2 D0 C1  &2458459.170 &1.25                        &2.28\\
2018-12-08  &4                      &04:33                       &08:46                         &A0 B2 D0 C1  &2458460.296 &0.61                        &3.76\\
2018-12-09  &4                      &06:06                       &09:03                         &K0 B2 D0 J3  &2458461.314 &0.46                        &4.41\\
2018-12-10  &6                      &04:48                       &08:40                         &K0 G2 D0 J3  &2458462.219 &0.54                        &8.12\\
2018-12-11  &2                      &07:23                       &08:24                         &K0 G2 D0 J3  &2458463.313 &0.60                        &12.66\\
2018-12-12  &3                      &05:07                       &08:20                         &A0 G1 D0 J3  &2458464.290 &0.71                        &4.82\\
2018-12-13  &4                      &04:58                       &08:11                         &A0 G1 J2 J3  &2458465.254 &0.87                        &5.63\\
2018-12-14  &2                      &05:17                       &06:19                         &A0 G1 J2 J3  &2458466.226 &0.74                        &4.87\\
2018-12-15  &2                      &04:52                       &07:45                         &A0 G1 J2 K0  &2458467.332 &0.53                        &5.49\\
           \hline    \end{tabular}  
\tablefoot{
\tablefootmark{a} {Number of data sets recorded at different hour angles per night. }
\tablefootmark{b} {UT time at the start of the first data set of the night .}    
\tablefootmark{c} {UT time at the start of the last data set of the night .} 
\tablefootmark{d} {Average visible seeing ($"$) during the night .}
\tablefootmark{e} {Average coherence time $\tau_{0}$ (ms) in the visible during the night .}        
}   
\end{table*} 

 \begin{table}                                                     
\tiny
 \caption{Interferometric calibrator stars (JMMC Stellar Diameters Catalog - JSDC, Version 2; \citealt{2017yCat.2346....0B}).}              
 \label{listA2}                                                       
 \centering                                              
 \begin{tabular}{rrrr}
 \hline\hline                                             
 name            &spectral type  &UDD-L$\tablefootmark{a}$ &UDD-N$\tablefootmark{b}$ \\         
 \hline                                                                                     
 7 Cet            &M1III       &5.50\,$\pm$0.55\,mas                         &5.52\,$\pm$0.55\,mas \\
 $\epsilon$ Lep       &K4III        &5.87\,$\pm$0.56\,mas                         &5.92\,$\pm$0.56\,mas \\
 V407 Pup      &M2Ia/ab   &2.60\,$\pm$0.25\,mas                         &2.60\,$\pm$0.25\,mas \\
 $\beta$ Col    &K1III CN  &3.61\,$\pm$0.41\,mas                         &3.64\,$\pm$0.41\,mas \\
 $\pi$ Eri         &M1III        &5.28\,$\pm$0.47\,mas                         &5.30\,$\pm$0.47\,mas \\
 $\theta$ CMa  &K3/4III     & 4.00\,$\pm$0.39\,mas                        &4.03\,$\pm$0.39\,mas \\
            \hline    \end{tabular}  
 \tablefoot{
 \tablefootmark{a} {$L$-band uniform disk diameter. }
 \tablefootmark{b} {$N$-band uniform disk diameter.}           
 }   
 \end{table}

\section{Estimation of the visibility and closure phase errors and selection of the calibrated interferometric data}  \label{datared}
\subsection{$L$-band} \label{dataredL}

\begin{figure*}
 \centering
    \hspace*{-7mm} \vspace*{0mm}
     \includegraphics[height=230mm,angle=0]{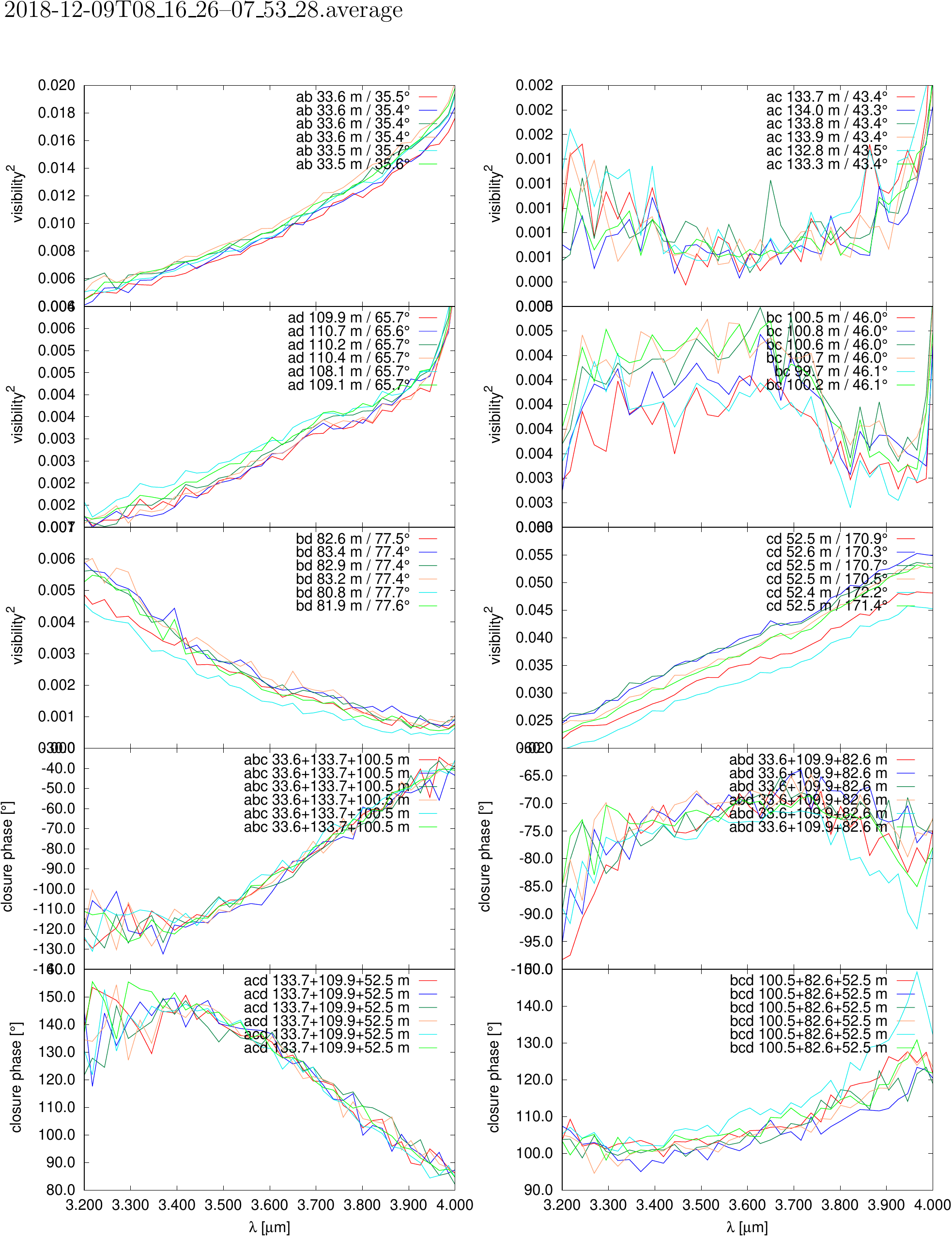}
      \caption[]{\small
              Examples of calibrated visibilities and CPs of FS~CMa $L$-band observation: all six interferometric data sets of an observation are shown (see Sect.\ref{datareductionL} for more details). 
    }
    \label{Lbanddata}
\end{figure*}

\begin{figure*}
 \centering
    \hspace*{-7mm} \vspace*{0mm}
     \includegraphics[height=91mm,angle=0]{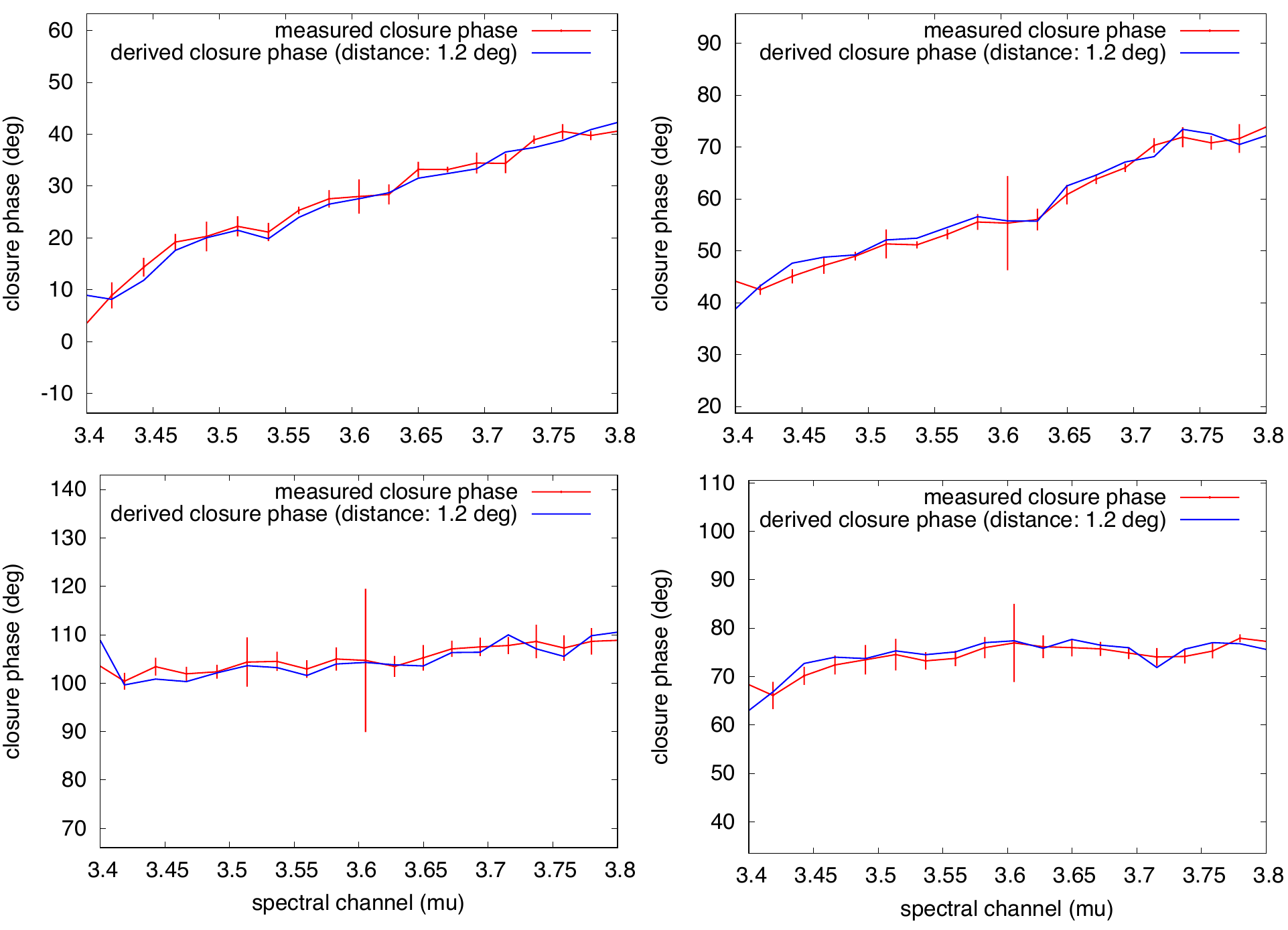}
      \caption[]{\small
              CP test:  good agreement between the measured CPs and the CPs calculated  by the other three CPs. 
    }
    \label{cptestgood}
\end{figure*}

\begin{figure*}
 \centering
    \hspace*{-7mm} \vspace*{0mm}
     \includegraphics[height=90mm,angle=0]{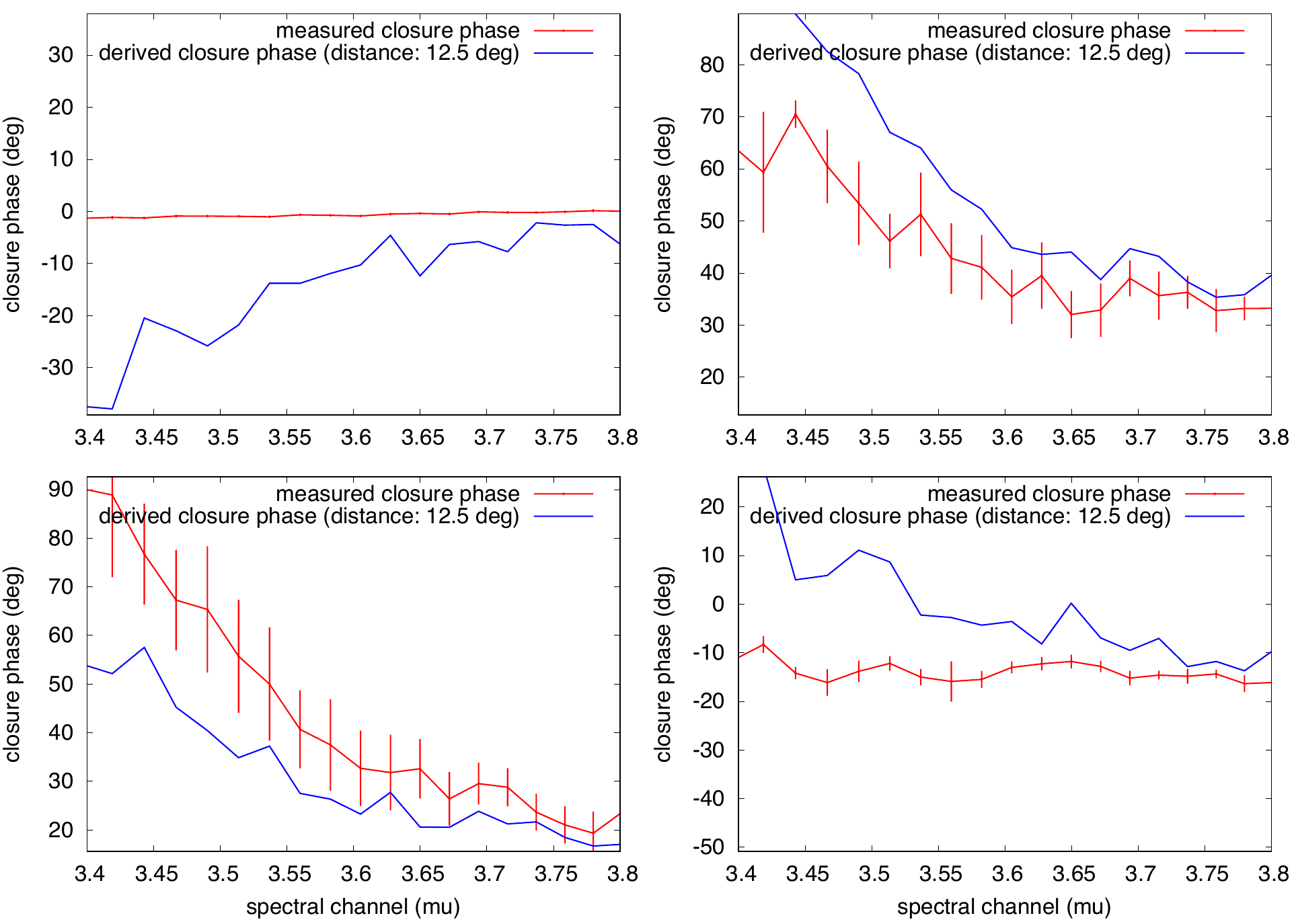}
      \caption[]{\small
              CP test:  strong mismatch between the measured CPs and the CPs  computed by the other three CPs. 
    }
    \label{cptestbad}
\end{figure*}

\begin{figure}
    \hspace{-17mm} \vspace{-5mm}
     \includegraphics[height=75mm,angle=0]{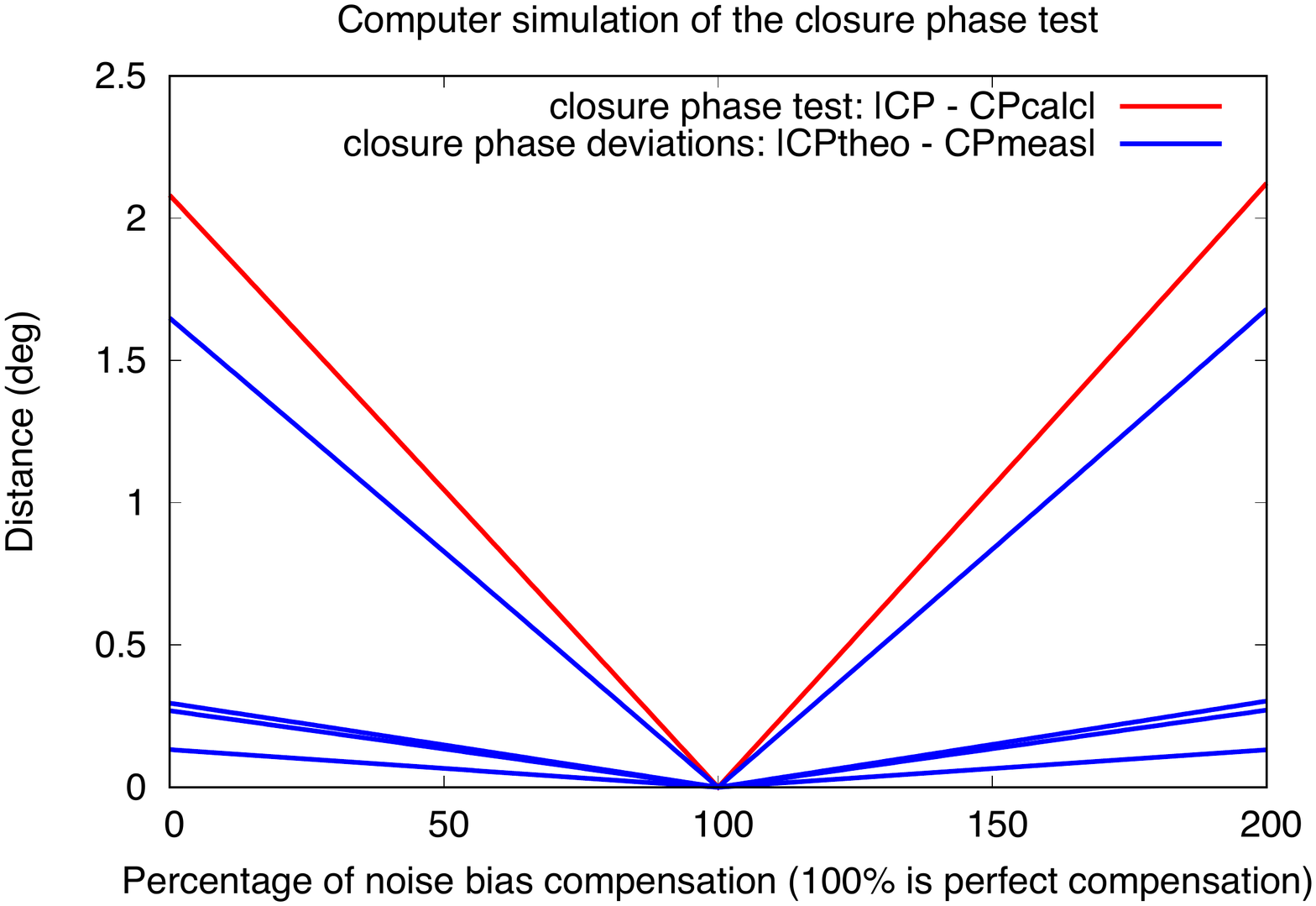}
      \caption[]{\small
              Computer simulation of the CP test, which shows the dependence of all CP distances on the percentage of noise bias compensation. The distance between the measured CP and the one derived from the three other CPs ( $\rm |CP - CPcalc|$; red line) is comparable to the  deviations between the four measured and theoretical CP ($\rm |CPtheo  - CPmeas|$; blue lines; see text in Appendix~\ref{dataredL} for details). 
    }
    \label{cptestsim}
\end{figure}

Figure~\ref{Lbanddata}  shows examples of our obtained calibrated visibilities and closure phases (CPs). All six data sets of an  observation discussed in Sect.~\ref{datareductionL}  are displayed.
To estimate the visibility and CP errors, we calculated the standard  deviations of the visibilities and CPs of the six calibrated interferometric data sets of each observation (see Sect.~\ref{datareductionL}) for each spectral channel. This is possible, because the six  calibrated interferometric data sets of each observation are recorded within about 15~min, and  therefore are covering nearly the same projected baseline vectors. The standard deviations were used as the common errors of the six  calibrated interferometric data sets of each observation.

To obtain a reliable data set for image reconstruction, we eliminated all CPs with errors >\,60$^{\circ}$ and all visibilities with a signal-to-noise ratio <\,0.5. In a second selection,
the reconstructed four CPs of each single observation were tested if they are affected by systematic errors. This test is called CP test and is discussed below in more detail. The four CPs measured simultaneously with a 4-telescope array are not independent, that is,  each one of the four CPs can be calculated by a linear combination of the three other CPs (see, for example, \citeads{2018ApJ...857...23C}). The distance between each measured CP and the corresponding one derived from the other three simultaneously measured CPs was calculated. This distance is a linear combination of the four CP deviations (e.g.,  the deviation between the measured CP and its true value).  In many cases this distance is small, that is, there is a good agreement between the measured CPs and the CPs  derived from the three other CPs. But there may be some observations where this distance is large. The reason for this could be, for example, a nonoptimal noise bias compensation in the average bispectrum. Figure~\ref{cptestgood}  shows an example observation, where the measured CPs and the CPs derived from the other three ones  are very similar with respect to their noise. In Fig.~\ref{cptestbad}  an observation is shown with a strong mismatch between the measured CPs and the derived CPs. In a second selection, we eliminated those observations where the mean deviation between the measured CPs and the derived CPs are larger than $\sim$\,8$^{\circ}$.

The feasibility of the aforementioned closure ~phase~test was investigated by a computer simulation.  The theoretical target was a 10~Jy source of similar structure, size, and sky position as FS~CMa.  The bispectrum of that target was simulated as if it were derived from data observed with the large AT configuration A0-G1-J2-K0 in $L$-band at low spectral resolution. The bispectrum was simulated with the noise bias terms caused by the noise of a 10~Jy source and  strong sky background, but without noise in the bispectrum.  Figure~\ref{cptestsim} shows such a closure~phase~test for the simulated bispectrum with different levels of  noise bias compensation (100\% corresponds to perfect compensation).  Because of the residulal noise bias, the reconstructed CPs  systematicaly deviate from their theoretical values. The deviations between the measured CPs and the theoretical ones is zero in the case of a perfect subtraction of the noise bias term. Figure~\ref{cptestsim} shows that the distance  between the measured CP and the same one derived from the three other CPs, as discussed above (e.g., the CP test, and denoted by $\rm |CP - CPcalc|$ in Fig.~\ref{cptestsim}) is comparable to the distances between the measured and theoretical (true) CPs  (denoted by $\rm |CPtheo  - CPmeas|$ in Fig.~\ref{cptestsim}). This experiment shows that systematic errors of the CP can roughly be determined by such a CP test.

\subsection{$N$-band} \label{dataredN}

\begin{figure*}
 \centering
    \hspace*{-7mm} \vspace*{0mm}
     \includegraphics[height=230mm,angle=0]{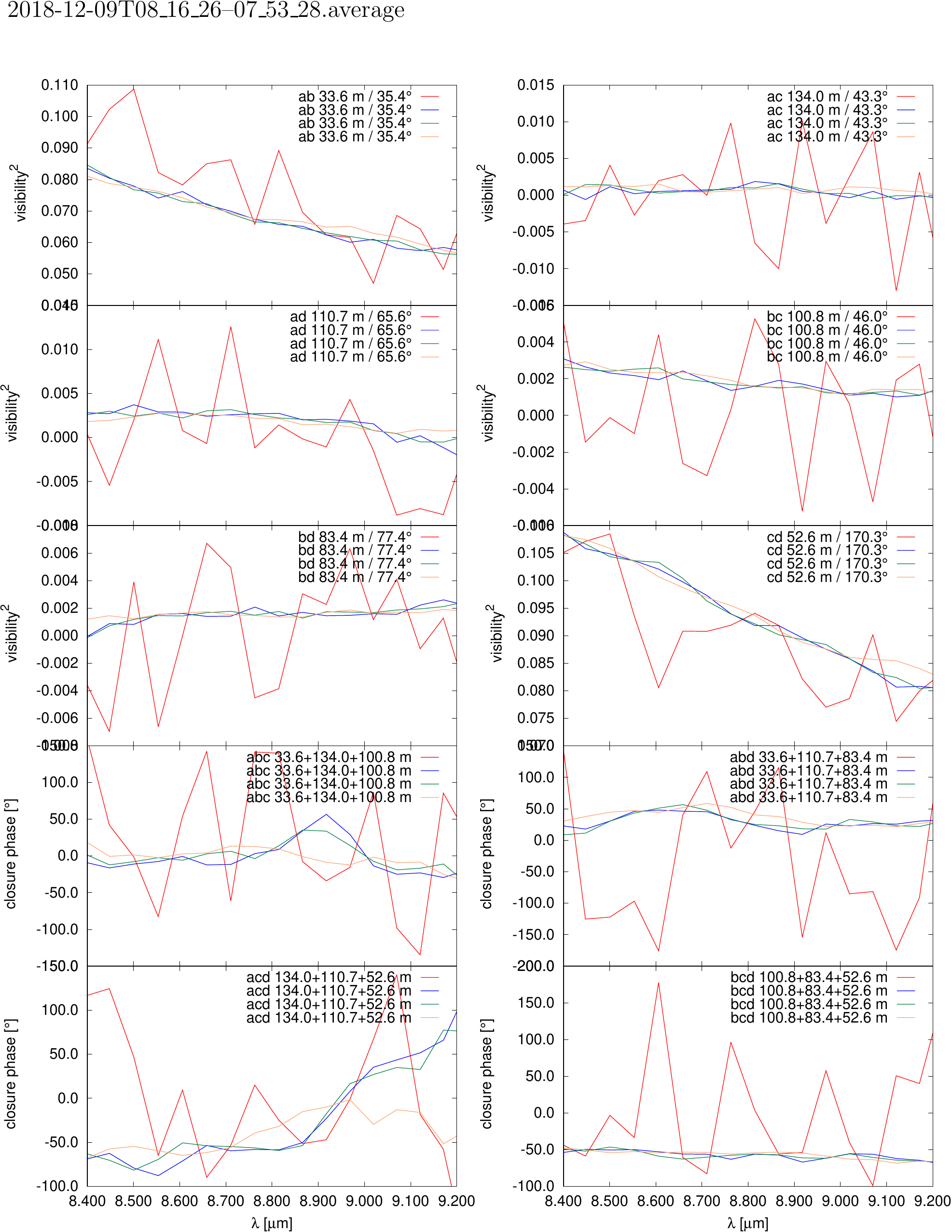}
      \caption[]{\small
              Example of calibrated visibilities and CPs of one $N$-band observation without and with spectral binning: red line denotes  data without spectral binning, and the blue, green and brown lines correspond to data  with spectral binnings of 9, 11 and 17, respectively.  Data processing without spectral binning (i.e., binning 1) is just shown for illustration, but is usually not used because of the severe noise (see Sect.\ref{datareductionN} for details).
    }
    \label{Nbanddata}
\end{figure*}

The visibility and CP errors for the calibrated interferometric $N$-band data, obtained with spectral binning of 11, were derived in the same way as in the $L$ band: the standard deviations derived from the four calibrated interferometric data sets (one for each of the four BCD configurations) of each observation (see Sect.~\ref{datareductionN}) were calculated for each spectral channel. The derived standard deviations were used as the common errors of the four interferometric data sets of each observation. 

The data set used for image reconstruction was selected in the same way as in the case of the $L$-band data. CPs with errors >~60$^{\circ}$ and visibilities (V$^2$) with signal-to-noise ratio <~0.5 were not used for the image reconstruction. In a second selection, the four CPs of each single observation were tested if they are affected by systematic errors. This was done with the CP test described in Appendix~\ref{dataredL} for the $L$-band data. In the second selection, we eliminated all observations where the mean deviation between the measured CPs and the derived ones are larger than $\sim$~1.7$\times$ the mean CP error of the corresponding observation. 

Figure~\ref{Nbanddata} shows examples of calibrated visibilities and CPs of a typical observation without and with different spectral binnings. The data without spectral binning (red lines) show very strong noise, which is strongly reduced by spectral binning (blue, green, and brown lines corrsponding to spectral binnings of 9, 11, and 17, respectively).

\section{Comparison of measured CPs and visibilities with the same ones derived from the reconstructions and from the models} \label{dataComparison}

\begin{figure*}
 \centering
    \hspace*{-7mm} \vspace*{0mm}
     \includegraphics[height=230mm,angle=0]{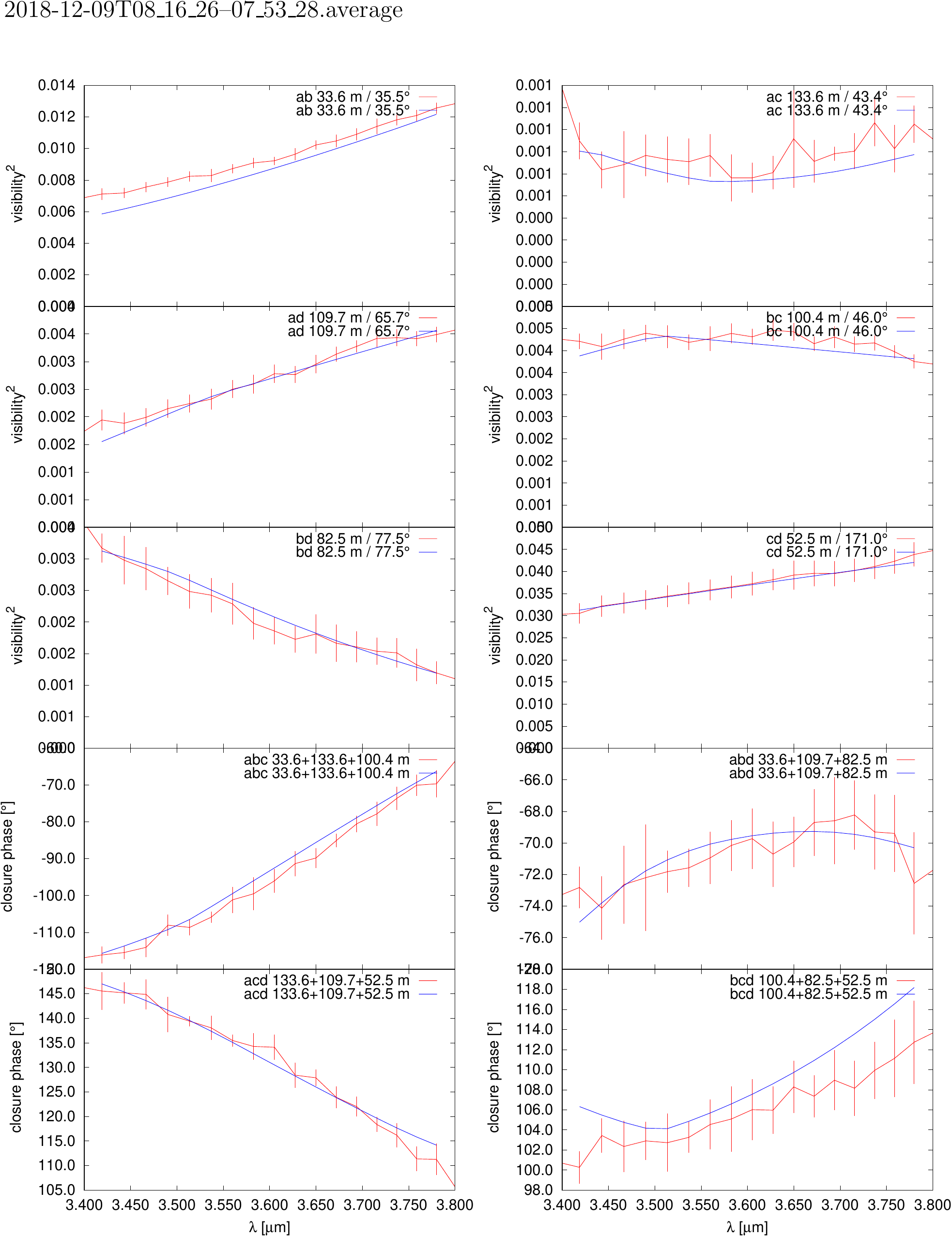}
      \caption[]{\small
              Comparison of measured  visibilities and CPs in $L$ band with those derived from the reconstructed image: (red) measured data,  and (blue) data derived from the reconstruction shown in Fig.~\ref{reconstructions} top. The data are averaged over all six data sets of each observation. 
    }
    \label{LbanddataRec}
\end{figure*}

\begin{figure*}
 \centering
    \hspace*{-7mm} \vspace*{0mm}
     \includegraphics[height=230mm,angle=0]{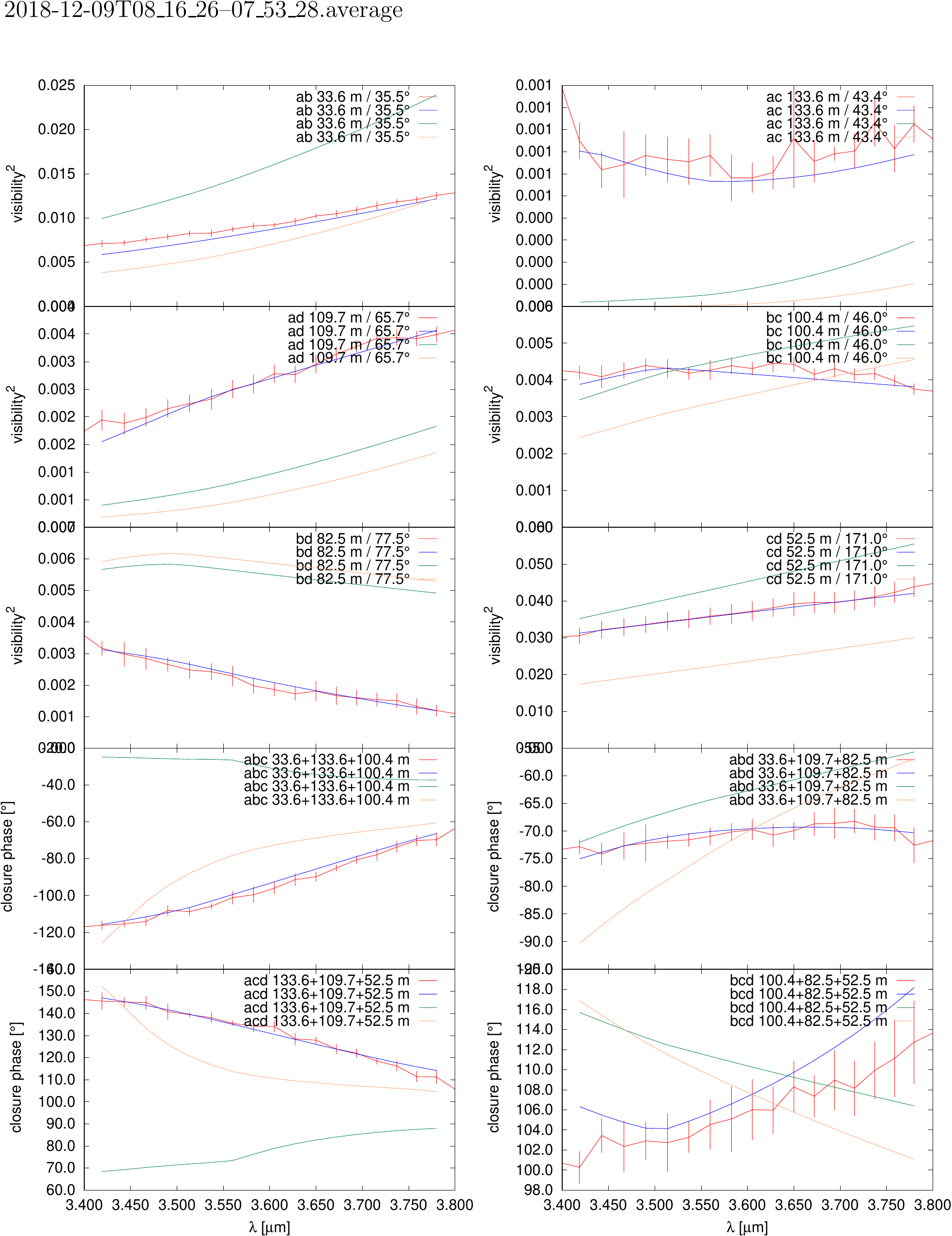}
      \caption[]{\small
              Comparison of various types of observed and model visibilities and CPs in the $L$ band: (red) measured data, (blue) data derived from the reconstruction shown in Fig~\ref{reconstructions} top, (green) data derived from the best-fit model 3 (Fig.~\ref{fig:bestfit_mod_51}, but with the central star scaled by factor 9.4), and (light brown) data from the best-fit model 4 (Fig.~\ref{fig:bestfit_mod_env}, but by scaling of the central star with factor 3.9). For more details see Sects.~\ref{subsec:cpvResiduals} and \ref{subsec:gasdisk}.The data are averaged over all six data sets of each observation. 
    }
    \label{LbanddataRecModel}
\end{figure*}

\begin{figure*}
 \centering
    \hspace*{-7mm} \vspace*{0mm}
     \includegraphics[height=230mm,angle=0]{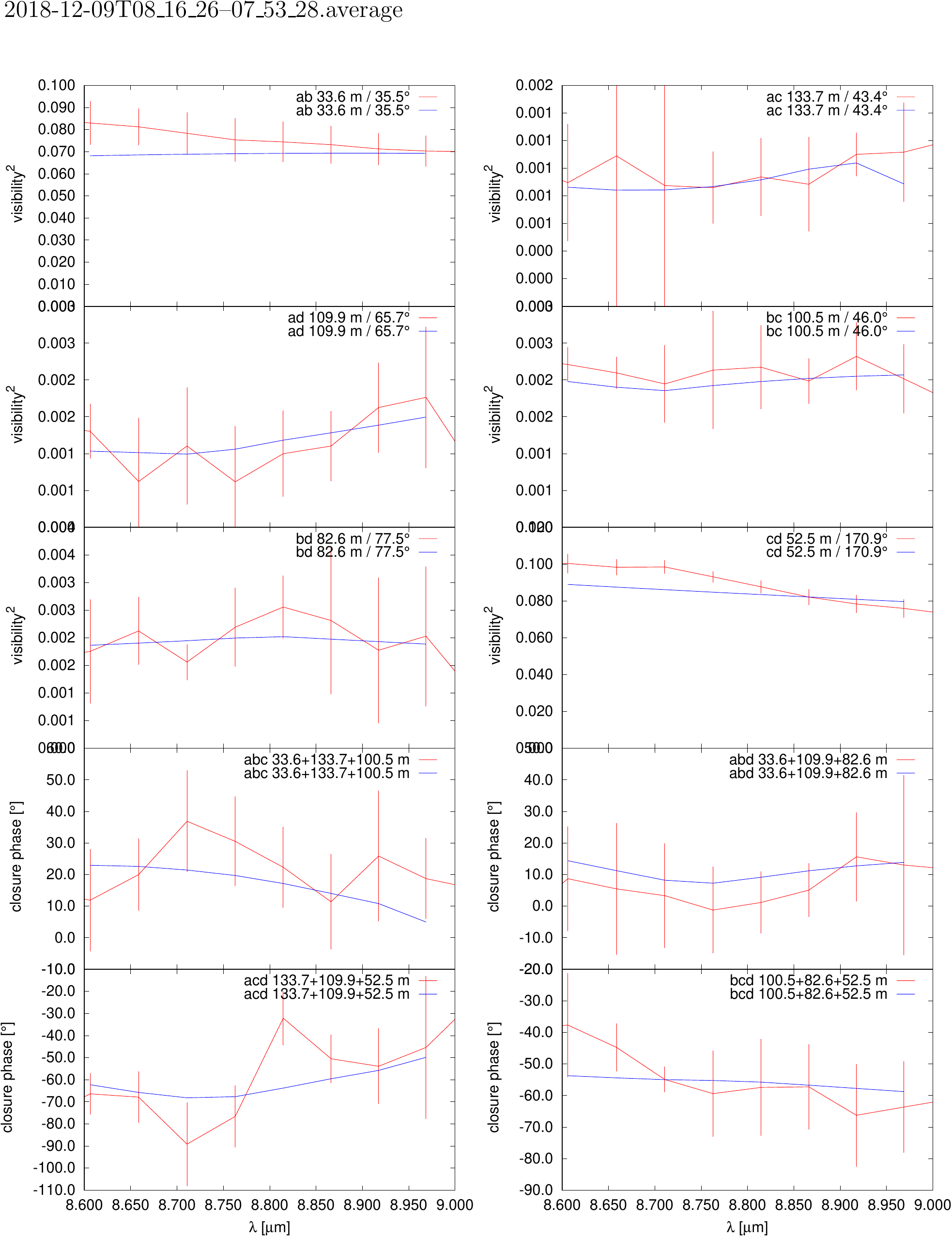}
      \caption[]{\small
              Comparison of measured visibilities and CPs in $N$ band (spectral binning 11) with those derived from the reconstructed image: (red) measured data, and (blue) data derived from the reconstruction shown in Fig.~\ref{reconstructions} bottom. The data are averaged over all four BCD configurations of each observation.  
    }
    \label{NbanddataRec}
\end{figure*}

\begin{figure*}
 \centering
    \hspace*{-7mm} \vspace*{0mm}
     \includegraphics[height=230mm,angle=0]{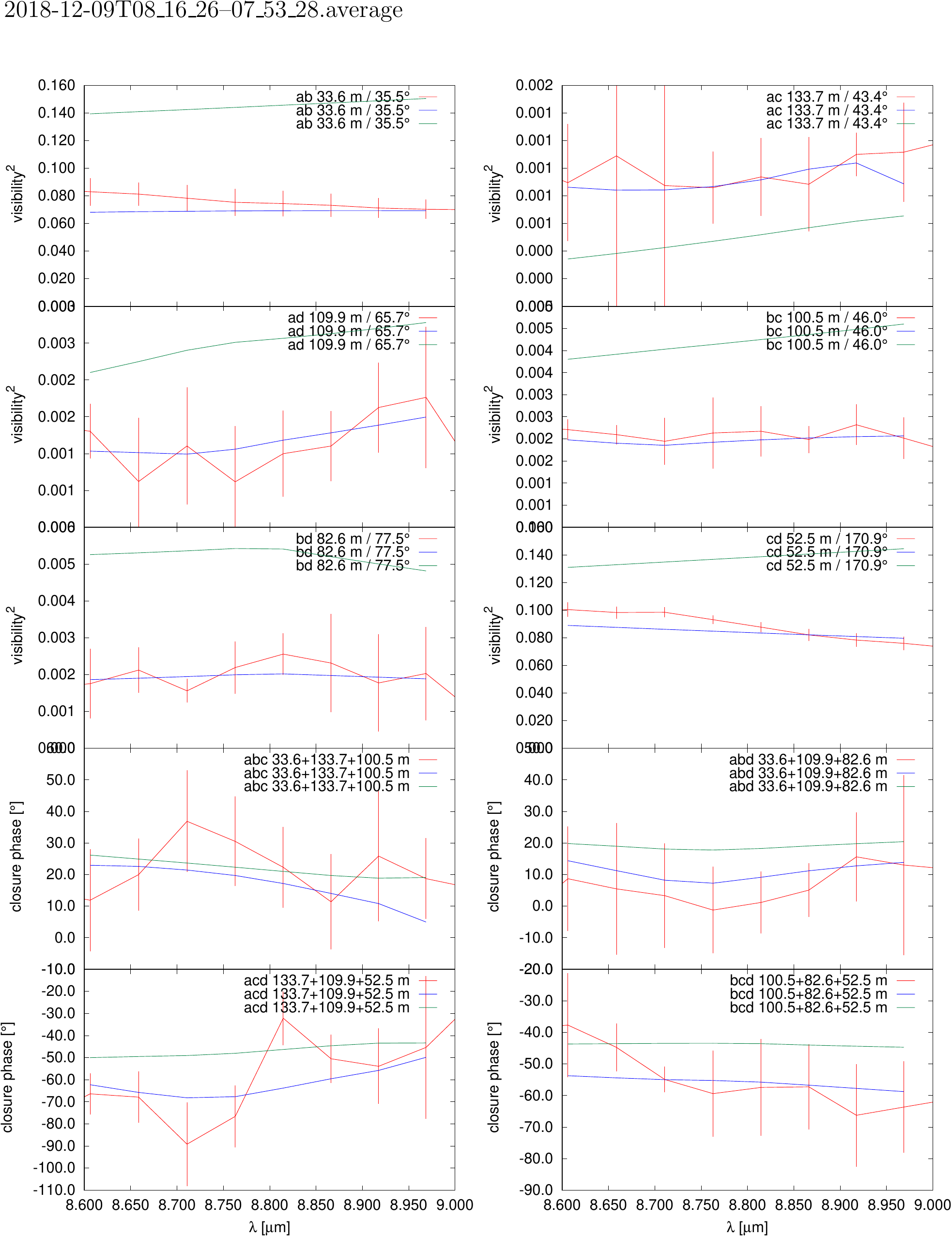}
      \caption[]{\small
              Comparison of various types of observed and model visibilities and CPs in the $N$ band (spectral binning 11):  (red) measured data, (blue) data derived from the reconstruction shown in Fig.~\ref{reconstructions} bottom, and (green) data derived from the best-fit $N$-band model 3 (Fig.~\ref{fig:bestfit_mod_nband}, but with the central star scaled by factor 9.2). The data are averaged over all four BCD configurations of each observation.  
    }
    \label{NbanddataRecModel}
\end{figure*}

\section{Image reconstruction}  \label{imarec}
The current version of the MATISSE image reconstruction package IRBis (mat$\_$cal$\_$imarec) consists of two optimization routines: a) a large-scale, bound-constrained nonlinear optimization routine called ASA$\_$CG (engine 1; \cite{hager2005},  \cite{hager2006}), and b) the new built-in optimization routine, a limited memory algorithm for bound-constrained optimization called L-BFGS-B (engine 2; \cite{byrd1995}, \cite{Zhu1997}).

The version of IRBis described in  \cite{2014A&A...565A..48H} contains two cost functions: $Q_{1}[o_k({\bf x})]$ (cost fct. 1) corresponding to the $\chi^2$ function of the bispectrum, and $Q_{2}[o_k({\bf x})]$ (cost fct. 2) corresponding to the $\chi^2$ function of the bispectrum phasors plus the bispectrum moduli. $o_k({\bf x})$ denotes the current iterated image and $\bf x$ is a two-dimensional vector in image space. The actual version of IRBis additionally contains a third cost function $Q_{3}[o_k({\bf x})]$ (cost fct. 3), which is the  $\chi^2$ function of the bispectrum phasors plus the squared visibilities, given by
\begin{eqnarray}
Q_3[o_k({\bf x})]  :=  \hspace{7cm} \nonumber \\
= \int \limits_{{\bf f_1},{\bf f_2} \in {\rm M}}   \frac{w_d({\bf f_1},{\bf f_2})}{\sigma^2_{\beta}({\bf f_1},{\bf f_2})}\cdot |\gamma_0\,\exp{[i\,\beta_k({\bf f_1},{\bf f_2})]} - \hspace{3.0cm} \nonumber \\
- \exp{[i\,\beta({\bf f_1},{\bf f_2})]}|^2\,d{\bf f_1}\,d{\bf f_2} + \hspace{1.9cm} \nonumber \\
+ f_0 \cdot \int \limits_{{\bf f} \in {\rm M_p}} \frac{w^p_d({\bf f})}{\sigma^2_p({\bf f})} \cdot | \gamma_0\, |O_k({\bf f})|^2 - |O({\bf f})|^2  |^2 \, d{\bf f}.  \hspace{2.8cm}
\label{equ:imagerec:equ6b}
\end{eqnarray}
Here, $\beta({\bf f_1},{\bf f_2})$ and $\beta_k({\bf f_1},{\bf f_2})$ denote the measured CPs and the CPs of the current iterated image, respectively.
The functions $\sigma_{\beta}({\bf f_1},{\bf f_2})$ and $\sigma_p({\bf f})$ are the errors of the bispectrum phasors and the squared visibilities, respectively. $O_k({\bf f})$ denotes the Fourier transform of the iterated image $o_k({\bf x})$. ${\bf f_1}$, ${\bf f_2}$, and  ${\bf f}$ are two-dimensional spatial frequency vectors. The functions $w_d({\bf f_1},{\bf f_2})$ and $w^p_d({\bf f})$ are the
$uv$ point density weights of the bispecrum and the squared visibilities $|O({\bf f})|^2$, respectively.
The factor $\gamma_0$ scales the bispectrum phasor and the squared visibilities of the iterated image to minimize the value of $Q_3[o_k({\bf x})]$.
The variable $f_0$ is the relative weight between the CP term and the squared visibility term in $Q_3[o_k({\bf x})]$.

\section{Estimation of the error maps of the reconstructed images}  \label{errormaps}
\begin{figure}
 \centering
    \hspace*{-0mm} \vspace*{-0mm}
     \includegraphics[height=225mm,angle=0]{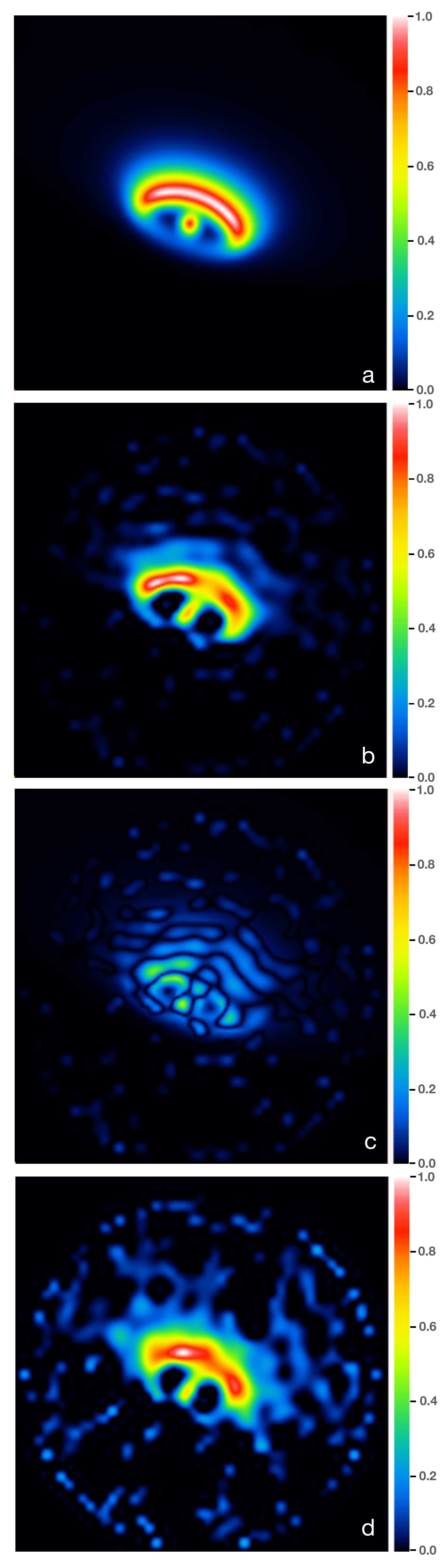}
      \caption[]{\small
              Estimation of the error map of the $L$-band reconstruction (Fig.~\ref{reconstructions} top) from synthetic data:  a) theoretical object used for generating the synthetic data (CPs and visibilities with the same noise and $uv$ coverage as in the measured data (see text for more details), b) image reconstructed from the synthetic data, c) derived error map, d) reconstruction shown in Fig.~\ref{reconstructions} (top). The FOV is 64~mas. North is up and east to the left.  
    }
    \label{Lerrormap}
\end{figure}

\begin{figure}
 \centering
    \hspace*{-0mm} \vspace*{-0mm}
     \includegraphics[height=225mm,angle=0]{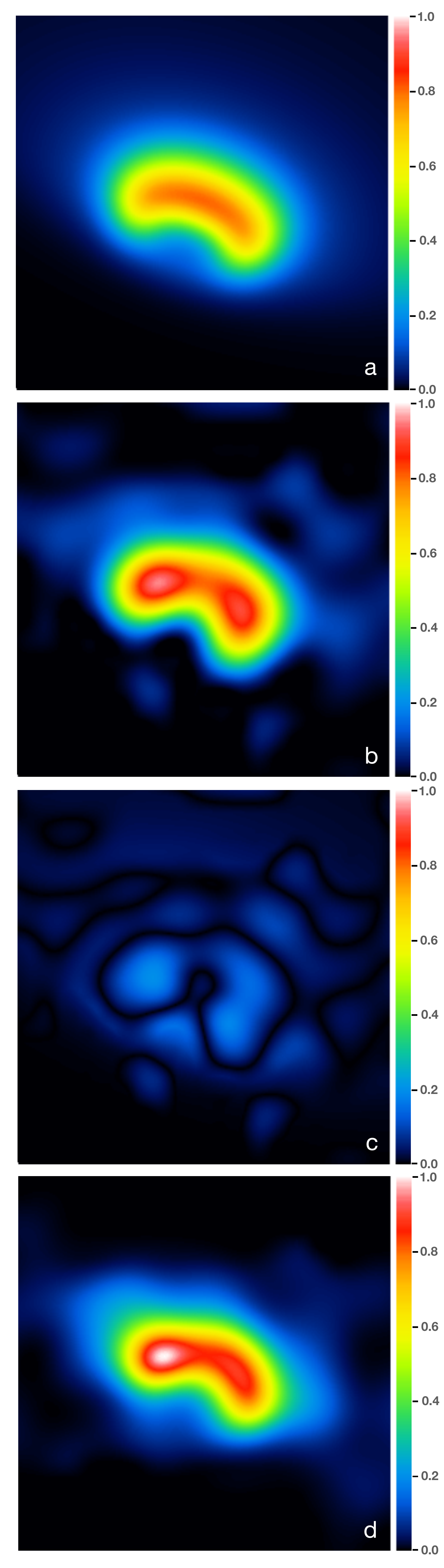}
      \caption[]{\small
              Estimation of the error map of the $N$-band reconstruction (Fig.~\ref{reconstructions} bottom) from synthetic data:  a) theoretical object, b) image reconstructed  from the synthetic data, c) derived error map, d) reconstruction shown in Fig.~\ref{reconstructions} (bottom). The FOV is 64~mas. North is up and east to the left. 
    }
    \label{Nerrormap}
\end{figure}

In this section, we describe the building of the error maps of the reconstructed $L$- and $N$-band images shown in Fig.~\ref{reconstructions}. For this goal, we performed computer simulations. In order to cover the noise introduced by the gaps
in the $uv$ coverage and by the noise in the measured data, we a) used the same $uv$ coverage as the one of the MATISSE observations, and b) degraded the computed visibilities and CPs
by simulated noise corresponding to the noise in the MATISSE observations, respectively.
As computer simulation targets, we have chosen the models similar to those shown in Figs.~\ref{fig:bestfit_mod_51} and \ref{fig:bestfit_mod_nband}, but with the central star intensity increased to the central star intensity of the reconstructed images in Fig.~\ref{reconstructions}.
The image reconstructions were performed with IRBis and the same parameters were used as for the reconstructed images shown in Fig.~\ref{reconstructions} and listed in Table~\ref{list1}.
The error map of each of the reconstructed images (Fig.~\ref{reconstructions}) is the absolute value of the difference between the computer simulation reconstruction and the computer simulation target.
Before calculating the difference, the reconstruction was a) scaled to the total intensity and
b) shifted to the center of the computer simulation target. To assign the reconstructions in Fig.~\ref{reconstructions} to the above produced error maps, each of the two reconstructions  was a) normalized to the same total intensity as the computer simulation target, and
b) shifted to the center of the computer simulation target.
Figures~\ref{Lerrormap} and \ref{Nerrormap} present the computer simulation experiments for deriving the error maps of the $L$- and $N$-band reconstructions in Fig.~\ref{reconstructions}, respectively. In all four images shown in Fig.~\ref{Lerrormap} (and \ref{Nerrormap}), the same intensity is coded with the same color.
Figures~\ref{Lerrormap}a and \ref{Nerrormap}a are the $L$- and $N$-band computer simulation targets, which are convolved with point spread functions corresponding to sizes of 2.7~mas and
6.6~mas, respectively. Figures~\ref{Lerrormap}b and \ref{Nerrormap}b present the reconstructed images obtained from the simulated data. Figures~\ref{Lerrormap}c and \ref{Nerrormap}c show the error maps for the $L$- and $N$-band reconstrcutions (Fig.~\ref{reconstructions}), derived from the computer simulation
targets (Figs.~\ref{Lerrormap}a and \ref{Nerrormap}a) and the computer simulation reconstructions (Figs.~\ref{Lerrormap}b and \ref{Nerrormap}b). Finally, Figures~\ref{Lerrormap}d and \ref{Nerrormap}d are the reconstructions shown in  Fig.~\ref{reconstructions}.

\section{Comparison of the measured CPs and visibilities with the ones derived from the model intensity distribution using the $\chi^2$ method} \label{chi2method}

The model-observation $\chi^2$ values (reduced $\chi^2$) of the CPs and visibilities of the $L$-band model image displayed in Fig.~\ref{fig:bestfit_mod_51} are 738 and 1163, respectively.
These big  $\chi^2$ values are mainly due to the fact that the model flux ratio of the central object and the disk disagree with the observations as discussed in Sect.~\ref{subsec:FluxRatio}. These
model-observation $\chi^2$ values are also very big in comparison to the reconstruction-observation $\chi^2$ values between the visibilities and CPs of the reconstructed
$L$-band image
(Fig.~\ref{reconstructions} top) and the observed visibilities and CPs. The CP $\chi^2$ value is 2.88 and the visibility $\chi^2$ value is 3.97.
The reconstruction-observation $\chi^2$ values of the reconstructed $N$-band image (Fig.~\ref{reconstructions} bottom) are 1.06 for the CPs and 4.94 for the visibilities.

\begin{figure}
    \hspace{0mm} \vspace{0mm}
     \includegraphics[height=90mm,angle=270]{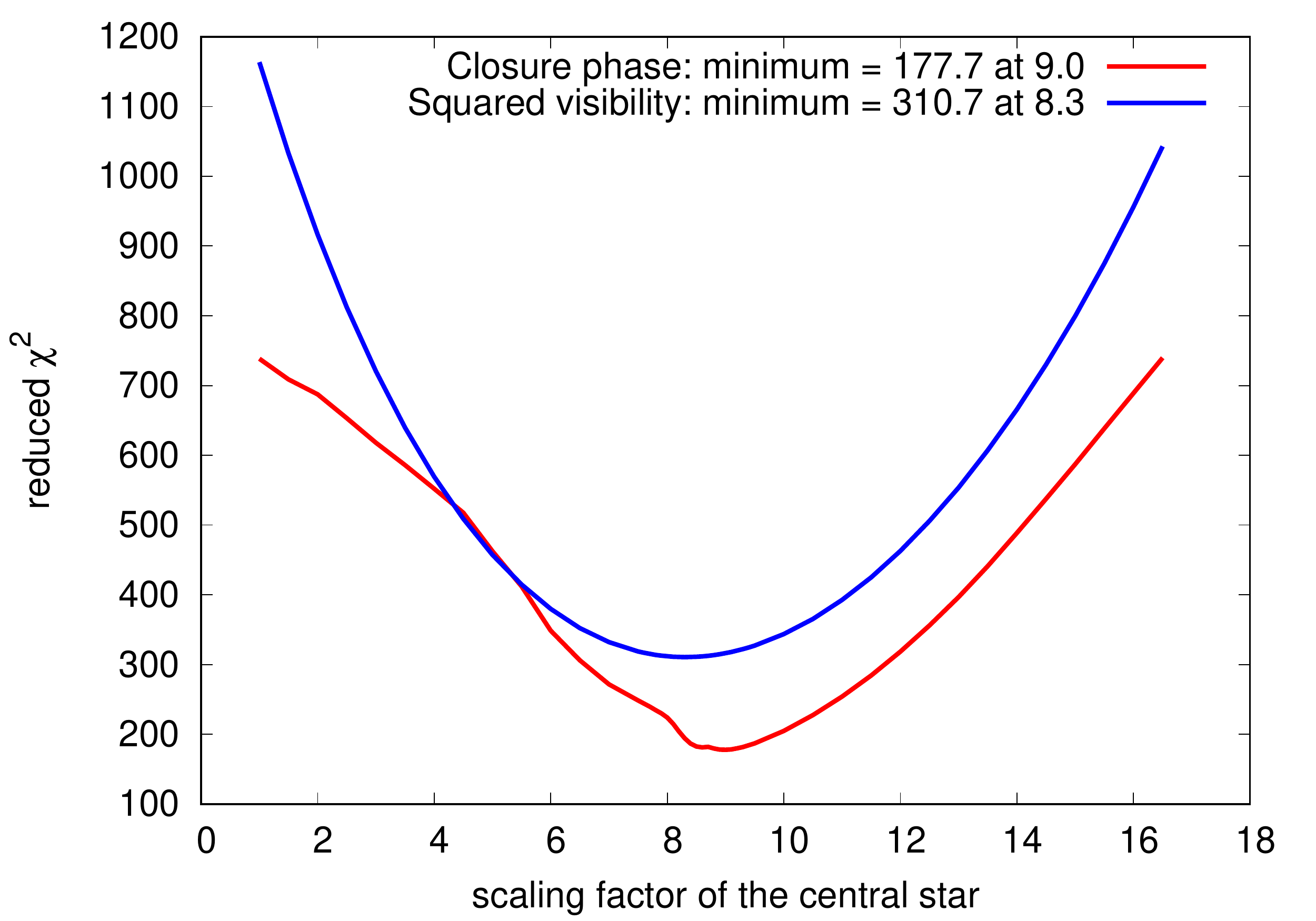}
      \caption[]{\small
               CP and visibility $\chi^2$ values of the best-fit $L$-band model 3 shown in Fig.~\ref{fig:bestfit_mod_51} as a function of the central star scaling factor.    }
    \label{fig:Chi2LbandMod51deg.scale}
\end{figure}

Figure~\ref{fig:Chi2LbandMod51deg.scale} shows the CP and visibility $\chi^2$ values of model 3 (Fig.~\ref{fig:bestfit_mod_51}) as a function of different
scaling factors of the model central star. The minimum CP $\chi^2$ value of 178 is achieved with a scaling factor of 9.0, which is similar to 9.4
obtained with the residual method (see Fig.~\ref{fig:ResidualsLbandMod51deg.scale}).

\begin{figure}
    \hspace{0mm} \vspace{0mm}
     \includegraphics[height=90mm,angle=270]{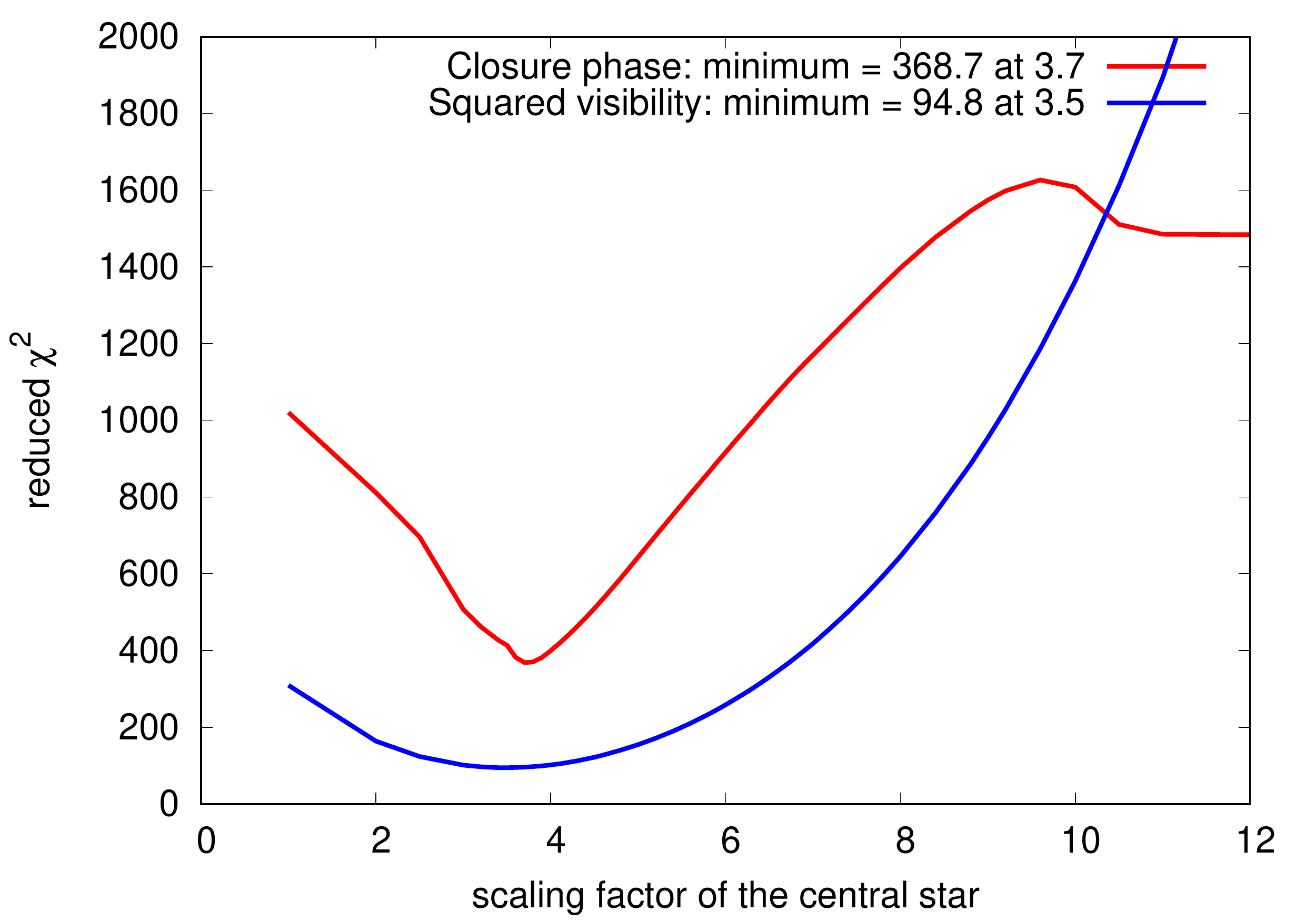}
      \caption[]{\small
              CP and visibility $\chi^2$ values of the modified best-fit $L$-band model 4 shown in Fig.~\ref{fig:bestfit_mod_env} as a function of the of the central star scaling factor.
    }
    \label{fig:Chi2LbandMod51degEnv.scale}
\end{figure}

Figure~\ref{fig:Chi2LbandMod51degEnv.scale} shows $\chi^2$ values
of the model 4 (Fig.~\ref{fig:bestfit_mod_env}) as a function of the scaling factors.
In this case the minimum CP $\chi^2$ value of 369 is obtained with a scaling factor of 3.7, which is similar to 3.9 derived with the residual method
(see Fig.\ref{fig:ResidualsLbandMod51degEnv.scale}).

\begin{figure}
    \hspace{0mm} \vspace{0mm}
     \includegraphics[height=90mm,angle=270]{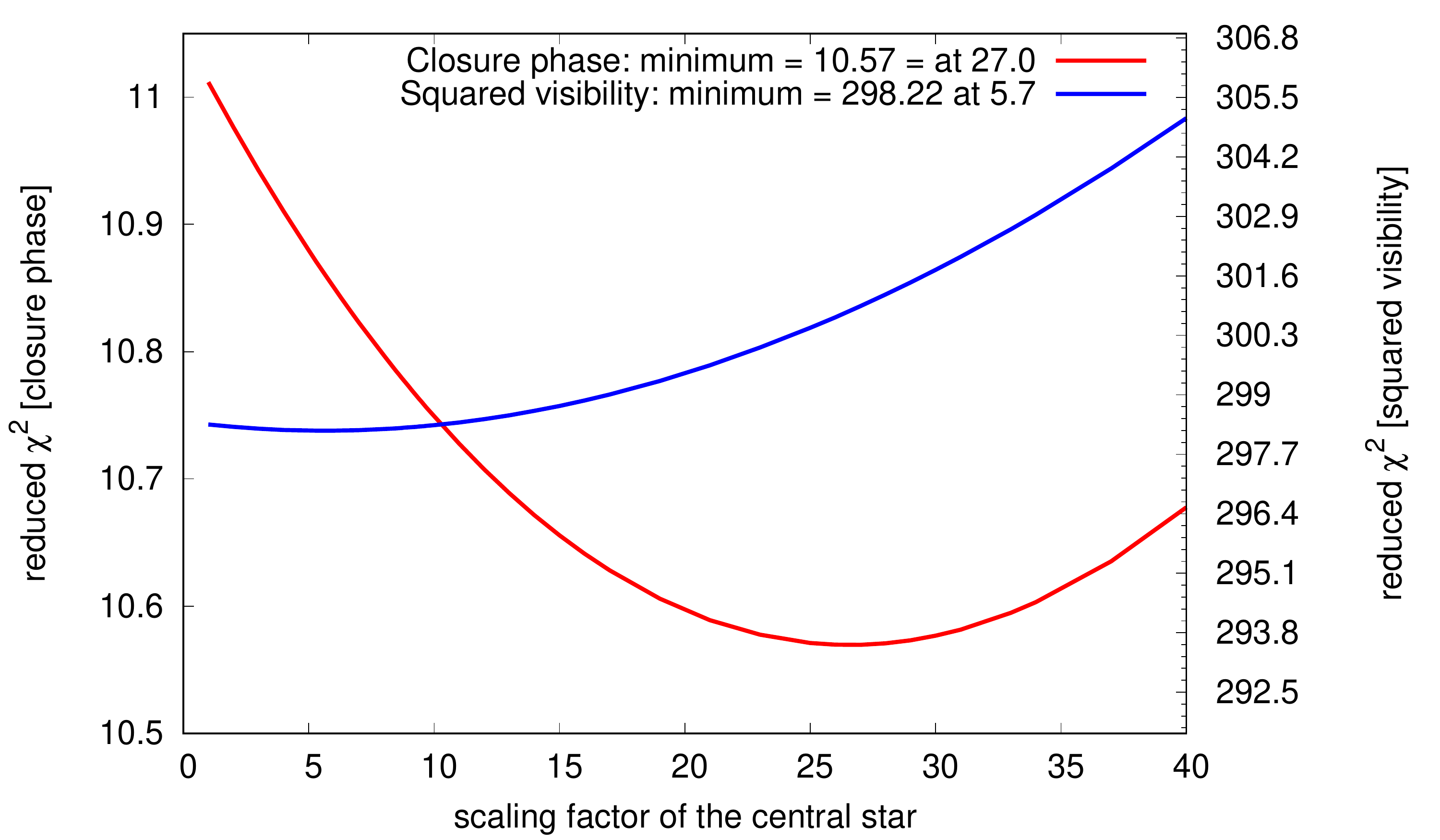}
      \caption[]{\small
             CP and visibility $\chi^2$ values of the modified best-fit $N$-band model shown in Fig.~\ref{fig:bestfit_mod_nband} as a function of the of the central star scaling factor.    }
    \label{fig:Chi2NbandMod51deg.scale}
\end{figure}

Figure~\ref{fig:Chi2NbandMod51deg.scale} presents the CP and visibility $\chi^2$ values of the $N$-band model image displayed in
Fig.~\ref{fig:bestfit_mod_nband} as a function of the different scaling factors of the model central star. The minimum CP $\chi^2$ value of 10.6 is obtained
with a scaling factor of 27, whereas the CP $\chi^2$ value in the case of no scaling is 11.0. The difference between scaling and no scaling of the
central model star is very small because of its very low flux (i.e., 0.03\% of the total flux of the best-fit $\it N$-band model). The very low flux of the central star is
responsible also for the different optimal scaling factors obtained with CP residuals (optimal scaling factor is equal to 9.2) and CP $\chi^2$ values (optimal
scaling factor is 27 in this case).

\begin{figure}
    \hspace{0mm} \vspace{0mm}
     \includegraphics[height=90mm,angle=270]{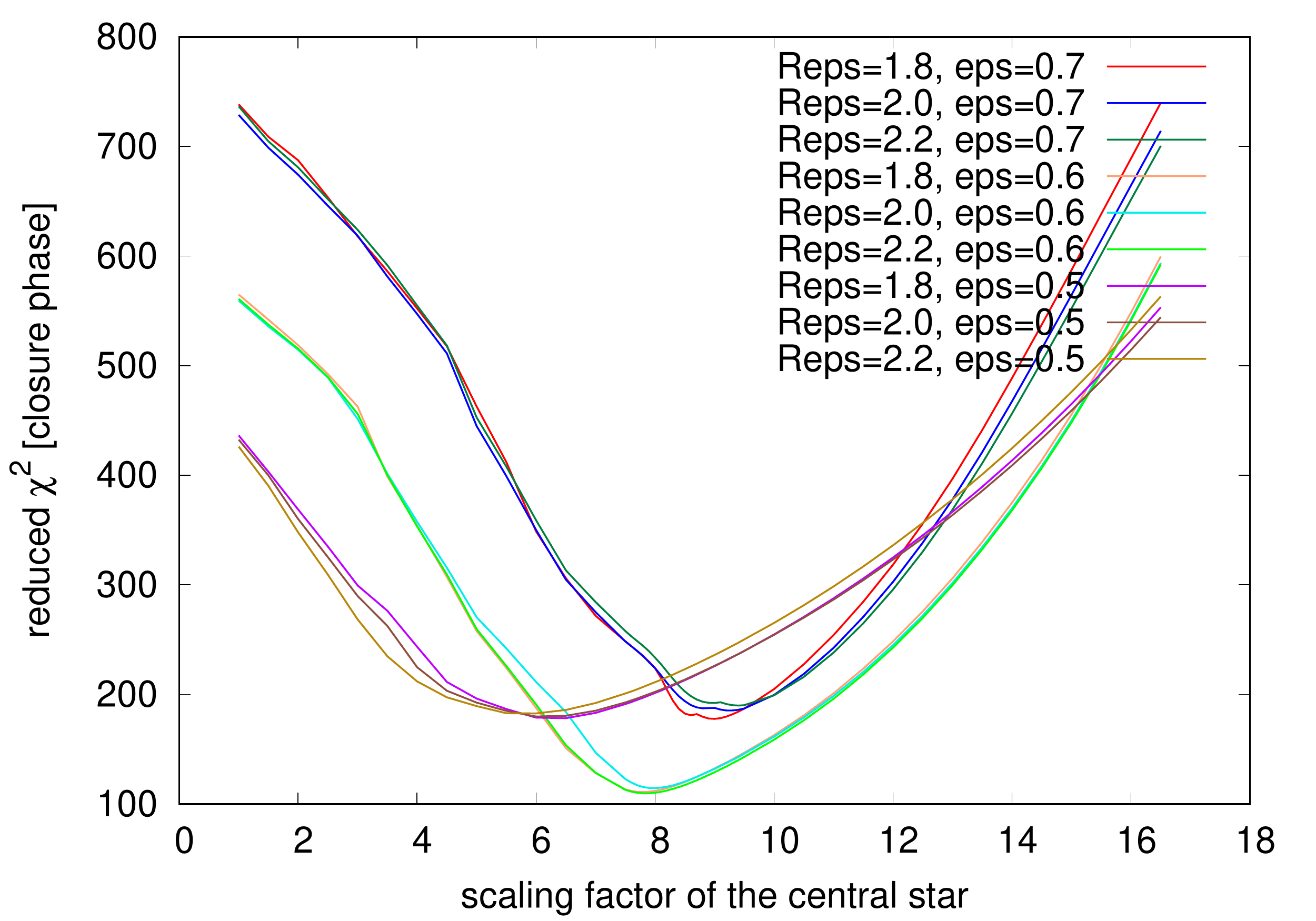}
      \caption[]{\small
               CP $\chi^2$ values of  $L$-band models (different versions of model 3) with different inner rim parameters $\varepsilon$ and $R_{\varepsilon}$  as function of the central star scaling factor. The best-fit models are those two with $\epsilon$ = 0.6 and $R_{\epsilon}$ = 1.8 and 2.2 $R_{\rm in}$ at the central star scaling factor 7.8. }
    \label{fig:Chi2Lbandmodel.scale}
\end{figure}

\begin{figure}[!ht]
    \centering
    \includegraphics[width=.4\textwidth]{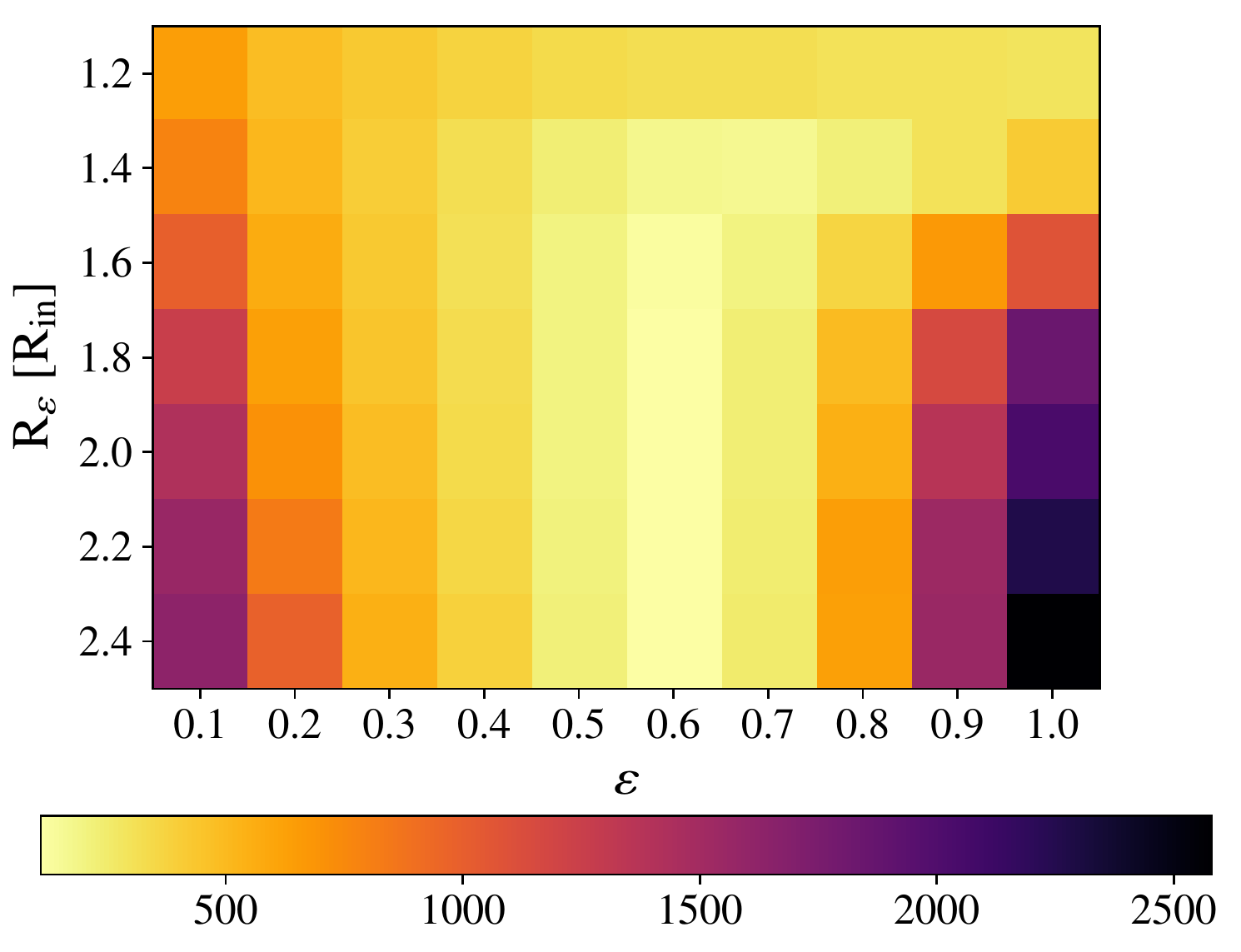}
    \caption{Model-observation CP $\chi^2$ values of $L$-band models (different versions of model 3) for different inner rim parameters $\varepsilon$ and $R_{\varepsilon}$ at the best-fit scaling factor 7.8  of the central star.}
    \label{fig:CPchi2grid}
\end{figure}

Figure~\ref{fig:Chi2Lbandmodel.scale} shows the model-obseravtion CP $\chi^2$ values for several versions of model 3
(i.e., $\rm R_{\epsilon}$--$\epsilon$ pairs with $\rm R_{\epsilon}$ = 1.8, 2.0, 2.2 and  $\epsilon$ = 0.5, 0.6, 0.7) as a function of the central star scaling factor.
The best-fit models, with the smallest CP $\chi^2$ values of 110 and 111, have a central star scaling factor of 7.8, which is similar to 8.4 obtained with
the residual method (see Fig.~\ref{fig:Lbandmodel.scale}), and $\epsilon$ = 0.6 and $\rm R_{\epsilon}$ = 1.8 and 2.2, which is identical to the results of the residual method used
in Sect.~\ref{subsec:cpvResiduals}. For this central star scaling factor of 7.8, Fig.~\ref{fig:CPchi2grid} shows the same type of $\rm R_{\epsilon}$--$\epsilon$ plot as in Fig.~\ref{fig:CPresidualgrid},
but using the model-observation CP $\chi^2$ values instead of the CP residuals. The results of the two fits provide, in both cases, best-fit models with
$\epsilon$ = 0.6 and $\rm R_{\epsilon}$ = 1.8 and 2.2.

 \end{appendix}  

\end{document}